\documentclass[12pt, letterpaper, preprint, comicneue]{aastex63}
\usepackage[T1]{fontenc}
\usepackage{color}
\usepackage{amsmath}
\usepackage{natbib}
\usepackage{ctable}
\usepackage{bm}
\usepackage[normalem]{ulem} 
\usepackage{xspace}
\usepackage{csvsimple} 

\usepackage{graphicx}
\usepackage{pgfkeys, pgfsys, pgfcalendar}

\let\oldbibliography\thebibliography 
\renewcommand{\thebibliography}[1]{%
  \oldbibliography{#1}%
  \setlength{\itemsep}{0pt}%
  \setlength{\parsep}{0pt}%
  \setlength{\parskip}{0pt}%
  \setlength{\bibsep}{0ex}
  \raggedright
}
\newcommand{\given}{\,|\,}

\newcommand{\bfi}[1]{\textbf{\textit{#1}}}

\newcommand{\eg}{\emph{e.g.}}

\newcommand{\foreign}[1]{\textsl{#1}}
\newcommand{\etal}{\foreign{et~al.}}

\newcommand{\bitem}{\begin{itemize}}
\newcommand{\eitem}{\end{itemize}}
\newcommand{\beq}{\begin{equation}}
\newcommand{\eeq}{\end{equation}}

\definecolor{orange}{rgb}{1,0.5,0}

\begin{document} \sloppy\sloppypar\frenchspacing 

\title{PROVABGS: The Probabilistic Stellar Mass Function of the BGS One-Percent
Survey} 
\suppressAffiliations

\newcounter{affilcounter}
\author[0000-0003-1197-0902]{ChangHoon Hahn}
\altaffiliation{changhoon.hahn@princeton.edu}
\affil{Department of Astrophysical Sciences, Princeton University, Peyton Hall, Princeton NJ 08544, USA} 
 
\author{Jessica Nicole Aguilar}
\affil{Lawrence Berkeley National Laboratory, 1 Cyclotron Road, Berkeley, CA 94720, USA}

\author[0000-0002-3757-6359]{Shadab Alam}
\affil{Tata Institute of Fundamental Research, Homi Bhabha Road, Mumbai 400005, India}

\author[0000-0001-6098-7247]{Steven Ahlen}
\affil{Physics Dept., Boston University, 590 Commonwealth Avenue, Boston, MA 02215, USA}

\author{David Brooks}
\affil{Department of Physics \& Astronomy, University College London, Gower Street, London, WC1E 6BT, UK}

\author[0000-0002-5954-7903]{Shaun Cole}
\affil{Institute for Computational Cosmology, Department of Physics, Durham University, South Road, Durham DH1 3LE, UK}

\author[0000-0002-1769-1640]{Axel de la Macorra}
\affil{Instituto de F\'{\i}sica, Universidad Nacional Aut\'{o}noma de M\'{e}xico,  Cd. de M\'{e}xico  C.P. 04510,  M\'{e}xico}

\author{Peter Doel}
\affil{Department of Physics \& Astronomy, University College London, Gower Street, London, WC1E 6BT, UK}

\author[0000-0002-3033-7312]{Andreu A. Font-Ribera}
\affil{Institut de F\'{i}sica d’Altes Energies (IFAE), The Barcelona Institute of Science and Technology, Campus UAB, 08193 Bellaterra Barcelona, Spain}

\author[0000-0002-2890-3725]{Jaime E. Forero-Romero}
\affil{Departamento de F\'isica, Universidad de los Andes, Cra. 1 No. 18A-10, Edificio Ip, CP 111711, Bogot\'a, Colombia}
\affil{Observatorio Astron\'omico, Universidad de los Andes, Cra. 1 No. 18A-10, Edificio H, CP 111711 Bogot\'a, Colombia}

\author[0000-0003-3142-233X]{Satya Gontcho A Gontcho}
\affil{Lawrence Berkeley National Laboratory, 1 Cyclotron Road, Berkeley, CA 94720, USA}

\author{Klaus Honscheid}
\affil{Center for Cosmology and AstroParticle Physics, The Ohio State University, 191 West Woodruff Avenue, Columbus, OH 43210, USA}
\affil{Department of Physics, The Ohio State University, 191 West Woodruff Avenue, Columbus, OH 43210, USA}

\author{Song Huang}
\affil{Department of Astronomy, Tsinghua University, 30 Shuangqing Road, Haidian District, Beijing, China, 100190}

\author[0000-0003-3510-7134]{Theodore Kisner}
\affil{Lawrence Berkeley National Laboratory, 1 Cyclotron Road, Berkeley, CA 94720, USA}

\author[0000-0001-6356-7424]{Anthony Kremin}
\affil{Lawrence Berkeley National Laboratory, 1 Cyclotron Road, Berkeley, CA 94720, USA}

\author[0000-0003-1838-8528]{Martin Landriau}
\affil{Lawrence Berkeley National Laboratory, 1 Cyclotron Road, Berkeley, CA 94720, USA}

\author[0000-0003-4962-8934]{Marc Manera}
\affil{Institut de F\'{i}sica d’Altes Energies (IFAE), The Barcelona Institute of Science and Technology, Campus UAB, 08193 Bellaterra Barcelona, Spain}

\author[0000-0002-1125-7384]{Aaron Meisner}
\affil{NSF's NOIRLab, 950 N. Cherry Ave., Tucson, AZ 85719, USA}

\author{Ramon Miquel} 
\affil{Instituci\'{o} Catalana de Recerca i Estudis Avan\c{c}ats, Passeig de Llu\'{\i}s Companys, 23, 08010 Barcelona, Spain}
\affil{Institut de F\'{i}sica d’Altes Energies (IFAE), The Barcelona Institute of Science and Technology, Campus UAB, 08193 Bellaterra Barcelona, Spain}

\author[0000-0002-2733-4559]{John Moustakas}
\affil{Department of Physics and Astronomy, Siena College, 515 Loudon Road, Loudonville, NY 12211, USA}

\author[0000-0001-6590-8122]{Jundan Nie}
\affil{National Astronomical Observatories, Chinese Academy of Sciences, A20 Datun Rd., Chaoyang District, Beijing, 100012, P.R. China}

\author{Claire Poppett}
\affil{Lawrence Berkeley National Laboratory, 1 Cyclotron Road, Berkeley, CA 94720, USA}
\affil{Space Sciences Laboratory, University of California, Berkeley, 7 Gauss Way, Berkeley, CA  94720, USA}
\affil{University of California, Berkeley, 110 Sproul Hall \#5800 Berkeley, CA 94720, USA}

\author{Graziano Rossi}
\affil{Department of Physics and Astronomy, Sejong University, Seoul, 143-747, Korea}

\author[0000-0003-4357-3450]{Am\'{e}lie Saintonge}
\affil{Department of Physics \& Astronomy, University College London, Gower Street, London, WC1E 6BT, UK}

\author[0000-0002-9646-8198]{Eusebio Sanchez}
\affil{CIEMAT, Avenida Complutense 40, E-28040 Madrid, Spain}

\author[0000-0002-0408-5633]{Christoph Saulder}
\affil{Korea Astronomy and Space Science Institute, 776, Daedeokdae-ro, Yuseong-gu, Daejeon 34055, Republic of Korea}

\author{Michael Schubnell}
\affil{Department of Physics, University of Michigan, Ann Arbor, MI 48109, USA}
\affil{University of Michigan, Ann Arbor, MI 48109, USA}

\author[0000-0002-6588-3508]{Hee-Jong Seo}
\affil{Department of Physics \& Astronomy, Ohio University, Athens, OH 45701, USA}

\author[0000-0002-2949-2155]{Małgorzata Siudek}
\affil{Institute of Space Sciences, ICE-CSIC, Campus UAB, Carrer de Can Magrans s/n, 08913 Bellaterra, Barcelona, Spain}

\author{Federico Speranza}
\affil{Department of Physics \& Astronomy, University College London, Gower Street, London, WC1E 6BT, UK}

\author[0000-0003-1704-0781]{Gregory~Tarl\'{e}}
\affil{University of Michigan, Ann Arbor, MI 48109, USA}

\author{Benjamin A. Weaver}
\affil{NSF's NOIRLab, 950 N. Cherry Ave., Tucson, AZ 85719, USA}

\author[0000-0003-2229-011X]{Risa H. Wechsler} 
\affil{Kavli Institute for Particle Astrophysics and Cosmology, Stanford University, Menlo Park, CA 94305, USA}
\affil{Physics Department, Stanford University, Stanford, CA 93405, USA}
\affil{SLAC National Accelerator Laboratory, Menlo Park, CA 94305, USA}

\author[0000-0002-5992-7586]{Sihan Yuan}
\affil{Kavli Institute for Particle Astrophysics and Cosmology, Stanford University, Menlo Park, CA 94305, USA}
\affil{SLAC National Accelerator Laboratory, Menlo Park, CA 94305, USA}

\author[0000-0002-4135-0977]{Zhimin Zhou}
\affil{National Astronomical Observatories, Chinese Academy of Sciences, A20 Datun Rd., Chaoyang District, Beijing, 100012, P.R. China}

\author[0000-0002-6684-3997]{Hu Zou}
\altaffiliation{The authors' affiliations are shown in Appendix \ref{sec:affil}.}
\affil{National Astronomical Observatories, Chinese Academy of Sciences, A20 Datun Rd., Chaoyang District, Beijing, 100012, P.R. China}

\begin{abstract}
    We present the probabilistic stellar mass function (pSMF) of galaxies in
    the DESI Bright Galaxy Survey (BGS), observed during the One-Percent
    Survey. 
    The One-Percent Survey was one of DESI's survey validation programs
    conducted from April to May 2021, before the start of the main survey. 
    It used the same target selection and similar observing strategy 
    as the main survey and successfully observed the spectra and redshifts of 
    143,017 galaxies in the $r < 19.5$ magnitude-limited BGS Bright sample and 95,499 
    galaxies in the fainter surface brightness and color selected BGS Faint 
    sample over $z < 0.6$.
    We derive pSMFs from posteriors of stellar mass, $M_*$, inferred from
    DESI photometry and spectroscopy using the \cite{hahn2022} PRObabilistic 
    Value-Added BGS (PROVABGS) Bayesian SED modeling framework. 
    We use a hierarchical population inference framework that statistically
    and rigorously propagates the $M_*$ uncertainties. 
    Furthermore, we include correction weights that account for the selection
    effects and incompleteness of the BGS observations. 
    We present the redshift evolution of the pSMF in BGS as well as the pSMFs
    of star-forming and quiescent galaxies classified using average specific
    star formation rates from PROVABGS. 
    Overall, the pSMFs show good agreement with previous stellar mass function 
    measurements in the literature. 
    Our pSMFs showcase the potential and statistical power of BGS, which in its
    main survey will observe >100$\times$ more galaxies.
    Moreover, we present the statistical framework for subsequent population 
    statistics measurements using BGS, which will characterize the global 
    galaxy population and scaling relations at low redshifts with unprecedented 
    precision. 
\end{abstract}


\section{Introduction} \label{sec:intro} 
The galaxy population can largely be characterized by a small number of scaling relations~\citep[see][for a review]{blanton2009}. This empirical network of scaling relations has been established through large galaxy surveys at all wavelengths, with major optical spectroscopic campaigns such as the Sloan Digital Sky
Survey~\citep[SDSS;][]{york2000}, Galaxy and Mass Assembly
survey~\citep[GAMA;][]{driver2011}, 
PRIsm MUlti-object Survey~\citep[PRIMUS;][]{coil2011}, 
and many others playing a crucial role.
Population statistics also encode important information, for instance the stellar mass function (SMF)  precisely characterizes the overall distribution of the stellar mass ($M_*$) of the galaxy population and its evolution across cosmic history ~\citep{li2009, marchesini2009, moustakas2013,
muzzin2013, leja2019a, driver2022}.


The relationship between the stellar masses and star formation rates (SFRs) of
galaxies reveal a bi-modality in the galaxy population with star-forming
galaxies lying on a tightly correlated ``star forming
sequence''~\citep[SFS;][]{noeske2007, daddi2007, salim2007, speagle2014, hahn2019}. 
The SFS evolves with redshift, with the SFR of galaxies at fixed mass steadily 
increasing from $z=0$ up to at least $z\sim 4$. 
This is also accompanied by a change in the fraction of quiescent galaxies caused 
by the quenching of star formation in some galaxies~\citep[\emph{e.g.}][]{ilbert2013,muzzin2013,santini2022}.
Additional scaling relations invoking quantities such as 
metallicity~\citep[\emph{e.g.}][]{tremonti2004,mannucci2010}, atomic and molecular gas 
mass~\citep[see reviews in][]{tacconi2020,saintonge2022}, and galaxy 
size~\citep[\emph{e.g.}][]{franx2008,vanderwel2014} have also been firmly established. 
These scaling relations and associated population statistics, along with 
careful measurements of their scatter and redshift evolution, are powerful 
tools to shed further light on galaxy formation and evolution. 

For one, they have the potential to reveal new trends among galaxies undetected
by previous observations and open new discovery space.
They can also be used to test galaxy formation models spanning 
empirical models~\citep[\emph{e.g.} {\sc UniverseMachine};][]{behroozi2019}, 
semi-analytic models~\citep[\emph{e.g.}][]{benson2012, henriques2015,
somerville2015a}, and 
hydrodynamical simulations~\citep[see][for a review]{somerville2015a}. 
Empirical models, for example, have been used to measure the timescale of star 
formation quenching~\citep{wetzel2013, hahn2017, tinker2017} or the dust content
of galaxies~\citep{hahn2021}. 

Furthermore, galaxy observables and their scaling relations can also be used to infer parameters that
dictate the physical processes in semi-analytic
models~\citep[\emph{e.g.}][]{henriques2009, lu2014, henriques2015}. 
Although full parameter exploration is currently computationally prohibitive 
for hydrodynamical simulations, they have been extensively compared to
observations~\citep[\emph{e.g.}][]{genel2014, dave2017a, trayford2017, dickey2021,donnari2021}.
Soon, machine learning techniques for accelerating and emulating simulations
will enable us to go beyond such comparisons and broadly explore parameter
space and galaxy formation
models~\citep[\emph{e.g.}][]{villaescusa-navarro2022a, jamieson2022}.
While many different approaches are available for expanding our understanding
of galaxies, they all require statistically powerful galaxy samples with
well-controlled systematics and well-understood selection functions. 

One survey that will provide galaxy samples with unprecedented statistical power
is the Dark Energy Spectroscopic
Instrument~\citep[DESI;][]{levi2013, desicollaboration2016, desicollaboration2016a,
abareshi2022}. 
Over its five-year operation, DESI will observe galaxy spectra using the 4-meter
Mayall telescope at Kitt Peak National Observatory with a focal plane filled
with 5000 robotically actuated fibers that direct the light to ten optical
spectrographs~\citep{schubnell2016, silber2023, miller2023}.
It will observe $\sim$40 million galaxy spectra over $360 < \lambda < 980$ nm
with spectral resolution of $2000 < \lambda/\Delta \lambda < 5500$ over 
${\sim}14,000~{\rm deg}^2$, a third of the sky.
In addition, DESI galaxies will also have photometry from the Legacy Imaging
Surveys Data Release 9~\citep[LS;][]{dey2019, schlegel2023}. 
The LS is a combination of three public projects (Dark Energy Camera Legacy Survey,
Beijing-Arizona Sky Survey~\citep{zou2017}, and Mayall $z$-band Legacy Survey) that jointly
imaged the DESI footprint in three optical bands ($g$, $r$, and $z$). 
DESI began observing its main survey on May 14, 2021. 

As part of its core observations, DESI is conducting the Bright Galaxy
Survey~\citep[BGS;][]{hahn2022c}.
BGS spans the same 14,000\,${\rm deg}^2$ footprint and will include low redshift
$z< 0.6$ galaxies that can be observed during bright time, when the night sky
is ${\sim}2.5\times$ brighter than nominal dark conditions.
BGS will provide two galaxy samples: the BGS Bright sample, a $r < 19.5$
magnitude-limited sample of ${\sim}10$ million galaxies, and the BGS Faint
sample, a fainter $19.5 < r < 20.175$ sample of ${\sim 5}$ million galaxies
selected using a surface brightness and color. 
The selection and completeness of the BGS samples are characterized in detail
in \cite{hahn2022c} (see also \citealt{myers2023}). 
Compared to the seminal SDSS main galaxy survey, BGS will provide a galaxy
sample two magnitudes deeper, over twice the sky, and double the median
redshift $z{\sim}0.2$. 
It will observe a broader range of galaxies than previous surveys  and provide
an opportunity to measure galaxy population statistics with unprecedented
precision.

BGS will also be accompanied by a value-added catalog: the Probabilistic
Value-Added BGS~\citep[PROVABGS;][]{hahn2022, kwon2022}.  
For every BGS galaxy, PROVABGS will provide physical properties including stellar
mass ($M_*$), average star formation rate ($\overline{\rm SFR}$), stellar
metallicity ($Z_*$), stellar age ($t_{\rm age}$), and dust content. 
These galaxy properties will be derived from state-of-the-art Spectral Energy
Distribution (SED) modeling of both DESI photometry and spectroscopy in a full
Bayesian inference framework. 
The SED model is designed to minimize model mis-specification by using highly
flexible non-parametric star formation and metallicity histories as well as a
flexible dust attenuation model.
Furthermore, the properties will be inferred using a fully Bayesian inference
framework and will thus provide statistically rigorous estimates of 
uncertainties and degeneracies among the properties.  
Ultimately, PROVABGS will provide consistently measured galaxy properties that
will enable analyses to take full advantage of the statistical power of BGS
with new techniques and approaches. 

A key application for PROVABGS will be measuring population statistics using 
a statistically rigorous methodology that correctly propagates the uncertainties
in galaxy property measurements. 
Current population statistics are by and large derived from simply binning
best-fit point estimates of galaxy properties. 
\cite{malz2020} demonstrated, in the context of inferring redshift distributions
from individual photometric redshift measurements, that using point estimates
is statistically incorrect and can lead to biased redshift distributions. 
Similarly, the point estimate approach can also lead to biased population 
statistics. 

Instead, we can estimate population statistics from combining individual
PROVABGS posteriors of galaxy properties using population inference in a
hierarchical Bayesian framework~\citep[\emph{e.g.}][]{hogg2010,
foreman-mackey2014, baronchelli2020}.
This approach correctly propagates the uncertainties in the galaxy properties 
from the individual posteriors of galaxies. 
As a result, they significantly improve the accuracy of population statistics
measurements and will enable more accurate measurements of key galaxy scaling
relations. 
In this work, we present the first such population statistics measurement for
BGS: the probabilistic stellar mass function (pSMF). 

In particular, we present the pSMF of BGS galaxies observed during the DESI
One-Percent Survey, a survey validation program conducted before the main
survey operations. 
We also present the statistical methodology for the population inference as
well as our methods for accounting for observational systematics and 
incompleteness. 
We begin in Section~\ref{sec:edr} with an overview of the BGS galaxies observed
during the DESI One-Percent Survey. 
Then, in Section~\ref{sec:provabgs}, we briefly summarize the PROVABGS SED
modeling framework used to infer the physical properties of the BGS galaxies.
Afterward, we present the pSMF inferred from the BGS observations in
Section~\ref{sec:results}. 
We summarize and discuss our results in Section~\ref{sec:summary}.
Throughout this work, we assume AB magnitudes and a flat $\Lambda$CDM cosmology
described by the final Planck results~\citep{planckcollaboration2014a}: 
$\Omega_m = 0.307$, $\Omega_b = 0.0483$, $H_0=67.8\,{\rm km\,s^{-1} Mpc^{-1}}$,
$A_s=2.19\times10^{-9}$, $n_s=0.9635$.

\section{The DESI Bright Galaxy Survey: One-Percent Survey}  \label{sec:edr}
DESI began its five years of operations in May 14, 2021~\citep{ops, kirbky2023}.
Before its start, DESI conducted the Survey Validation (SV) campaign to verify
that the survey will meets its scientific and performance
requirements. 
The SV campaign was divided into two main programs: the first, SV1,
characterized the survey's performance for different observing conditions and
was used to optimize sample selection. 
The second, the One-Percent Survey (or SV3), observed a dataset that can be
used for representative clustering measurements and deliver a ‘truth’ sample
with high completeness over an area at least 1\% of the expected main survey
footprint.
We refer readers to \cite{sv} and \cite{edr} for details on the DESI SV programs.
For details on how DESI targets are selected, we refer readers to
\cite{cooper2022, zhou2023, raichoor2023}, and \cite{chaussidon2023}\footnote{See also \cite{allendo2020, raichoor2020, ruiz2020, yeche2020, zhou2020} for details on preliminary target selection 
and \cite{vi, viqso} for the visual inspection of spectra used to inform target selection.}.
In this work, we focus on BGS galaxies observed during the One-Percent Survey.

The One-Percent Survey was observed on 38 nights from April 2021 to the end of 
May 2021.
During this time, DESI observed 288 bright time exposures that cover 214 BGS
`tiles', planned DESI pointings. 
The tiles were arranged so that a set of 11 overlapping tiles has their centers 
arranged around a 0.12 deg circle, forming a ‘rosette’ completeness pattern. 
In total, the One-Percent Survey observed 20 rosettes covering 180 
${\rm deg}^2$ spanning the northern galactic cap (see Figure 1 in
\citealt{hahn2022c}).

All BGS spectra observed during the One-Percent Survey are reduced using the
`fuji' version of the DESI spectroscopic data reduction
pipeline~\citep{guy2023}. 
First, spectra are extracted from the spectrograph CCDs using the 
{\em spectroperfectionism} algorithm of \cite{bolton2010}.
Then, fiber-to-fiber variations are corrected by flat-fielding and a sky model,
empirically derived from sky fibers, is subtracted from each spectrum.
Afterwards, the fluxes in the spectra are calibrated using stellar model fits
to standard stars. 
The final processed spectra is then derived by co-adding the calibrated spectra
across exposures of the same tile. 
In total, DESI observed spectra of 155,022 BGS Bright and 109,418 BGS Faint 
targets during the One-Percent Survey. 

For each spectrum, redshift is measured using 
{\em Redrock}\footnote{https://redrock.readthedocs.io}~\citep{redrock2023, redrockqso2023}, 
a redshift fitting algorithm that uses $\chi^2$ minimization computed from a
linear combination of Principal Component Analysis (PCA) basis spectral
templates in three template classes (``stellar'',  ``galaxy'', and ``quasar'').
{\em Redrock} also provides measures of redshift uncertainty, $\mathtt{ZERR}$
and redshift confidence, $\Delta\chi^2$, which corresponds to the difference
between the $\chi^2$ values of the best-fit model and the next best-fit model.
We restrict our sample to galaxy targets with reliable redshift measurements, 
as defined in \cite{hahn2022c} and \cite{sv}:
we only keep targets with spectra classified as galaxy spectra by 
{\em Redrock}, no {\em Redrock} warning flags, $\Delta\chi^2 > 40$,
and {\em Redrock} redshift uncertainty $\mathtt{ZERR} < 0.0005 (1 + z)$.
We also exclude any targets observed using malfunctioning fiber positioners.
Lastly, we impose a redshift range of $0 < z < 0.6$.  
After these cuts, our One-Percent Survey BGS sample includes 143,074 BGS Bright
galaxies and 96,771 BGS Faint galaxies.

\section{PROVABGS SED Modeling} \label{sec:provabgs}
For each BGS galaxy, we derive its $M_*$ and other physical properties, 
such as $\overline{\rm SFR}$, mass-weighted metallicity ($Z_{\rm MW}$), and 
mass-weighted stellar age ($t_{\rm age, MW}$), from DESI
photometry and spectroscopy using the PROVABGS SED modeling
framework~\citep{hahn2022}.  
PROVABGS models galaxy SEDs using stellar population synthesis with
a non-parametric star-formation history (SFH) with a starburst, a non-parametric
metallicity history (ZH) that evolves with time, and a flexible dust
attenuation prescription.
The non-parameteric SFH and ZH prescriptions are derived from SFHs and ZHs of
simulated galaxies in the Illustris hydrodynamic
simulation~\citep{vogelsberger2014, genel2014, nelson2015} and provide compact 
and flexible representations of SFHs and ZHs.
For the stellar population synthesis, PROVABGS uses the Flexible Stellar
Population Synthesis~\citep[FSPS;][]{conroy2009, conroy2010b} model with MIST
isochrones~\citep{paxton2011, paxton2013, paxton2015, choi2016, dotter2016},
\cite{chabrier2003} initial mass function (IMF), and a combination of
MILES~\citep{sanchez-blazquez2006} and BaSeL~\citep{lejeune1997, lejeune1998,
westera2002} spectral libraries.
The PROVABGS SED model excludes emission lines by masking the wavelength 
ranges of emission lines. 
For dust, PROVABGS uses the two component \cite{charlot2000}
attenuation model with birth cloud and diffuse dust components and 
does not include re-radiated dust emission.  

Furthermore, PROVABGS provides a Bayesian inference framework that infers
full posterior probability distributions of the SED model parameter:
$p(\theta\given {\bfi X}^{\rm photo}, {\bfi X}^{\rm spec})$, where 
${\bfi X}^{\rm photo}$ represents the photometry and ${\bfi X}^{\rm spec}$ 
represents the spectroscopy. 
In total, $\theta$ has 13 parameters: $M_*$, 6 parameters specifying the SFH
($\beta_1, \beta_2, \beta_3, \beta_4, f_{\rm burst}, t_{\rm burst}$), 2
parameters specifying ZH ($\gamma_1, \gamma_2$), 3 parameters specifying
dust attenuation ($\tau_{\rm BC}, \tau_{\rm ISM}, n_{\rm dust}$), and a
nuisance parameter for the fiber aperture effect. 
Posteriors accurately estimate uncertainties and degeneracies among galaxy
properties. 
Furthermore, they are essential for rigorous hierarchical population 
inference as we later demonstrate.

In practice, accurately estimating a 13-dimensional posterior requires a large
number ($\gtrsim$100,000) of SED model evaluations, which requires prohibitive
computational resources --- roughly 10 CPU hours per galaxy. 
To address this challenge, PROVABGS samples the posterior using the
\cite{karamanis2020} ensemble slice Markov Chain Monte Carlo (MCMC) sampling
with the {\sc zeus} Python package\footnote{https://zeus-mcmc.readthedocs.io/}.
PROVABGS further accelerates the inference by using neural emulators for the
SED models. 
The emulators are accurate to subpercent level and $>100\times$ faster than the
original SED model based on FSPS~\citep{kwon2022}. 
With {\sc zeus} and neural emulation, deriving a posterior takes $\sim$5 min
per galaxy with PROVABGS.
Moreover, \cite{hahn2022} used forward-modeled synthetic BGS observations to 
demonstrate PROVABGS can accurately infer $M_*$ over the full expected $M_*$ 
range of BGS. 

\begin{figure}
\begin{center}
    \includegraphics[width=0.6\textwidth]{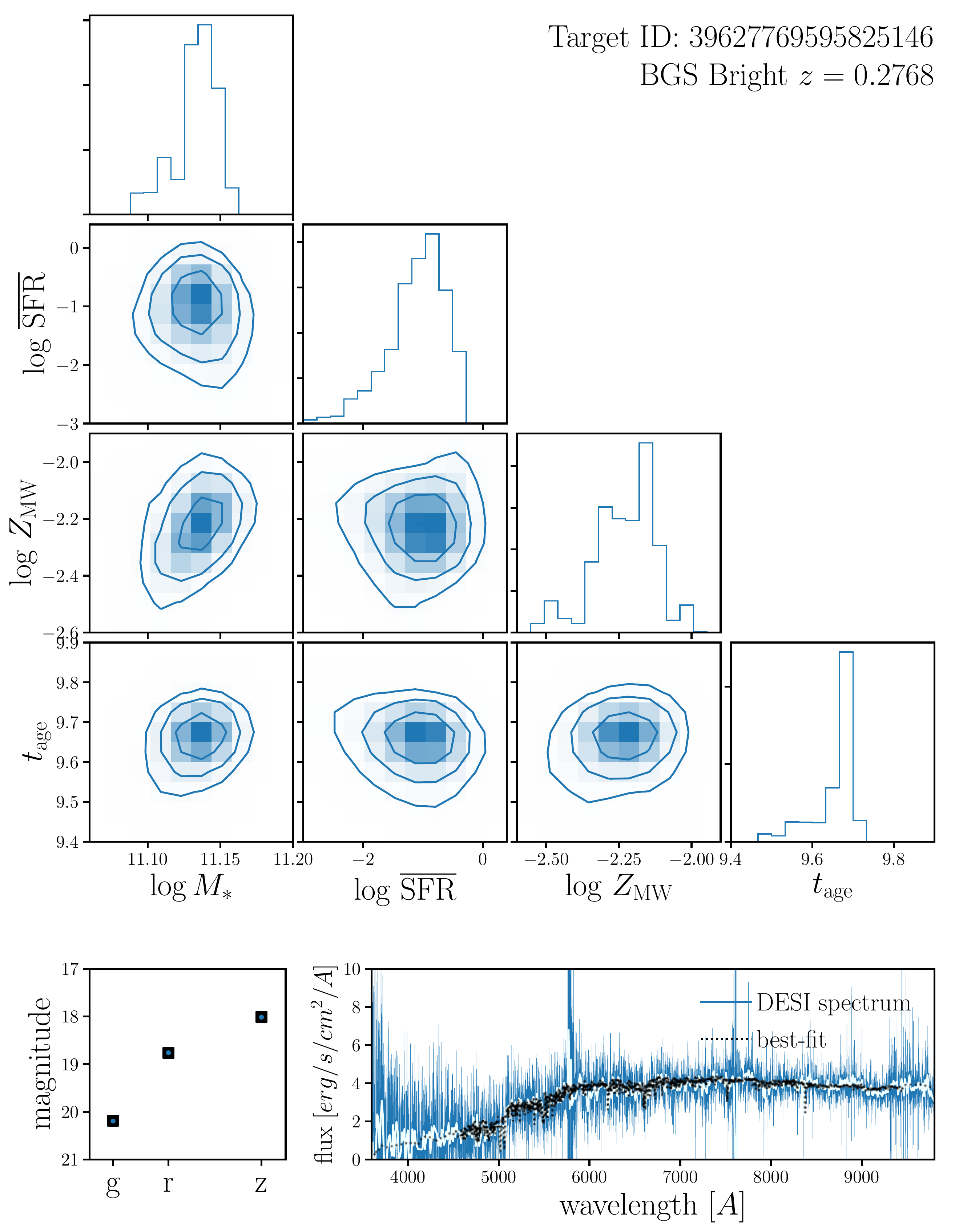}
    \caption{
        {\em Top panels}: 
        Posteriors of galaxy properties, $M_*$, $\overline{\rm SFR}$, 
        $Z_{\rm MW}$, and $t_{\rm age, MW}$, for a randomly selected BGS 
        Bright galaxy with $z=0.2768$ (target ID: 39627769595825146) inferred
        using the PROVABGS SED modeling framework from DESI photometry and
        spectroscopy. 
        The contours mark the 12, 40, 68, and 86 percentiles of the posterior. 
        With the PROVABGS posteriors, we accurately estimate the galaxy
        properties, their uncertainties, and any degeneracies among them. 
        {\em Bottom panels}: 
        Comparison of the best-fit PROVABGS SED model prediction (black) to
        observations (blue). 
        We compare the $g$, $r$, and $z$-band photometry in the left panel and 
        spectra in the right panel. 
        We include the observed spectra with coarser binning for clarity (azure). 
        We use PROVABGS to infer the posterior of galaxy properties for every 
        BGS galaxy in the DESI One-Percent Survey.
    }\label{fig:posterior}
\end{center}
\end{figure}

\begin{figure}
\begin{center}
    \includegraphics[width=0.6\textwidth]{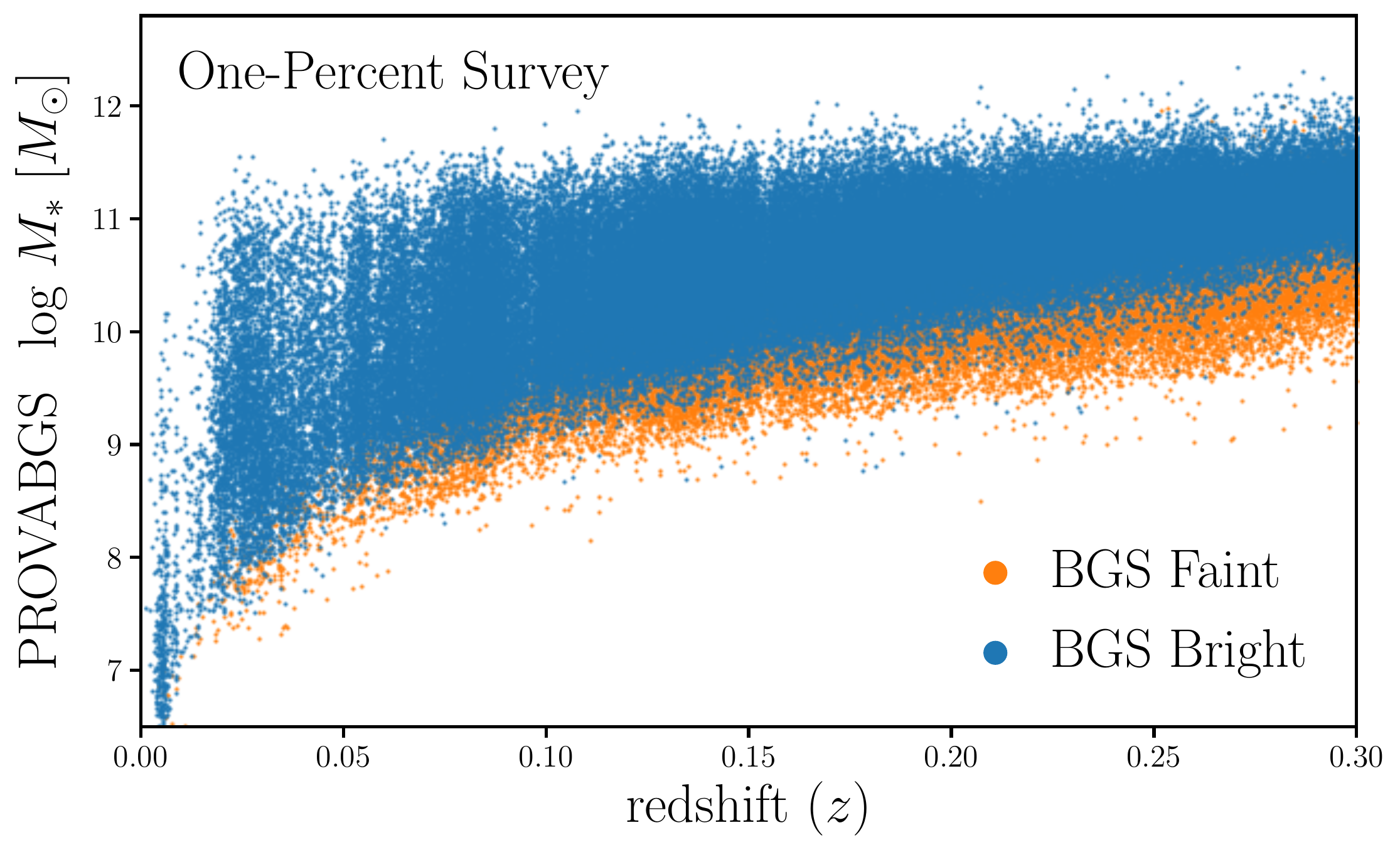}
    \caption{
        $M_*$ as a function of $z$ of BGS Bright (blue) and Faint (orange)
        galaxies in the DESI One-Percent Survey. 
        For $M_*$, we use the best-fit values derived using PROVABGS. 
        BGS Bright is a magnitude-limited sample to $r < 19.5$ while BGS Faint
        includes fainter galaxies $19.5 < r < 20.175$ selected using surface 
        brightness ($r_{\rm fiber}$) and color~\citep{hahn2022c}. 
        In total, we infer the posteriors of 143,017 BGS Bright and 95,499 BGS
        Faint galaxies across $z < 0.6$ in the DESI One-Percent Survey. 
    }\label{fig:mstar_z}
\end{center}
\end{figure}

In Figure~\ref{fig:posterior}, we demonstrate the PROVABGS SED modeling
framework for a randomly selected BGS Bright galaxy with $z=0.2768$ 
(target ID: 39627769595825146).
In the top panels, we present the posteriors of galaxy properties, $M_*$, 
$\overline{\rm SFR}$, $Z_{\rm MW}$, and $t_{\rm age, MW}$, inferred from DESI
photometry and spectroscopy. 
We mark the 12, 40, 68, and 86 percentiles of the posterior with the contours. 
The posteriors illustrate that we can precisely measure the properties of BGS
galaxies from DESI photometry and spectroscopy. 
Furthermore, with the full posterior, we accurately estimate the uncertainties
on the galaxy properties and the degeneracies among them (\emph{e.g.} $M_*$ and
$\overline{\rm SFR}$). 
In the bottom panels, we compare the PROVABGS SED model prediction using the
best-fit parameter values (black) to DESI observations (blue). 
The bottom left panel compares the optical $g$, $r$, and $z$-band photometry 
while the right panel compares the spectra. 
For reference, we also include the observed spectra with coarser wavelength 
binning (azure). 
The comparison shows good agreement between the best-fit model and the
observations. 

We derive a PROVABGS posterior (\emph{e.g.} Figure~\ref{fig:posterior}) for
every galaxy in the DESI One-Percent Survey. 
In Figure~\ref{fig:mstar_z}, we present the best-fit $M_*$ measurements as a
function of $z$ for the BGS galaxies in the DESI One-Percent Survey. 
We mark the galaxies in the BGS Bright sample in blue and the ones in the BGS
Faint sample in orange. 

\section{Results} \label{sec:results}
From the posteriors of galaxy properties inferred using
PROVABGS~(Section~\ref{sec:provabgs}), we derive the marginalized 1D
posterior of $M_*$,  $p(M_* \given {\bfi X_i})$, from
observed spectrophotometry ${\bfi X_i}$ of galaxy $i$.
Using these posteriors, we can estimate the probabilistic SMF (pSMF) of BGS
galaxies using population inference in a hierarchical Bayesian 
framework~\citep[\emph{e.g.}][]{hogg2010, foreman-mackey2014, baronchelli2020}.
In other words, we can infer $p(\phi\given\{{\bfi X_i}\})$, the probability
distribution of $\phi$ given the BGS observations, $\{{\bfi X_i}\}$. 
$\phi$ is the set of population hyperparameters that describe the pSMF,
$\Phi(M_*; \phi)$.
This approach is statistically rigorous and enables us to correctly propagate 
the uncertainties in our $M_*$ measurements to the pSMF. 


In this work, we estimate the pSMF using a Gaussian Mixture
Model~\citep[GMM;][]{press1992, mclachlan2000, blanton2003}, which provides a highly flexible
description of the $M_*$ distribution: 
\begin{equation}
    \Phi(M_*; \phi) = \sum\limits_{j=1}^{k} \mathcal{N}(M_*; \phi_j).
\end{equation} 
$k$ is the number of Gaussian components and  $\phi_j$ is the mean and 
standard deviation of the $j^{\rm th}$ Gaussian component of the GMM. 
Previous works have used parametric functions (\emph{e.g.} the Schechter function)
to describe the pSMF~\citep[\emph{e.g.}][]{leja2019a}. 
We opt for GMMs in order to produce a non-parametric measurement of the pSMF.
In a subsequent work, Speranza~\etal~(in prep.), we will present the BGS pSMF
measured using a parametric model with continuous redshift evolution.

To infer $p(\phi\given\{{\bfi X_i}\})$, we follow the same approach described
in \cite{hahn2022}:
\begin{align}
p(\phi \given \{{\bfi X_i}\}) 
    =&~\frac{p(\phi)~p( \{{\bfi X_i}\} \given \phi)}{p(\{{\bfi X_i}\})}\\
    =&~\frac{p(\phi)}{p(\{{\bfi X_i}\})}\int p(\{{\bfi X_i}\} \given \{\theta_i\})~p(\{\theta_i\} \given \phi)~{\rm d}\{\theta_i\}.\\
    =&~\frac{p(\phi)}{p(\{{\bfi X_i}\})}\prod\limits_{i=1}^N\int p({\bfi X_i} \given \theta_i)~p(\theta_i \given \phi)~{\rm d}\theta_i\label{eq:popinf} \\
    =&~\frac{p(\phi)}{p(\{{\bfi X_i}\})}\prod\limits_{i=1}^N\int \frac{p(\theta_i \given {\bfi X_i})~p({\bfi X_i})}{p(\theta_i)}~p(\theta_i \given \phi)~{\rm d}\theta_i\\
    =&~p(\phi)\prod\limits_{i=1}^N\int \frac{p(\theta_i \given {\bfi X_i})~p(\theta_i \given \phi)}{p(\theta_i)}~{\rm d}\theta_i. 
\end{align}
\noindent We can estimate the integral using $S_i$ Monte Carlo samples from
the individual posteriors $p(\theta_i \given {\bfi X_i})$: 
\begin{align}
    p(\phi \given \{{\bfi X_i}\}) \approx~p(\phi)\prod\limits_{i=1}^N\frac{1}{S_i}\sum\limits_{j=1}^{S_i}
    \frac{p(\theta_{i,j} \given \phi)}{p(\theta_{i,j})}.
\end{align} 
$p(\phi)$ and $p(\theta_{i,j})$ are the priors on the 
population hyperparameters and the SED model parameters. 
We use uniform distributions, $p(\phi) = 1$ and  
$p(\theta_{i,j}) = 1$, for both priors in this work. 

Since the sample of BGS galaxies is not volume limited and complete as a
function of $M_*$, we must account for selection effects and incompleteness
when estimating the pSMF. 
Thus, we include weights derived from $z^{\rm max}_i$, the maximum redshift that
galaxy $i$ could have and still be included in the BGS samples. 
We derive $z^{\rm max}_i$ for each galaxy by redshifting the SED
predicted by the best-fit (maximum likelihood) parameters and determining the 
maximum $z$ that the galaxy could be placed before it falls out of the survey selection. 
We use the best-fit parameters, rather than, \eg, samples drawn from the posteriors. 
However, this does not have a significant effect because BGS spans a relatively 
narrow redshift
range and BGS galaxies are primarily selected using $r$-band magnitudes.
The $r$-band bandpass lies at the center of DESI spectral wavelength range so the
SED models used to calculate $z_{\rm max}$ would be well constrained by the observed 
spectrum.
We then derive the comoving volume, $V^{\rm max}_i$, out to $z^{\rm max}_i$, and
include a factor of $1/V^{\rm max}_i$ in the galaxy weight $w_i$. 

Next, we include correction weights for spectroscopic incompleteness driven by
fiber assignment and redshift failures. 
The incompleteness from fiber assignment is due to the fact that DESI is not able to
assign fibers to all galaxies included in the BGS target selection. 
Furthermore, there is significant variation
in the assignment probability due to the clustering of galaxies. 
Meanwhile, incompleteness from redshift failure is caused by the fact that we
do not successfully measure the redshift for every spectrum. 
The redshift failure rate depends significantly on the surface brightnesses of the galaxies
and the signal-to-noise ratio of the spectra. 
We describe how we derive  the incompleteness correction weights for fiber
assignment and redshift failures, $w_{i, {\rm FA}}$ and $w_{i, {\rm ZF}}$, in
Appendix~\ref{sec:spec_comp}. 
Each BGS galaxy is assigned a weight of 
$w_i = (w_{i, {\rm FA}}\times w_{i, {\rm ZF}})/V^{\rm max}_i$.

We modify Eq.~\ref{eq:popinf} to include galaxy weights, $w_i$: 
\begin{align}
p(\phi \given \{{\bfi X_i}\}) 
    \approx&~\frac{p(\phi)}{\prod\limits_{i=1}^N p({\bfi X_i})^{w_i}} 
    \prod\limits_{i=1}^N \left(\int p({\bfi X_i} \given \theta_i)~p(\theta_i \given \phi)~{\rm d}\theta_i \right)^{w_i} \\ 
    \approx&~\frac{p(\phi)}{\prod\limits_{i=1}^N p({\bfi X_i})^{w_i}} 
    \prod\limits_{i=1}^N \left( \sum\limits_{j=1}^{S_i}
    \frac{p(\theta_{i,j} \given \phi)}{p(\theta_{i,j})} \right)^{w_i} \\
    \approx&~\frac{p(\phi)}{\prod\limits_{i=1}^N p({\bfi X_i})^{w_i}} 
    \prod\limits_{i=1}^N \left( \sum\limits_{j=1}^{S_i}
    \frac{\Phi(\theta_{i,j}; \phi)}{p(\theta_{i,j})} \right)^{w_i}.
\end{align} 
The weights are included in the exponent so, for example, a 
galaxy with $w_i = 2$ would have the same contribution to 
$p(\phi\given\{{\bfi X_i}\})$ as two galaxies with $w_i = 1$. 

In practice, we do not derive the full posterior 
$p(\phi \given \{{\bfi X_i}\})$. 
Instead we derive the maximum a posteriori (MAP) hyperparameter 
$\phi_{\rm MAP}$ that maximizes $p(\phi \given \{{\bfi X_i}\})$ or 
$\log p(\phi \given \{{\bfi X_i}\})$.
We expand, 
\begin{align}
\log p(\phi \given \{{\bfi X_i}\}) 
    \approx&~\log p(\phi) + 
    \sum\limits_{i=1}^N w_i \log \left(\sum\limits_{j=1}^{S_i} \frac{\Phi(\theta_{i,j}; \phi)}{p(\theta_{i,j})} \right).
\end{align} 
Since the first two terms are constant, we derive $\phi_{\rm MAP}$ by
maximizing 
\begin{equation}
    \max_\phi~~\sum\limits_{i=1}^N w_i \log \left(\sum\limits_{j=1}^{S_i} \frac{\Phi(\theta_{i,j}; \phi)}{p(\theta_{i,j})} \right),
\end{equation}
using the {\sc Adam} optimizer~\citep{kingma2017}.  
We derive $\phi_{\rm MAP}$ for BGS galaxies in redshift bins of width 
$\Delta z = 0.04$, starting from $z =0.01$, in order to examine the redshift
evolution of the SMF within BGS. 



\begin{figure}
\begin{center}
    \includegraphics[width=0.9\textwidth]{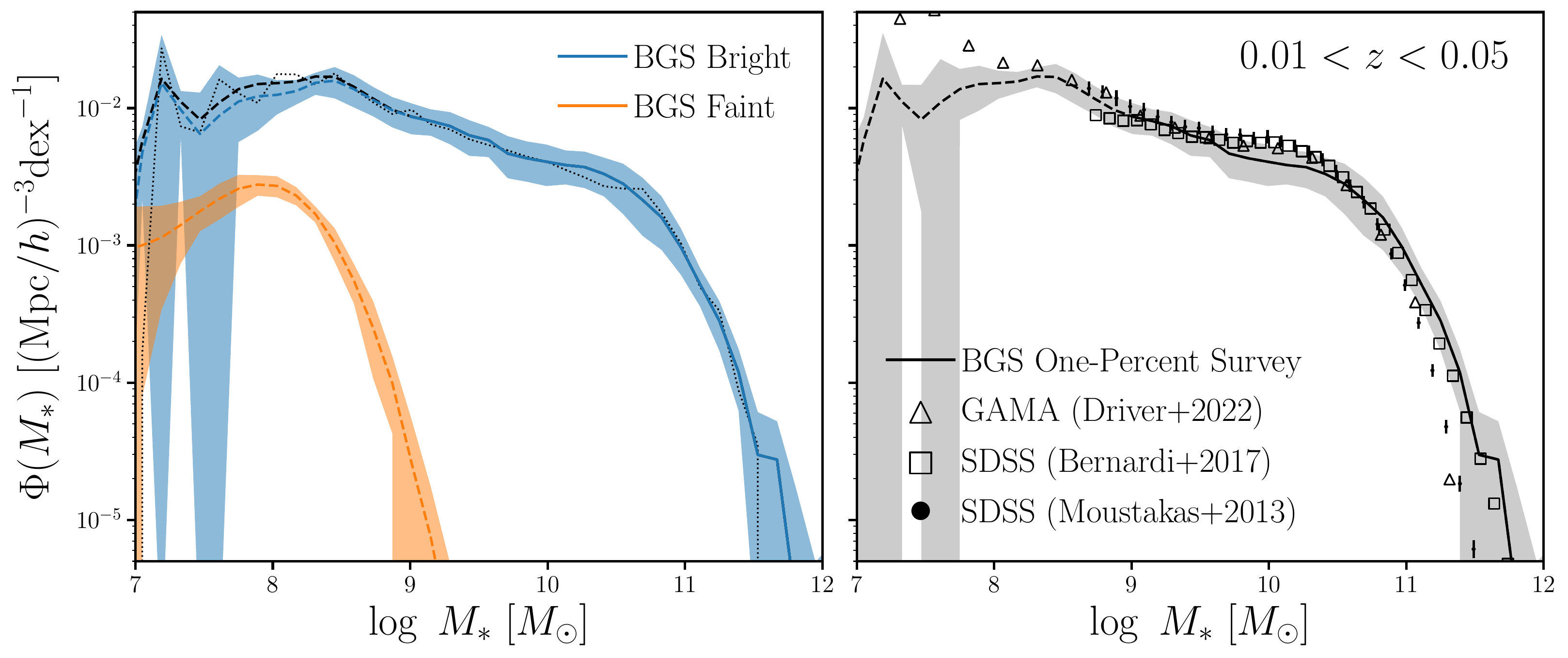} 
    \caption{
        The probabilistic SMF (pSMF) of BGS galaxies in the One-Percent Survey
        at $0.01 < z < 0.05$ (black line). 
        In the left panel, we present the pSMFs of BGS Bright (blue) and
        Faint (orange) galaxies, separately. 
        We also include the SMF measured using the standard point estimate 
        approach (black dotted), which underestimates the SMF at the low and 
        high mass ends. 
        We represent uncertainties on the pSMF, estimated using a standard
        jackknife technique (Appendix~\ref{sec:jack}), in the shaded regions.
        The solid line represents the pSMF above the completeness limit 
        $M_* > M_{\rm lim} = 10^{8.975}M_\odot$ (Appendix~\ref{sec:mscomp}).
        In the right panel, we include SMF measurements from previous
        spectroscopic surveys for comparison: SDSS~\citep{moustakas2013,
        bernardi2017} and GAMA~\citep{driver2022}. 
        Overall, the pSMF of BGS are in good agreement with SMF
        measurements from previous surveys.  
    }\label{fig:psmf}
\end{center}
\end{figure}

\begin{figure}
\begin{center}
    \includegraphics[width=0.5\textwidth]{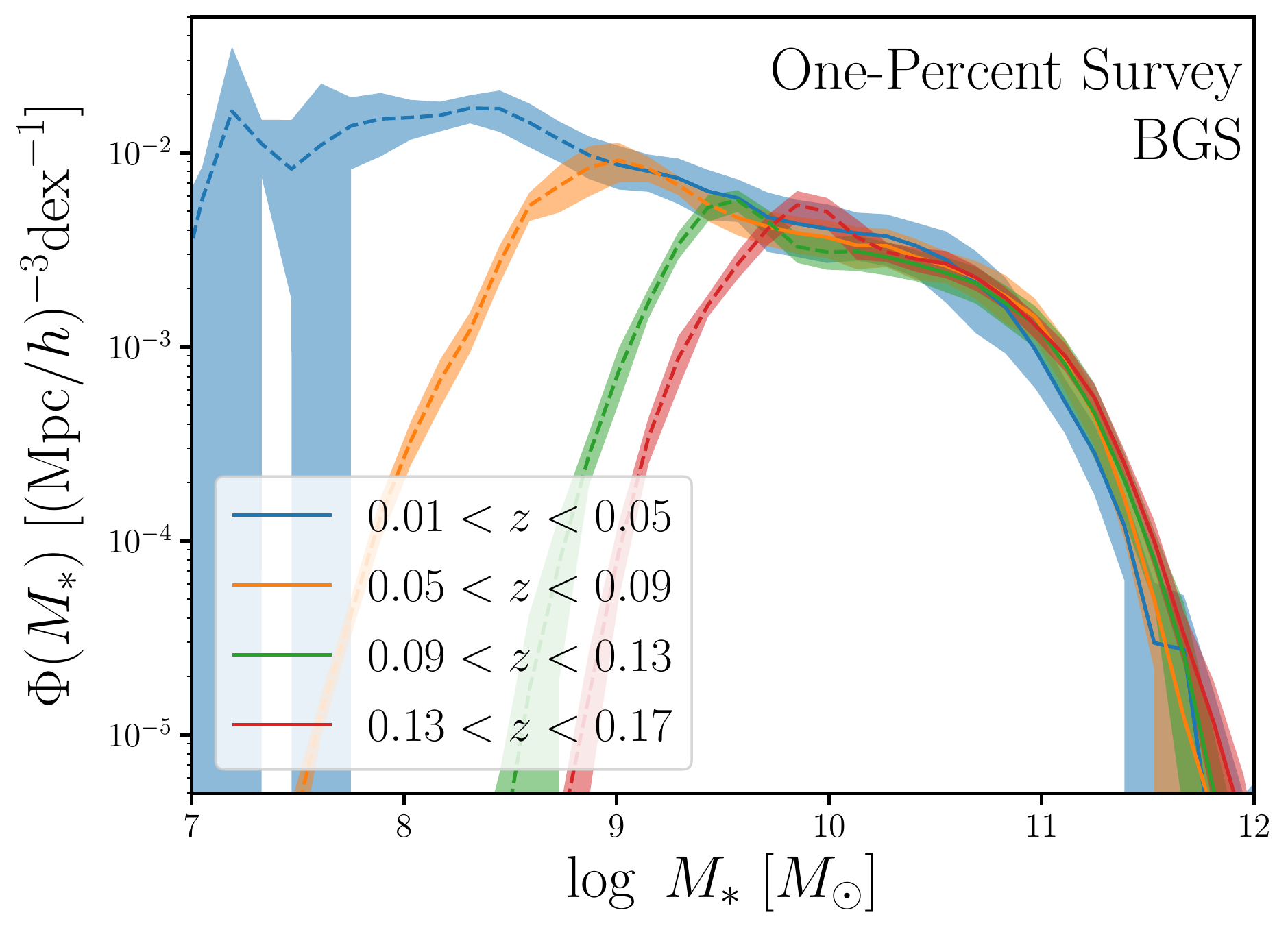} 
    \caption{
        The BGS pSMF over the redshift range $0.01 < z < 0.17$ in bins of
        $\Delta z = 0.04$. 
        The shaded regions represent the uncertainties on the pSMF, estimated
        using a standard jackknife technique.
        The solid lines represent the pSMF above the completeness limit 
        $M_* > M_{\rm lim}$ while the dashed lines represent the pSMF below
        the limit.
        We find no significant redshift evolution of the pSMFs over this range. 
        The main BGS survey will observe $>100\times$ more galaxies than the
        One-Precent Survey and will characterize the pSMF redshift evolution 
        even more precisely. 
    }\label{fig:psmfz}
\end{center}
\end{figure}

\subsection{The Probabilistic Stellar Mass Function} \label{sec:psmf}
We present the pSMF of BGS galaxies in the One-Percent Survey 
in Figure~\ref{fig:psmf} (black line) at $0.01 < z < 0.05$. 
In the left panel, we also present the pSMFs of the BGS Bright (blue) and Faint
(orange) galaxies, separately.  
BGS Bright galaxies are selected using a $r < 19.5$ magnitude limit. 
As a result, the BGS Bright sample is $M_*$ complete above $M_{\rm lim} >
10^{8.975}M_\odot$. 
We derive $M_{\rm lim}$ in Appendix~\ref{sec:mscomp} and mark the pSMF above
the completeness limit in solid and below the limit in dashed. 
Meanwhile, the BGS Faint sample is selected using a surface brightness and
color selection.   
It includes fainter galaxies, $19.5 < r < 20.175$, with overall lower $M_*$
than BGS Bright.
The shaded regions represent the uncertainties of the pSMF,
which we derive using a standard jackknife technique
(Appendix~\ref{sec:jack}). 
The jackknife uncertainties conservative estimate the
combined uncertainties from the population inference as 
well as cosmic variance~\citep{norberg2009}. 

We also include the SMF estimated using the standard point estimate approach (black dotted), 
derived using the maximum likelihood $M_*$ point estimates with the same galaxy weights, $w_i$. 
At intermediate $M_*$ range, $10^9 < M_* < 10^{11}M_\odot$, we find good
agreement with the pSMF. 
However, the standard approach significantly underestimates the SMF outside
this $M_*$ range. 
These discrepancies are due to the fact that point estimates of $M_*$
ignore  the uncertainties in the inferred $M_*$. 
The impact is significant at the most massive end of the SMF with fewer
galaxies. 
It is also significant at the least massive end where the observations
have lower signal-to-noise so the individual $M_*$ posteriors are broader.
The discrepancies are present in all other redshift bins and underscore the
importance of correctly propagating the $M_*$ uncertainties. 

In the right panel, we compare the BGS pSMF to SMF measurements from previous 
spectroscopic surveys: SDSS~\citep[black circle and square;][]{moustakas2013, bernardi2017} 
and GAMA~\citep[black triangle][]{driver2022}.
We note that there is significant variance in SMF measurements in the
literature, especially at the high $M_*$ end. 
This is partly due to the different modeling methodologies used to derive
$M_*$, which can contribute >0.1 dex discrepancies~\citep{pacifici2023}. 
Furthermore, there are also discrepancies due to photometric corrections
applied to SDSS photometry, assumptions on the stellar populations, and
dust~\citep{bernardi2017}.
We also do not account for differences in the IMF and cosmology. 
In a subsequent work, we will present a detailed comparison of BGS $M_*$
measurements using different methods. 
Overall, we find good agreement with previous SMF measurements, especially in
the intermediate $M_*$ range where we precisely infer the pSMF.  

In Figure~\ref{fig:psmfz}, we present the redshift evolution of the pSMF over 
$0.01 < z < 0.17$ in redshift bins of width $\Delta z = 0.04$. 
The shaded region represent the jackknife uncertainties for the pSMF.
The solid line represents the pSMF above $M_{\rm lim}$ while the dashed lines
represent the pSMF below the limit. 
We only include 4 redshift bins, since $M_{\rm lim} > 10^{10.5}M_\odot$ for 
$z > 0.17$ (Table~\ref{tab:mscomp}).
The pSMFs in Figure~\ref{fig:psmfz} do not reveal a significant redshift
dependence given their uncertainties. 
We note that the large uncertainties for the $0.01 < z < 0.05$ pSMF is driven
by large-scale structure at RA $\sim 195$ deg, Dec $\sim 28$ deg, and 
$z\sim0.0244$. 
The main BGS survey will observe $>100\times$ more BGS galaxies than the
One-Percent Survey with comparable signal-to-noise,\footnote{
The One-Percent Survey was observed using exposure times 1.2$\times$ longer 
than the main survey.}
and enable pSMF measurements with unprecedented precision. 

\begin{figure}
\begin{center}
    \includegraphics[width=0.5\textwidth]{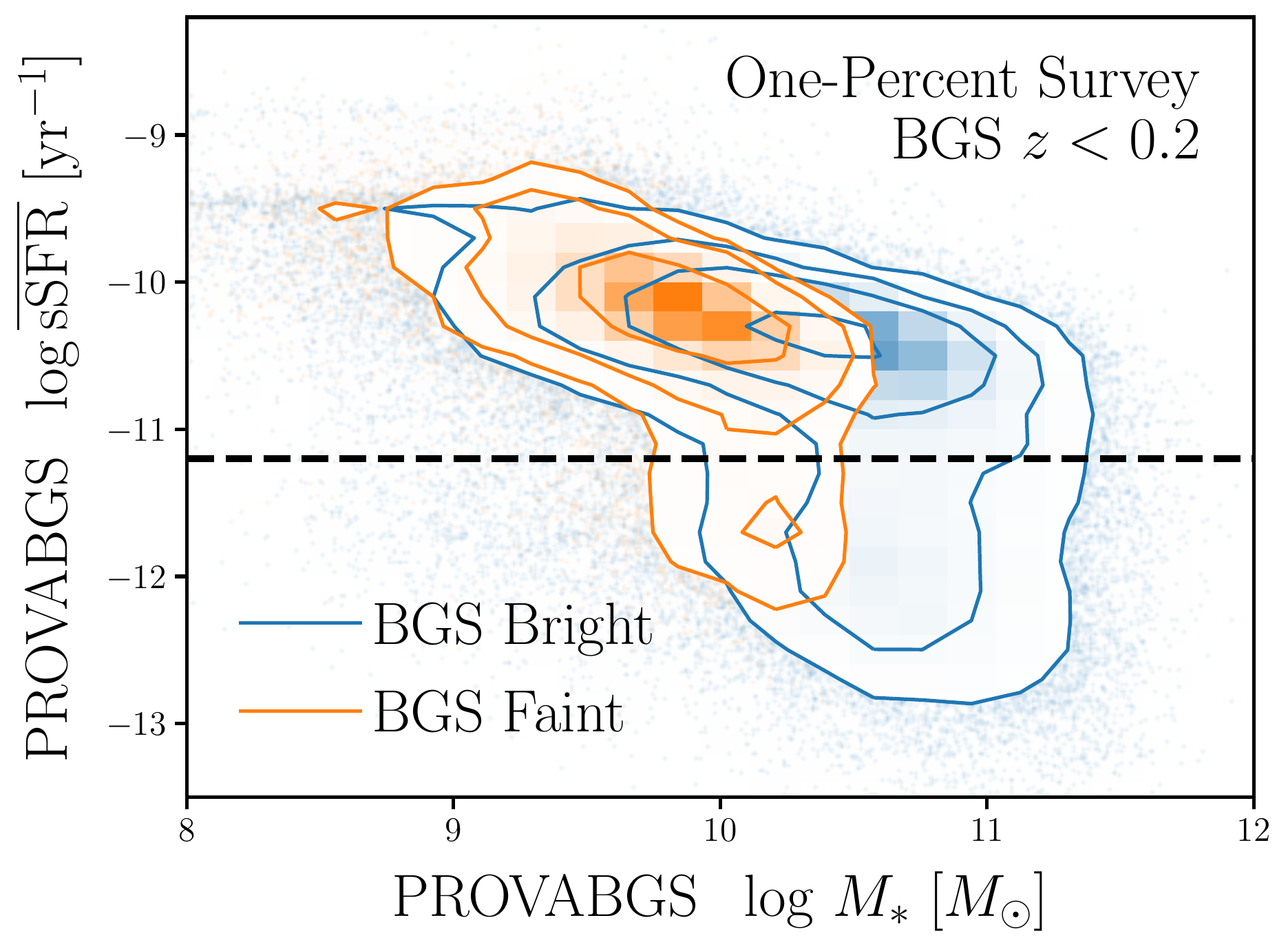} 
    \caption{
        The $M_*-\overline{\rm sSFR}$ distribution of BGS galaxies at $z < 0.2$. 
        sSFR is derived using $\overline{\rm SFR}$, average SFR over the last 1
        Gyr, inferred using PROVABGS. 
        The $M_*-\overline{\rm sSFR}$ distribution is bimodal with star-forming galaxies
        lying on the star-forming sequence.
        We classify galaxies with $\overline{\rm sSFR} > 10^{-11.2}\,{\rm yr}^{-1}$ as
        star-forming galaxies and $\overline{\rm sSFR} < 10^{-11.2}\,{\rm yr}^{-1}$ as
        quiescent. 
    }\label{fig:sfq}
\end{center}
\end{figure}

\begin{figure}
\begin{center}
    \includegraphics[width=0.9\textwidth]{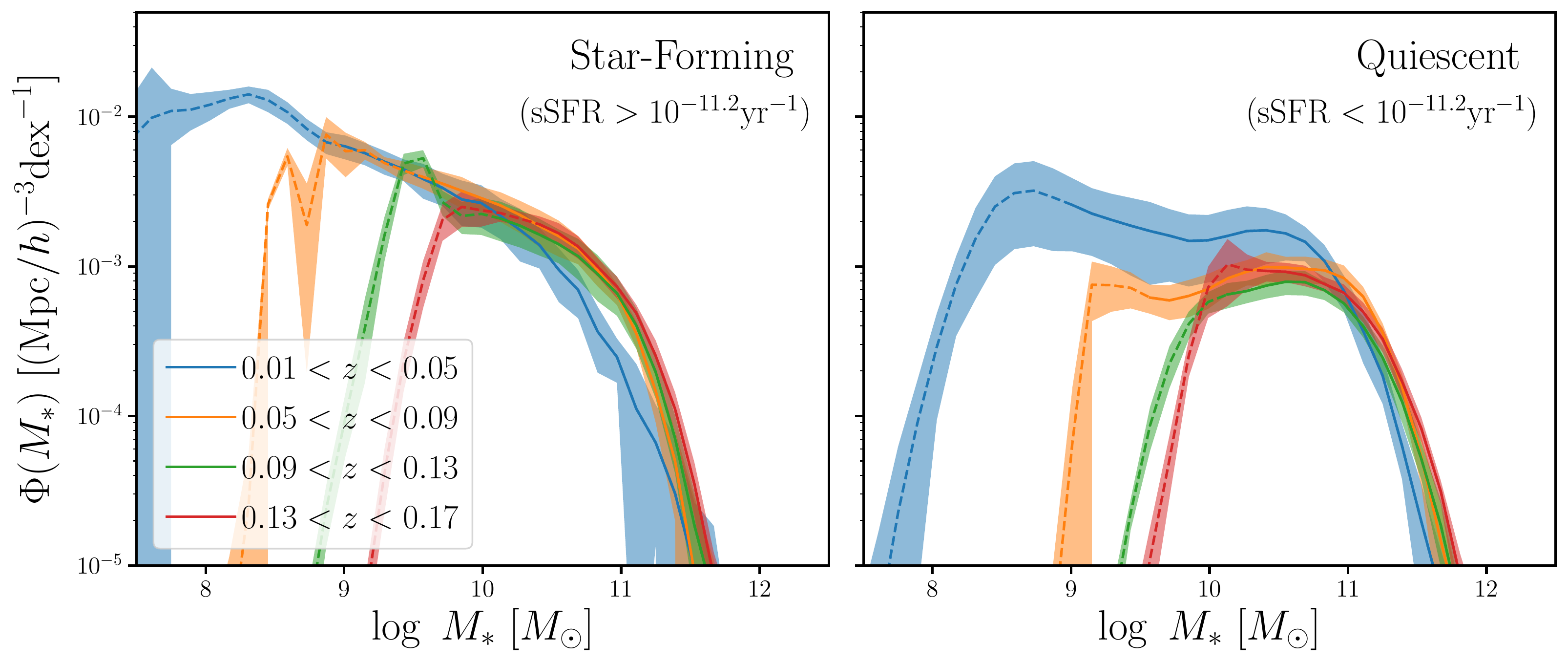} 
    \caption{
        The pSMF of star-forming (left) and quiescent (right) BGS Bright
        galaxies over $0.01 < z < 0.17$ in bins of $\Delta z = 0.04$. 
        Star-forming and quiescent galaxies are classified using an empirically
        determined ${\rm sSFR} = 10^{-11.2}{\rm yr}^{-1}$ cut. 
        We represent the uncertainties for the pSMF in the shaded regions and 
        the pSMFs above/below the $M_*$ completeness limits in solid/dashed
        lines.
        We find little overall evolution of the pSMFs over the redshift range investigated here. 
    }\label{fig:sfqsmf}
\end{center}
\end{figure}

\subsection{Star-Forming and Quiescent Galaxies in the BGS} \label{sec:sfq}
In addition to the pSMF of the full galaxy population, we can also examine the
pSMF of the star-forming and quiescent subpopulations using 
$\overline{\rm SFR}$, average SFR over the last 1 Gyr, inferred with PROVABGS. 
In Figure~\ref{fig:sfq}, we present the distribution of $M_*$ versus average
specific SFR, $\overline{\rm sSFR} = \overline{\rm SFR}/M_*$, for BGS Bright
(blue) and Faint (orange) galaxies at $z < 0.2$. 
The $M_*-{\rm sSFR}$ distribution of the BGS galaxies reveal a clear
bi-modality with star-forming galaxies lying on the SFS and quiescent galaxies
lying $\gtrsim$ 1 dex below the sequence. 
Figure~\ref{fig:sfq} also confirms that BGS Faint galaxies have overall lower
$M_*$ than BGS Bright galaxies and are primarily star-forming galaxies. 
This is due to the fact that the $(z - W1)-1.2(g-r)+1.2$ color used to select
BGS Faint galaxies is a proxy for H$\alpha$ and H$\beta$ emission lines.  

To further examine the star-forming and quiescent galaxy populations, we
classify BGS Bright galaxies as star-forming or quiescent using a 
$\overline{\rm sSFR} = 10^{-11.2}\,{\rm yr}^{-1}$ cut. 
We determine this cut empirically based roughly on the $\overline{\rm sSFR}$ 
of the ``green valley'' between the SFS and the quiescent mode. 
We opt for a $\overline{\rm sSFR}$ cut rather than more sophisticated methods
in the literature~\citep[\emph{e.g.}][]{hahn2019, donnari2019} for simplicity. 
In Figure~\ref{fig:sfqsmf}, we present the pSMF of star-forming and quiescent
BGS Bright galaxies at $0.01 < z < 0.17$ in bins of $\Delta z = 0.04$.
The shaded regions represent the jackknife uncertainties for the pSMF. 
The solid lines represent the pSMFs above the completeness limit while the
dashed lines represent the pSMFs below the limit. 
Overall, the pSMFs show little evolution over these $M_*$ and redshift ranges 
except for a possible decline at the massive end of the star-forming pSMF. 

\begin{figure}
\begin{center}
    \includegraphics[width=0.45\textwidth]{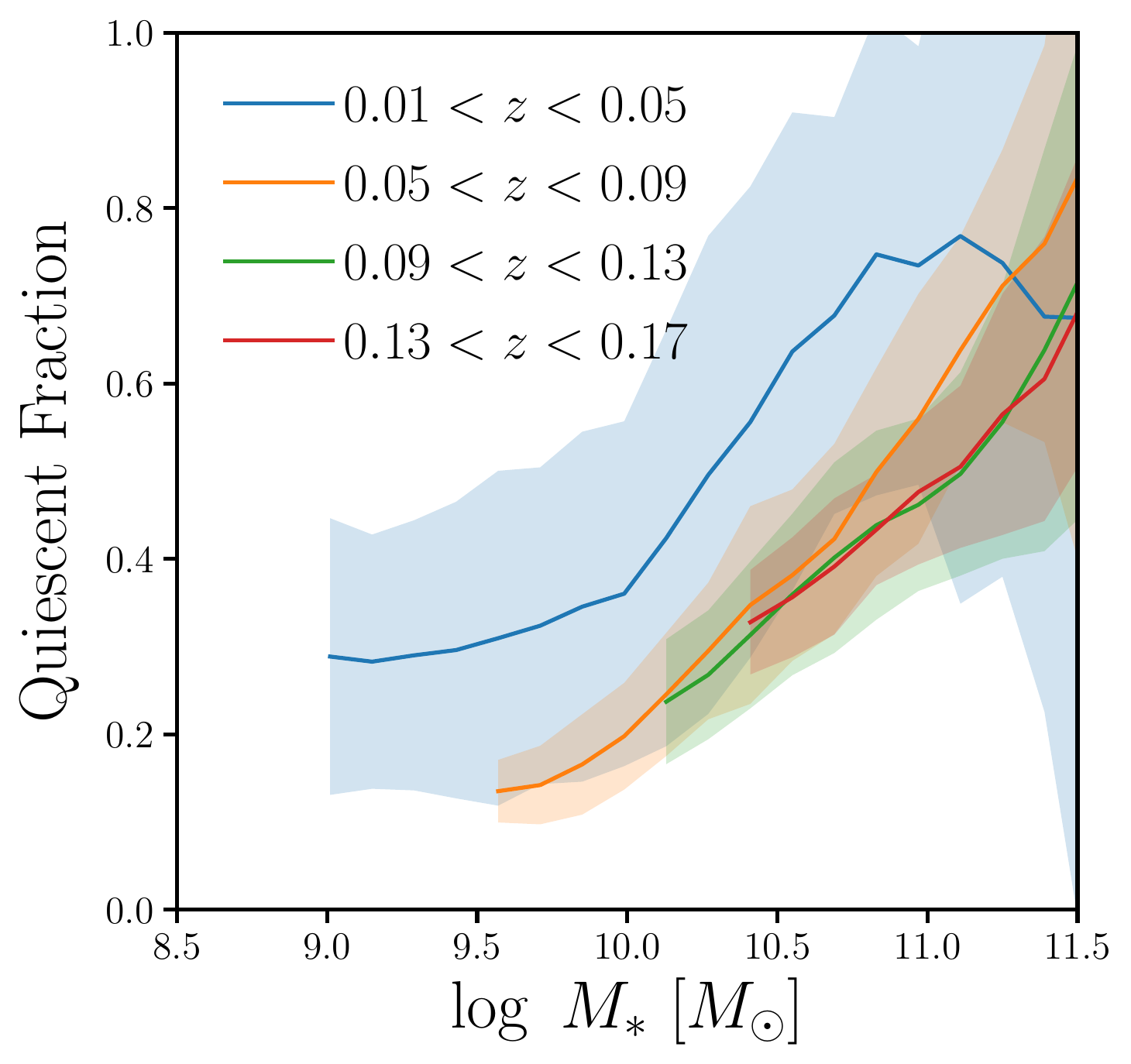} 
    \caption{
        The quiescent fraction of BGS Bright galaxies over $0.01 < z < 0.17$ in
        bins of $\Delta z =0.04$.
        We present the uncertainties in the shaded region and only include  
        the quiescent fraction above the $M_*$ completeness limit. 
        The quiescent fractions increase with $M_*$ at all redshifts. 
        Furthermore, the quiescent fractions suggest an overall increase in the
        quiescent population with lower redshift.
        We find no significant redshift evolution of the quiescent fraction 
        over our redshift range. 
    }\label{fig:qf}
\end{center}
\end{figure}

Next, we present the fraction of quiescent galaxies in BGS Bright as a function
of $M_*$ over $0.01 < z < 0.17$ in Figure~\ref{fig:qf}.
The quiescent fraction is derived by taking the ratio of the pSMFs of quiescent
galaxies over all galaxies and measured for each $\Delta z =0.04$ bin.
The shaded region represent the uncertainties derived from propagating the
jackknife uncertainties of the pSMFs. 
We focus on the quiescent fraction of BGS Bright galaxies above the $M_*$
completeness limit: $M_* > M_{\rm lim}$. 
At each redshift bin, the quiescent fraction increases with $M_*$ to $\sim$1 at
$M_*\sim10^{11.5}M_\odot$. 
We find no significant redshift evolution of quiescent fraction across 
$0.01 < z < 0.17$. 
Although the significant statistical uncertainties obfuscate a clear trend, 
the quiescent fraction evolution is in good qualitative agreement with previous
works~\citep[\emph{e.g.}][]{baldry2006, iovino2010, peng2010, hahn2015}.
Upcoming observations from the DESI main survey will increase the number of BGS
galaxies by >100$\times$ and enable precise comparisons of the quiescent
fraction measurements. 

\section{Summary and Discussion} \label{sec:summary}
Over its five-year operation, starting on May 2021, the DESI Bright Galaxy
Survey (BGS) will observe the spectra of $\sim$15 million galaxies out to 
$z < 0.6$ over 14,000 ${\rm deg}^2$.  
BGS will produce two main galaxy samples: a $r < 19.5$ magnitude-limited BGS
Bright sample and a fainter $19.5 < r < 20.175$ surface brightness and color
selected BGS Faint sample. 
Compared to the SDSS main galaxy survey, the BGS galaxy samples will be over
two magnitudes deeper, twice the sky, and double the median redshift
$z\sim0.2$. 
They will include diverse galaxy subpopulations that have the potential to
reveal new trends among galaxies that were previously undetectable and open new
discovery space. 

In addition, each galaxy in BGS will have measurements of its detailed physical
properties (\emph{e.g.} $M_*$, SFR, $Z_*$, $t_{\rm age}$) from PROVABGS. 
These properties will be inferred from DESI spectrophotometry using
state-of-the-art  SED modeling in a fully Bayesian inference framework. 
PROVABGS will provide statistically rigorous estimates of uncertainties and
degeneracies among the properties.
With these measurements alongside its statistical power, BGS will provide 
a powerful galaxy sample with which to measure scaling relations and population 
statistics that will characterize the global galaxy population and test galaxy 
formation models with unprecedented precision.

In this work, we showcase the potential of BGS by presenting the probabilistic
stellar mass function (pSMF) using $\sim$250,000 BGS galaxies observed solely
during one of DESI's survey validation program.
The pSMF are derived using a hierarchical population inference framework that 
statistically and rigorously propagates uncertainties on $M_*$ and provide improved
estimates of the SMF at the lowest and highest $M_*$ regimes. 
We also describe how we account for selection effects and incompleteness in the
BGS observations (Appendix~\ref{sec:spec_comp}).  
Overall, we find good agreement between our pSMF and previous SMF measurements
in the literature. 
We also examine the pSMF of the star-forming and quiescent galaxy population
classified using a simple $\overline{\rm sSFR} = 10^{-11.2}{\rm yr}^{-1}$ cut
and find qualitative agreement with previous works. 

This work is first of a series of papers that will present population
statistics for BGS galaxies using PROVABGS. 
For the pSMF in this work, we used a flexible GMM to provide a non-parametric
measurement of the SMF.
In a subsequent work, Speranza~\etal~(in prep.), we will present the pSMF of BGS
measured using a parametric model with continuous redshift evolution. 
In another work, we will present in depth comparison of $M_*$ measured using
different methodologies and assumptions. 
We will also explore the color dependence of the fiber aperture effect and its 
impact on inferred galaxy properties in Ramos~\etal~(in prep.). 
Lastly, the hierarchical population inference framework presented in this work
can be extended to population statistics beyond the SMF. 
We will extend the framework to the SFR-$M_*$ distribution and present the
probabilistic SFR-$M_*$ distribution and quiescent fraction in future work. 

All of the pSMFs presented in this work are measured from BGS galaxies observed
from April to May of 2021 during the DESI One-Percent Survey.
Since then, DESI has already completed nearly two years of observations.
As of writing (May 2023), DESI has observed over 22 million galaxy spectra
in total and over 10 million BGS galaxy spectra.  
With three out of the five years of operation remaining, BGS has completed
completed $\sim$60\% of its observations and is ahead of schedule.  
BGS will also be further extended by additional 
low-redshift dwarf galaxies observed with the DESI 
low-z secondary program~\citep{darragh-ford2022}.
DESI observations will be publicly released periodically, starting with the
Early Data Release (EDR) later this year. 
The EDR will include observations from the One-Percent Survey used in this
work. 
An accompanying PROVABGS catalog will be released with each data release.
All of the pSMF measurements and data used to generate the figures presented
in this work is publicly available at \url{https://doi.org/10.5281/zenodo.8018936}.

\section*{Acknowledgements}
It's a pleasure to thank Alex I. Malz and Peter Melchior for helpful 
discussions. 
We also thank Rita Tojeiro and Rebecca Canning for their detailed review and feedback on an earlier version of this work.  
This work was supported by the AI Accelerator program of the Schmidt Futures
Foundation.

This material is based upon work supported by the U.S. Department of Energy (DOE), Office of Science, Office of High-Energy Physics, under Contract No. DE–AC02–05CH11231, and by the National Energy Research Scientific Computing Center, a DOE Office of Science User Facility under the same contract. Additional support for DESI was provided by the U.S. National Science Foundation (NSF), Division of Astronomical Sciences under Contract No. AST-0950945 to the NSF’s National Optical-Infrared Astronomy Research Laboratory; the Science and Technologies Facilities Council of the United Kingdom; the Gordon and Betty Moore Foundation; the Heising-Simons Foundation; the French Alternative Energies and Atomic Energy Commission (CEA); the National Council of Science and Technology of Mexico (CONACYT); the Ministry of Science and Innovation of Spain (MICINN), and by the DESI Member Institutions: \url{https://www.desi.lbl.gov/collaborating-institutions}.

The DESI Legacy Imaging Surveys consist of three individual and complementary projects: the Dark Energy Camera Legacy Survey (DECaLS), the Beijing-Arizona Sky Survey (BASS), and the Mayall z-band Legacy Survey (MzLS). DECaLS, BASS and MzLS together include data obtained, respectively, at the Blanco telescope, Cerro Tololo Inter-American Observatory, NSF’s NOIRLab; the Bok telescope, Steward Observatory, University of Arizona; and the Mayall telescope, Kitt Peak National Observatory, NOIRLab. NOIRLab is operated by the Association of Universities for Research in Astronomy (AURA) under a cooperative agreement with the National Science Foundation. Pipeline processing and analyses of the data were supported by NOIRLab and the Lawrence Berkeley National Laboratory. Legacy Surveys also uses data products from the Near-Earth Object Wide-field Infrared Survey Explorer (NEOWISE), a project of the Jet Propulsion Laboratory/California Institute of Technology, funded by the National Aeronautics and Space Administration. Legacy Surveys was supported by: the Director, Office of Science, Office of High Energy Physics of the U.S. Department of Energy; the National Energy Research Scientific Computing Center, a DOE Office of Science User Facility; the U.S. National Science Foundation, Division of Astronomical Sciences; the National Astronomical Observatories of China, the Chinese Academy of Sciences and the Chinese National Natural Science Foundation. LBNL is managed by the Regents of the University of California under contract to the U.S. Department of Energy. The complete acknowledgments can be found at \url{https://www.legacysurvey.org/}.

Any opinions, findings, and conclusions or recommendations expressed in this material are those of the author(s) and do not necessarily reflect the views of the U. S. National Science Foundation, the U. S. Department of Energy, or any of the listed funding agencies.

The authors are honored to be permitted to conduct scientific research on Iolkam Du’ag (Kitt Peak), a mountain with particular significance to the Tohono O’odham Nation.

\appendix
\section{Spectroscopic Completeness} \label{sec:spec_comp}
Spectroscopic galaxy surveys, such as BGS, do not successfully measure the
redshift for all of the galaxies they target. 
As a result, this spectroscopic incompleteness must be accounted for when
measuring galaxy population statistics such as the SMF.  
In this appendix, we present how we estimate the spectroscopic incompleteness
for BGS and derive the weights we use to correct for its impact on the SMF. 

For BGS, spectroscopic incompleteness is primarily driven by fiber assignment
and redshift failures.  
DESI uses 10 fiber-fed spectrographs with 5000 fibers but targets more galaxies
than available fibers. 
For instance, the BGS Bright and Faint samples have $\sim 860$ and 
$530\,{\rm targets}/{\rm deg}^2$, respectively. 
In addition, of the 5000, a minimum of 400 ‘sky’ fibers are dedicated to measuring
the sky background for accurate sky subtraction and an additional 100 fibers
are assigned to standard stars for flux calibration~\cite{guy2023}.

Furthermore, each fiber is controlled by a robotic fiber positioner on the
focal plane. 
These positioners can rotate on two arms and be positioned within a circular
patrol region of radius 1.48 arcmin~\citep{schubnell2016, desicollaboration2016, abareshi2022, silber2023}.
Although the patrol regions of adjacent positioners slightly overlap, the
geometry of the positioners cause higher incompleteness in regions with high
target density~\citep{smith2019}.
To mitigate the incompleteness from the fiber assignment, BGS will observe its  
footprint with four passes.
With this strategy, BGS achieves $\sim$80\% fiber assignment
completeness~\citep{hahn2022c}.

To estimate fiber assignment completeness, we run the fiber assignment
algorithm~\citep{fba} on BGS targets 128 separate times.
For each BGS galaxy, $i$, we count the total number of times out of 128 fiber 
assigned realizations that the galaxy is assigned a fiber: $N_{i, {\rm FA}}$. 
Then to correct for the fiber assignment incompleteness, we assign correction
weights
\begin{equation} \label{eq:w_fa}
    w_{i, \mathrm{FA}} = \frac{128}{N_{i, \mathrm{FA}}}
\end{equation}
to each BGS galaxy. 

\begin{figure}[h]
\begin{center}
    \includegraphics[width=0.5\textwidth]{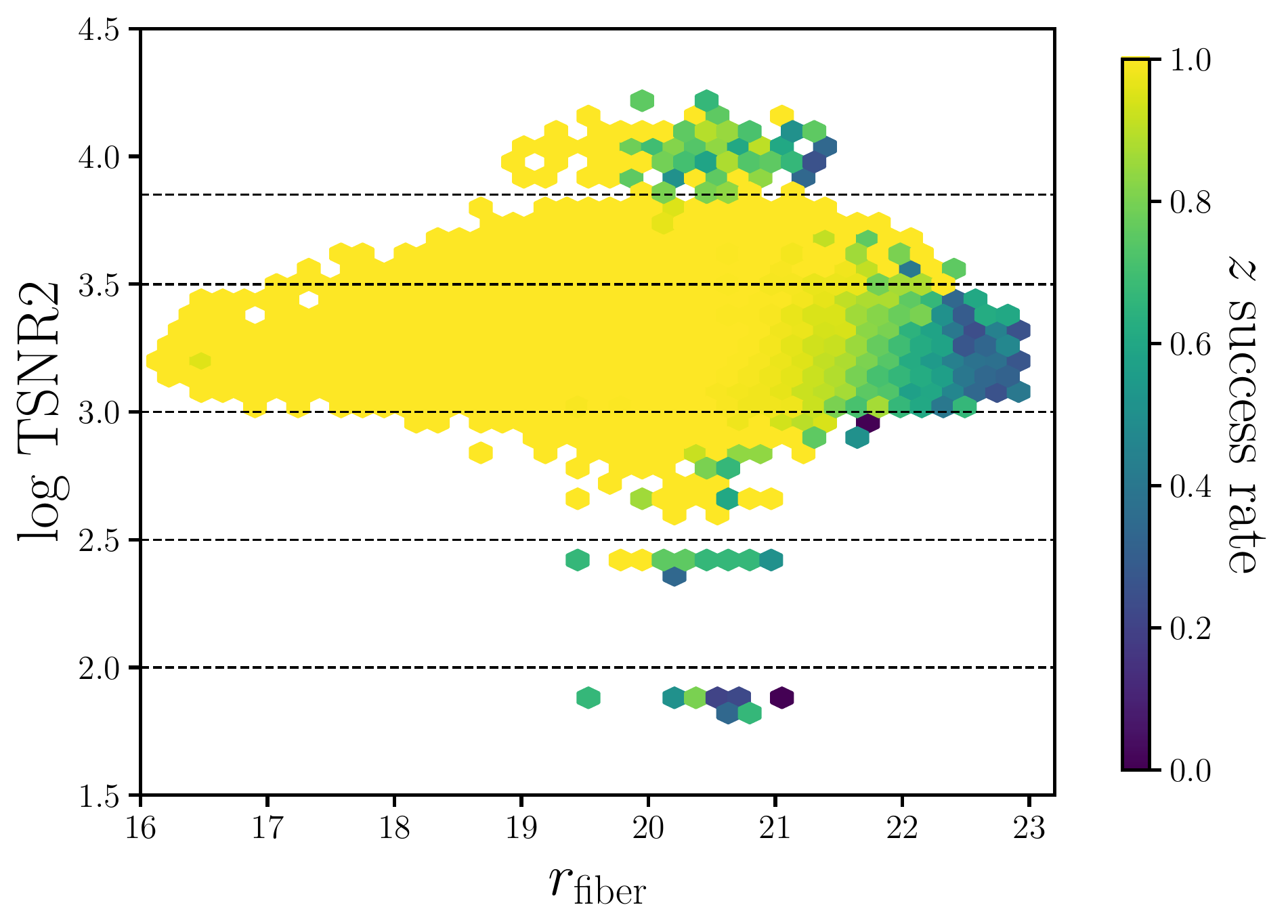} 
    \caption{
        Redsift success rate of BGS Bright galaxies as a function of 
        $r_{\rm fiber}$ and TSNR2.
        TSNR2 is a statistic that quantifies the signal-to-noise ratio of the
        observed spectrum. 
        The color map represents the mean redshift success rate in each hexbin.
        We mark the TSNR2 bins (black dashed) that we use to separately fit the
        redshift success rate as a function of $r_{\rm fiber}$ using
        Eq.~\ref{eq:zsucc}.
        In each TSNR2 bin, redshift success decreases as $r_{\rm fiber}$
        increases. 
    }\label{fig:zfail0}
\end{center}
\end{figure}

Although we measure a spectrum for each galaxy assigned a fiber, we do not
measure reliable redshifts for every spectra that meet the criteria specified
in Section~\ref{sec:edr}.
This redshift measurement failure significantly contributes to spectroscopic
incompleteness. 
For BGS, redshift failure of an observed galaxy spectrum depends mainly on
fiber magnitude and a statistic, TSNR2.
Fiber magnitude is the predicted flux of the BGS object within a 1.5\arcsec
diameter fiber; we use $r$-band fiber magnitude, $r_{\rm fiber}$.  
TSNR2 roughly corresponds to the signal-to-noise ratio of the spectrum and is 
the statistic used to calibrate the effective exposure times in DESI
observations~\citep{guy2023}.

\begin{figure}
\begin{center}
    \includegraphics[width=0.5\textwidth]{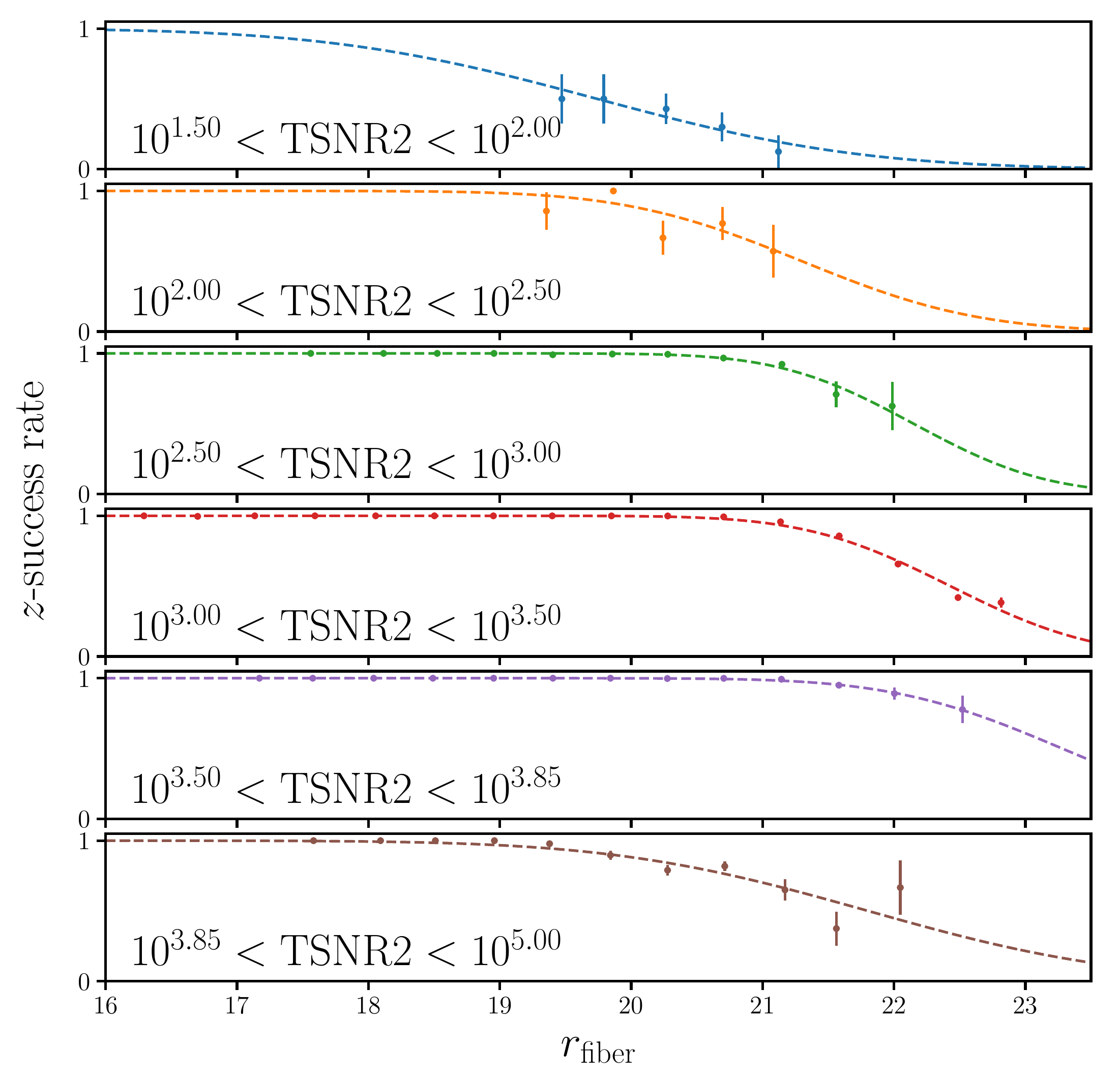}
    \caption{
        Redshift success rates of BGS Bright galaxies  as a function of 
        $r_{\rm fiber}$ in 6 TSNR2 bins. 
        The error bars represent the Poisson uncertainties.
        In each panel, we include the best-fit analytic (Eq.~\ref{eq:zsucc})
        approximation of the redshift success rate (dashed) derived from
        $\chi^2$ minimization. 
        We use this analytic approximation to calculate the galaxy weights to
        correct for spectroscopic incompleteness caused by failures to
        accurately measure redshifts from observed spectra.
    }\label{fig:zfail1}
\end{center}
\end{figure}

In Figure~\ref{fig:zfail0}, we present the redshift success rate of BGS
Bright galaxies as a function of $r_{\rm fiber}$ and TSNR2.
In each hexbin, the color map represents the mean $z$-success rate. 
We include all hexbins with more than 2 galaxies. 
Overall, the $z$-success rate depends significantly on $r_{\rm fiber}$:
galaxies with fainter $r_{\rm fiber}$ have lower $z$-success rates. 
However, the $r_{\rm fiber}$ dependence itself varies in bins of TSNR2. 
We mark the edges of the bins in black dashed: $\log {\rm TSNR2} = 2.0, 2.5,
3.0, 3.5, 3.85$.
Within each of the TSNR2 bins, the $r_{\rm fiber}$ dependence of the
$z$-success rate does not vary significantly. 
In Figure~\ref{fig:zfail1}, we present the $z$-success rate of BGS Bright
galaxies as a function of $r_{\rm fiber}$ for each of the 6 TSNR2 bins.
We mark the range of TSNR2 in the bottom left of each panel. 
The error bars represent the Poisson uncertainties of the $z$-success rate.

To correct for the effect of redshift failures, we include an additional
correction weight for each BGS galaxy: 
\begin{equation}
    w_{i, {\rm ZF}} = \frac{1}{f_{z-{\rm sucess}}(r_{{\rm fiber},i}, {\rm
    TSNR2}_i)}.
\end{equation} 
$f_{z-{\rm sucess}}(r_{{\rm fiber},i}, {\rm TSNR2}_i)$ is the $z$-success rate
as a function of $r_{\rm fiber}$ and TSNR2 of the galaxy. 
Galaxies with $f_{z-{\rm sucess}} = 1$ (100\% $z$-success) will have 
$w_{i, {\rm ZF}} = 1.0$ while galaxies with $f_{z-{\rm success}} = 0.1$ 
(10\% $z$-success) will have $w_{i, {\rm ZF}} = 10$.
For $f_{z-{\rm sucess}}(r_{{\rm fiber},i}, {\rm TSNR2}_i)$, we fit the
following functional form for each TSNR2 bin: 
\begin{equation} \label{eq:zsucc}
    f_{z-{\rm success}}(r_{\rm fiber}) = \frac{1}{2} \bigg(1-{\rm erf}(c_0
    (r_{\rm fiber} - c_1))\bigg).
\end{equation}
In Figure~\ref{fig:zfail1}, we present the best-fit 
$f_{z-{\rm success}}(r_{\rm fiber})$ for each of the TSNR2 bins in dashed. 
The best-fit coefficients, $c_0, c_1$, are derived from $\chi^2$ minimization.
We repeat this procedure independently for BGS Bright galaxies as well as the
BGS Faint galaxies with $(z - W1) - 1.2(g - r) + 1.2 \ge 0$,
and BGS Faint galaxies with $(z - W1) - 1.2(g - r) + 1.2 < 0$.
We list the best-fit values in bins of TSNR2 for each of the samples in
Table~\ref{tab:zfail}. 

\begin{table} 
    \caption{Best-fit coefficients of Eq.~\ref{eq:zsucc}, which describes the 
    $z$-success rate as a function of $r_{\rm fiber}$ for different TSNR2 bins
    for BGS Bright and Faint samples.} 
    \begin{center}
        \begin{tabular}{lcc} \toprule
            TSNR2 range & $c_0$ & $c_1$ \\[3pt]
            \multicolumn{3}{c}{BGS Bright} \\
            \hline 
            $10^{1.5}   - 10^{2}$       & 0.443 & 19.7 \\ 
            $10^{2}     - 10^{2.5}$       & 0.668 & 21.3 \\ 
            $10^{2.5}   - 10^{3}$       & 0.888 & 22.1 \\ 
            $10^{3}     - 10^{3.5}$       & 0.822 & 22.4 \\ 
            $10^{3.5}   - 10^{3.85}$    & 0.698 & 23.3 \\ 
            $10^{3.85}  - 10^{5}$      & 0.465 & 21.8 \\ 
            \hline            
            \hline 
            \multicolumn{3}{c}{BGS Faint}\\
            \multicolumn{3}{c}{$(z - W1) - 1.2(g - r) + 1.2 \ge 0$} \\
            \hline 
            $10^{1.5}   - 10^{2.5}$     & 1.67  & 21.1 \\ 
            $10^{2.5}   - 10^{3}$       & 1.65  & 21.8 \\ 
            $10^{3}     - 10^{3.1}$     & 1.49  & 22.1 \\ 
            $10^{3.1}   - 10^{3.2}$     & 1.32  & 22.3 \\ 
            $10^{3.2}   - 10^{3.3}$     & 1.33  & 22.4 \\ 
            $10^{3.3}   - 10^{3.5}$     & 0.907 & 23.1 \\ 
            $10^{3.5}   - 10^{3.85}$    & 1.03  & 23.0 \\ 
            $10^{3.85}  - 10^{5}$       & 0.924 & 21.6 \\ 
            \hline 
            \hline 
            \multicolumn{3}{c}{BGS Faint}\\
            \multicolumn{3}{c}{$(z - W1) - 1.2(g - r) + 1.2 < 0$} \\
            \hline 
            $10^{2.5}   - 10^{3}$       & 1.48  & 20.9 \\ 
            $10^{3}     - 10^{3.1}$     & 2.40  & 21.2 \\ 
            $10^{3.1}   - 10^{3.2}$     & 1.30  & 21.8 \\ 
            $10^{3.2}   - 10^{3.3}$     & 1.27  & 22.0 \\ 
            $10^{3.3}   - 10^{3.5}$     & 1.83  & 21.6 \\ 
            $10^{3.5}   - 10^{3.85}$    & 0.798 & 22.9 \\ 
            $10^{3.85}  - 10^{5}$       & 1.29  & 20.6 \\ 
            \hline 
        \end{tabular} \label{tab:zfail}
    \end{center}
\end{table}



\begin{figure}
\begin{center}
    \includegraphics[width=0.5\textwidth]{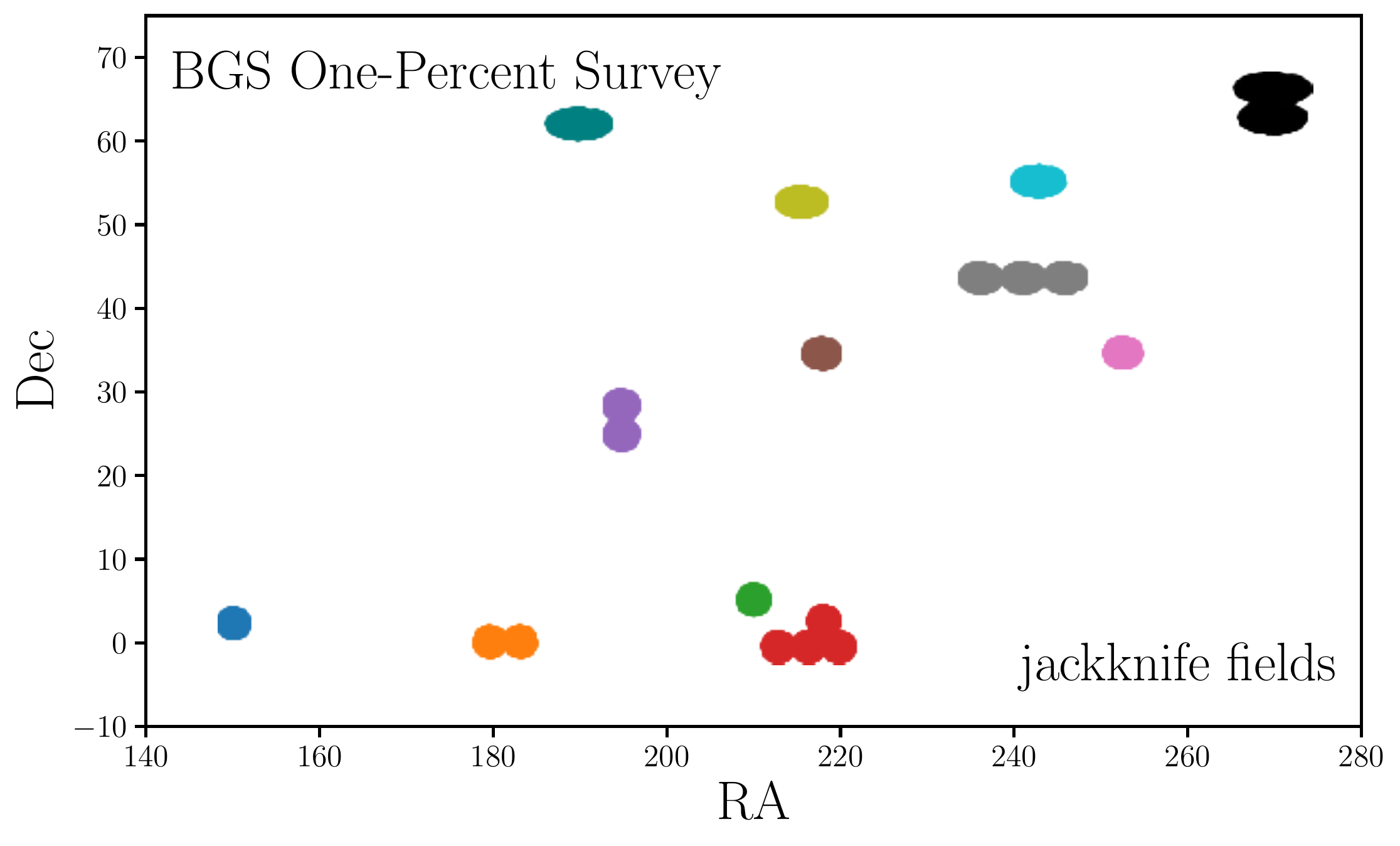} 
    \caption{
        The RA and Dec of the 12 jackknife fields of the BGS One-Percent Survey
        used to estimate the uncertainties on the SMF from sample variance. 
        We mark each field with a distinct color. 
    }\label{fig:jack}
\end{center}
\end{figure}

\section{Uncertainties on the SMF} \label{sec:jack}
We estimate the uncertainties of the pSMF from sample variance using the standard 
jackknife technique. 
We split the BGS sample into subsamples and then estimate uncertainties using the 
subsample-to-subsample variations:  
\begin{equation} \label{eq:jack} 
    \sigma_\Phi = \left(\frac{N_{\rm jack}-1}{N_{\rm jack}}
    \sum\limits_{k=1}^{N_{\rm jack}} (\Phi_k - \Phi)^2 \right).
\end{equation} 
$N_{\rm jack}$ is the number of jackknife subsamples and $\Phi_k$ represents
the SMF estimated from the BGS galaxies excluding the jackknife subsample $k$. 
In this work, we split the BGS sample into 12 jackknife fields based on the
angular positions of galaxies. 
We present the jackknife fields in Figure~\ref{fig:jack} with distinct colors. 

\section{Stellar Mass Completeness} \label{sec:mscomp}
In this appendix, we describe how we derive $M_{\rm lim}$, the $M_*$ limit
above which our BGS Bright sample is complete. 
Although there are various methods for estimating $M_{\rm lim}$ in the
literature, \emph{e.g.} based on estimating the mass-to-light
ratio~\citep{pozzetii2010, moustakas2013}, we adopt a simple approach that
takes advantage of the fact that BGS Bright is a magnitude-limited sample. 

\begin{figure}
\begin{center}
    \includegraphics[width=0.45\textwidth]{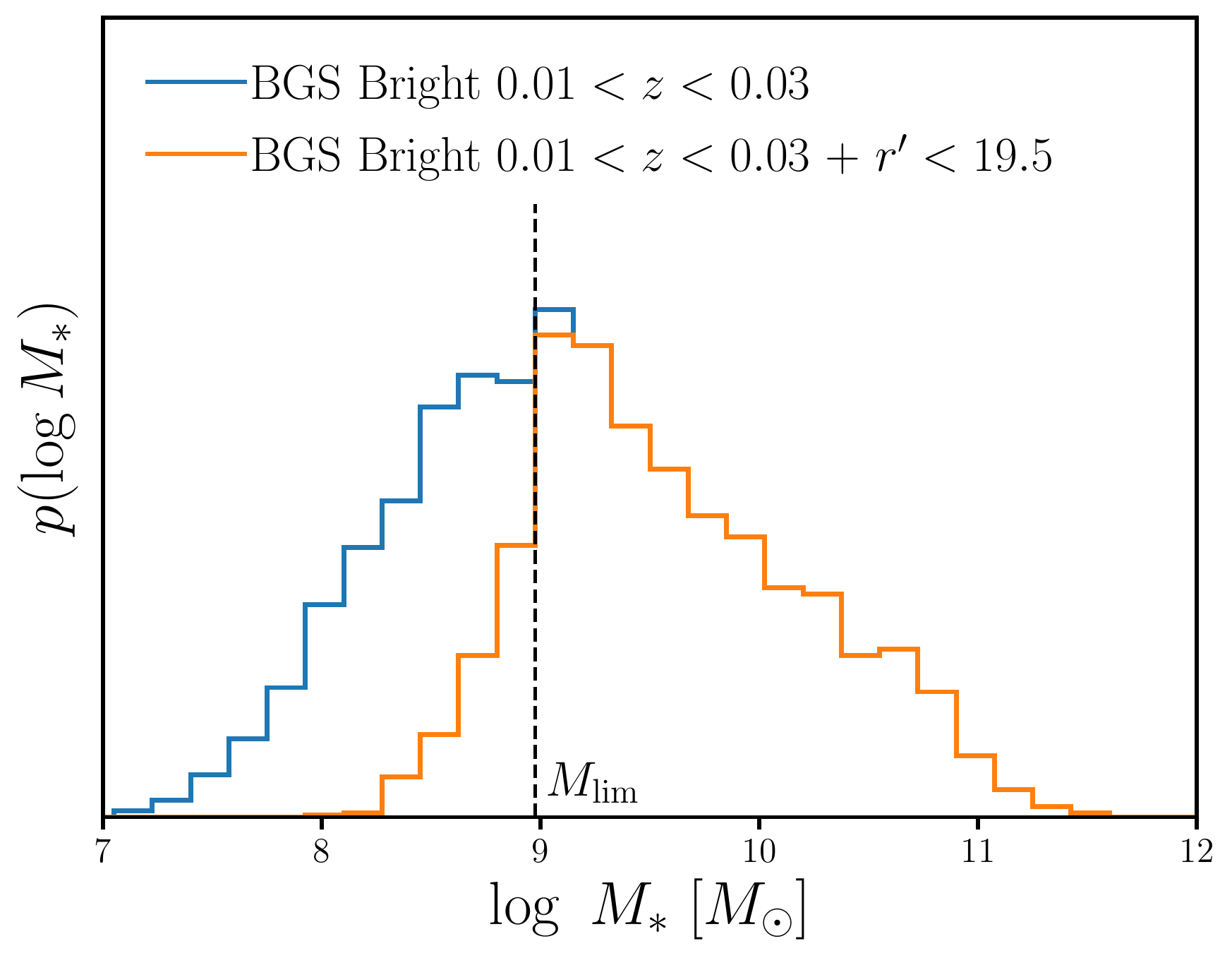}
    \caption{
        The $M_*$ distribution of BGS Bright galaxies with $0.01 < z < 0.03$
        (blue) and the $M_*$ distribution of same set of galaxies that would
        remain in the BGS Bright magnitude limit if they were redshifted to 
        $z' = z + 0.02$. 
        We set the stellar mass completeness limit, $M_{\rm lim}$, for $0.01 <
        z < 0.05$ to the $M_*$ where more than 10\% of galaxies are excluded in
        the latter distribution. 
        For each $\Delta z$ bin, we repeat this procedure to derive 
        $M_{\rm lim}$ values. 
    }\label{fig:ms_comp0}
\end{center}
\end{figure}

To derive $M_{\rm lim}$ in redshift bins of width $\Delta z=0.04$, we first
split the galaxy sample into narrower bins of $\Delta z/2$. 
For each narrower redshift bin, $i \Delta z/2 < z < (i+1) \Delta z/2$, we take
the best-fit {\sc PROVABGS} SEDs for all of the galaxies in the bin and
artificially redshift them to $z' = z + \Delta z/2$.
Then at $z'$, the galaxies would have fluxes of
\begin{equation}
    f_\lambda' = f_\lambda \frac{d_L(z)^2}{d_L(z')^2}.
\end{equation} 
$d_L(z)$ represents the luminosity distance at redshift $z$. 
Afterward, we calculate the $r$-band magnitudes, $r'$, for $f_\lambda'$ and
impose the $r' < 19.5$ magnitude limit of the BGS Bright. 
We then compare the $M_*$ distribution of all the galaxies in 
$i \Delta z/2 < z < (i+1) \Delta z/2$ to the galaxies in 
$(i+1) \Delta z/2 < z < (i+2) \Delta z/2$ with $r' < 19.5$.
For instance,  we present the $M_*$ distributions of all BGS
Bright galaxies in $0.01 < z < 0.03$ (blue) and the BGS Bright galaxies in
$0.01 < z < 0.03$ with $r' < 19.5$ (orange) in Figure~\ref{fig:ms_comp0}.

\begin{table} 
    \caption{Stellar mass completeness limit, $M_{\rm lim}$ for redshift bins
    of width $\Delta z = 0.04$.} 
    \begin{center}
        \begin{tabular}{lc} \toprule
            $z$ range & $\log_{10} M_{\rm lim}$ \\[3pt]
            \hline 
            $0.01 - 0.05$   & 8.975 \\ 
            $0.05 - 0.09$   & 9.500 \\ 
            $0.09 - 0.13$   & 10.20 \\ 
            $0.13 - 0.17$   & 10.38 \\ 
            $0.17 - 0.21$   & 10.72 \\ 
            \hline            
\end{tabular} \label{tab:mscomp}
\end{center}
\end{table}

\begin{figure}[h]
\begin{center}
    \includegraphics[width=0.6\textwidth]{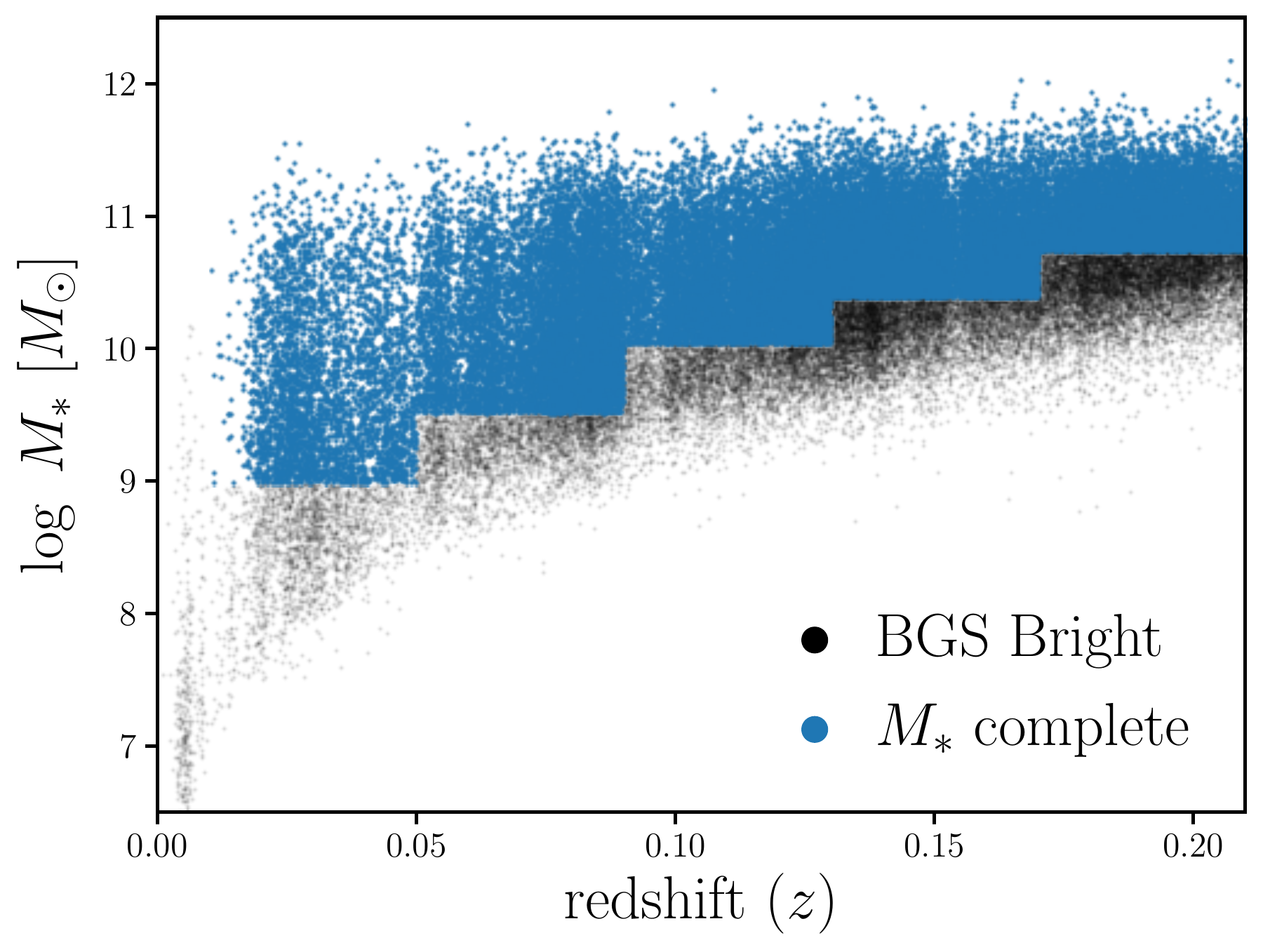}
    \caption{
        $M_*$ and redshift relation of BGS Bright galaxies in the One-Percent 
        Survey (black) and the galaxies within the stellar mass completeness 
        limit ($M_* < M_{\rm lim}$; blue). 
        $M_{\rm lim}$ is derived in redshift bins of width $\Delta z = 0.04$. 
        The lowest redshift bin ($0.01 < z < 0.05$) is complete down to 
        $M_* < 10^9 M_\odot$. 
    }\label{fig:ms_comp1}
\end{center}
\end{figure}

Since galaxies become fainter when they are placed at higher redshifts,
\emph{i.e.} $r' > r$, the $r' < 19.5$ sample has fewer low $M_*$ galaxies. 
We determine the $M_*$ at which, more than 10\% of galaxies are excluded in the
$r' < 19.5$ sample (black dashed) and set this limit as $M_{\rm lim}$ for the
redshift bins: $0.01 < z < 0.05$.
Our procedure for deriving $M_{\rm lim}$ takes advantage of the fact that
galaxy samples at lower redshifts are complete down to lower $M_*$ than at
higher redshifts. 
We repeat this procedure for all the $\Delta z = 0.04$ redshift bins that we
use to measure the SMF.
In Table~\ref{tab:mscomp}, we list $M_{\rm lim}$ values for each of the
redshift bins. 
Furthermore, we present the $M_*$ and redshift relation of BGS Bright galaxies
(black) and the stellar mass complete sample (blue) in
Figure~\ref{fig:ms_comp1}. 

\section{Author Affiliations} \label{sec:affil}
\noindent$^1$Department of Astrophysical Sciences, Princeton University, Peyton Hall, Princeton NJ 08544, USA\\ 
$^2$Lawrence Berkeley National Laboratory, 1 Cyclotron Road, Berkeley, CA 94720, USA\\
$^3$Tata Institute of Fundamental Research, Homi Bhabha Road, Mumbai 400005, India\\
$^4$Physics Dept., Boston University, 590 Commonwealth Avenue, Boston, MA 02215, USA
$^5$Department of Physics \& Astronomy, University College London, Gower Street, London, WC1E 6BT, UK\\
$^6$Institute for Computational Cosmology, Department of Physics, Durham University, South Road, Durham DH1 3LE, UK\\
$^7$Instituto de F\'{\i}sica, Universidad Nacional Aut\'{o}noma de M\'{e}xico,  Cd. de M\'{e}xico  C.P. 04510,  M\'{e}xico\\
$^8$Institut de F\'{i}sica d’Altes Energies (IFAE), The Barcelona Institute of Science and Technology, Campus UAB, 08193 Bellaterra Barcelona, Spain\\
$^9$Departamento de F\'isica, Universidad de los Andes, Cra. 1 No. 18A-10, Edificio Ip, CP 111711, Bogot\'a, Colombia\\
$^{10}$Observatorio Astron\'omico, Universidad de los Andes, Cra. 1 No. 18A-10, Edificio H, CP 111711 Bogot\'a, Colombia\\ 
$^{11}$Center for Cosmology and AstroParticle Physics, The Ohio State University, 191 West Woodruff Avenue, Columbus, OH 43210, USA\\
$^{12}$Department of Physics, The Ohio State University, 191 West Woodruff Avenue, Columbus, OH 43210, USA\\
$^{13}$Department of Astronomy, Tsinghua University, 30 Shuangqing Road, Haidian District, Beijing, China, 100190\\
$^{14}$NSF's NOIRLab, 950 N. Cherry Ave., Tucson, AZ 85719, USA\\
$^{15}$Instituci\'{o} Catalana de Recerca i Estudis Avan\c{c}ats, Passeig de Llu\'{\i}s Companys, 23, 08010 Barcelona, Spain\\
$^{16}$Department of Physics and Astronomy, Siena College, 515 Loudon Road, Loudonville, NY 12211, USA\\
$^{17}${National Astronomical Observatories, Chinese Academy of Sciences, A20 Datun Rd., Chaoyang District, Beijing, 100012, P.R. China\\
$^{18}$Space Sciences Laboratory, University of California, Berkeley, 7 Gauss Way, Berkeley, CA  94720, USA\\
$^{19}$University of California, Berkeley, 110 Sproul Hall \#5800 Berkeley, CA 94720, USA\\
$^{20}$Department of Physics and Astronomy, Sejong University, Seoul, 143-747, Korea\\
$^{21}$CIEMAT, Avenida Complutense 40, E-28040 Madrid, Spain\\
$^{22}$Korea Astronomy and Space Science Institute, 776, Daedeokdae-ro, Yuseong-gu, Daejeon 34055, Republic of Korea\\
$^{23}$Department of Physics, University of Michigan, Ann Arbor, MI 48109, USA\\
$^{24}$University of Michigan, Ann Arbor, MI 48109, USA\\
$^{25}$Department of Physics \& Astronomy, Ohio University, Athens, OH 45701, USA\\
$^{26}$Institute of Space Sciences, ICE-CSIC, Campus UAB, Carrer de Can Magrans s/n, 08913 Bellaterra, Barcelona, Spain\\
$^{27}$Kavli Institute for Particle Astrophysics and Cosmology, Stanford University, Menlo Park, CA 94305, USA\\
$^{28}$Physics Department, Stanford University, Stanford, CA 93405, USA\\
$^{29}$SLAC National Accelerator Laboratory, Menlo Park, CA 94305, USA

\bibliography{psmf} 

\begin{thebibliography}{}
\expandafter\ifx\csname natexlab\endcsname\relax\def\natexlab#1{#1}\fi
\providecommand{\url}[1]{\href{#1}{#1}}
\providecommand{\dodoi}[1]{doi:~\href{http://doi.org/#1}{\nolinkurl{#1}}}
\providecommand{\doeprint}[1]{\href{http://ascl.net/#1}{\nolinkurl{http://ascl.net/#1}}}
\providecommand{\doarXiv}[1]{\href{https://arxiv.org/abs/#1}{\nolinkurl{https://arxiv.org/abs/#1}}}

\bibitem[{Abareshi {et~al.}(2022)Abareshi, Aguilar, Ahlen, Alam, Alexander,
  Alfarsy, Allen, Prieto, Alves, Ameel, Armengaud, Asorey, Aviles, Bailey,
  {Balaguera-Antol{\'i}nez}, Ballester, Baltay, Bault, Beltran, Benavides,
  BenZvi, Berti, Besuner, Beutler, Bianchi, Blake, Blanc, Blum, Bolton, Bose,
  Bramall, Brieden, Brodzeller, Brooks, Brownewell, {Buckley-Geer}, Cahn, Cai,
  Canning, Rosell, Carton, Casas, Castander, {Cervantes-Cota}, Chabanier,
  Chaussidon, Chuang, Circosta, Cole, Cooper, {da Costa}, Cousinou, Cuceu,
  Davis, Dawson, {de la Cruz-Noriega}, {de la Macorra}, {de Mattia},
  Della~Costa, Demmer, Derwent, Dey, Dey, Dhungana, Ding, Dobson, Doel,
  {Donald-McCann}, Donaldson, Douglass, Duan, Dunlop, Edelstein, Eftekharzadeh,
  Eisenstein, {Enriquez-Vargas}, Escoffier, Evatt, Fagrelius, Fan, Fanning,
  Fawcett, Ferraro, Ereza, Flaugher, {Font-Ribera}, {Forero-Romero}, Frenk,
  Fromenteau, G{\"a}nsicke, {Garcia-Quintero}, Garrison, Gazta{\~n}aga,
  Gerardi, {Gil-Mar{\'i}n}, Gontcho, {Gonzalez-Morales}, {Gonzalez-de-Rivera},
  {Gonzalez-Perez}, Gordon, Graur, Green, Grove, Gruen, Gutierrez, Guy, Hahn,
  Harris, Herrera, {Herrera-Alcantar}, Honscheid, Howlett, Huterer, Ir{\v
  s}i{\v c}, Ishak, Jelinsky, Jiang, Jimenez, Jing, Joyce, Jullo, Juneau,
  Kara{\c c}ayl{\i}, Karamanis, Karcher, Karim, Kehoe, Kent, Kirkby, Kisner,
  Kitaura, Koposov, Kov{\'a}cs, Kremin, Krolewski, L'Huillier, Lahav, Lambert,
  Lamman, Lan, Landriau, Lane, Lang, Lange, Lasker, Guillou, Leauthaud,
  Van~Suu, Levi, Li, Magneville, Manera, Manser, Marshall, McCollam, McDonald,
  Meisner, Mezcua, Miller, Miquel, {Montero-Camacho}, Moon, Martini,
  {Meneses-Rizo}, Moustakas, Mueller, {Mu{\~n}oz-Guti{\'e}rrez}, Myers,
  Nadathur, Najita, Napolitano, Neilsen, Newman, Nie, Ning, Niz, Norberg,
  Noriega, O'Brien, Obuljen, {Palanque-Delabrouille}, Palmese, Zhiwei,
  Pappalardo, Peng, Percival, Perruchot, Pogge, Poppett, Porredon, Prada,
  Prochaska, Pucha, {P{\'e}rez-Fern{\'a}ndez}, {P{\'e}rez-R{\'a}fols},
  Rabinowitz, Raichoor, {Ramirez-Solano}, {Ram{\'i}rez-P{\'e}rez}, Ravoux,
  Reil, Rezaie, Rocher, Rockosi, Roe, Roodman, Ross, Rossi, Ruggeri,
  {Ruhlmann-Kleider}, Sabiu, Safonova, Said, Saintonge, Catonga, Samushia,
  Sanchez, Saulder, Schaan, Schlafly, Schlegel, Schmoll, Scholte, Schubnell,
  Secroun, Seo, Serrano, Sharples, Sholl, Silber, Silva, Sirk, Siudek, Smith,
  Sprayberry, Staten, Stupak, Tan, Tarl{\'e}, Tie, Tojeiro,
  {Ure{\~n}a-L{\'o}pez}, Valdes, Valenzuela, Valluri, {Vargas-Maga{\~n}a},
  Verde, Walther, Wang, Wang, Weaver, Weaverdyck, Wechsler, Wilson, Yang, Yu,
  Yuan, Y{\`e}che, Zhang, Zhang, Zhao, Zhou, Zhou, Zou, Zou, Zou, \&
  Zu}]{abareshi2022}
Abareshi, B., Aguilar, J., Ahlen, S., {et~al.} 2022, Overview of the
  {{Instrumentation}} for the {{Dark Energy Spectroscopic Instrument}},
  \dodoi{10.48550/arXiv.2205.10939}

\bibitem[{{Alexander} {et~al.}(2023){Alexander}, {Davis}, {Chaussidon},
  {Fawcett}, {X. Gonzalez-Morales}, {Lan}, {Y{\`e}che}, {Ahlen}, {Aguilar},
  {Armengaud}, {Bailey}, {Brooks}, {Cai}, {Canning}, {Carr}, {Chabanier},
  {Cousinou}, {Dawson}, {de la Macorra}, {Dey}, {Dey}, {Dhungana}, {Edge},
  {Eftekharzadeh}, {Fanning}, {Farr}, {Font-Ribera}, {Garcia-Bellido},
  {Garrison}, {Gazta{\~n}aga}, {A Gontcho}, {Gordon}, {Medellin Gonzalez},
  {Guy}, {Herrera-Alcantar}, {Jiang}, {Juneau}, {Kara{\c{c}}ayl{\i}}, {Kehoe},
  {Kisner}, {Kov{\'a}cs}, {Landriau}, {Levi}, {Magneville}, {Martini},
  {Meisner}, {Mezcua}, {Miquel}, {Camacho}, {Moustakas},
  {Mu{\~n}oz-Guti{\'e}rrez}, {Myers}, {Nadathur}, {Napolitano}, {Nie},
  {Palanque-Delabrouille}, {Pan}, {Percival}, {P{\'e}rez-R{\`a}fols},
  {Poppett}, {Prada}, {Ram{\'\i}rez-P{\'e}rez}, {Ravoux}, {Rosario},
  {Schubnell}, {Tarl{\'e}}, {Walther}, {Weiner}, {Youles}, {Zhou}, {Zou}, \&
  {Zou}}]{viqso}
{Alexander}, D.~M., {Davis}, T.~M., {Chaussidon}, E., {et~al.} 2023, \aj, 165,
  124, \dodoi{10.3847/1538-3881/acacfc}

\bibitem[{{Allende Prieto} {et~al.}(2020){Allende Prieto}, {Cooper}, {Dey},
  {G{\"a}nsicke}, {Koposov}, {Li}, {Manser}, {Nidever}, {Rockosi}, {Wang},
  {Aguado}, {Blum}, {Brooks}, {Eisenstein}, {Duan}, {Eftekharzadeh},
  {Gazta{\~n}aga}, {Kehoe}, {Landriau}, {Lee}, {Levi}, {Meisner}, {Myers},
  {Najita}, {Olsen}, {Palanque-Delabrouille}, {Poppett}, {Prada}, {Schlegel},
  {Schubnell}, {Tarl{\'e}}, {Valluri}, {Wechsler}, \&
  {Y{\`e}che}}]{allendo2020}
{Allende Prieto}, C., {Cooper}, A.~P., {Dey}, A., {et~al.} 2020, Research Notes
  of the American Astronomical Society, 4, 188,
  \dodoi{10.3847/2515-5172/abc1dc}

\bibitem[{{Bailey et al.}(2023)}]{redrock2023}
{Bailey et al.} 2023

\bibitem[{Baldry {et~al.}(2006)Baldry, Balogh, Bower, Glazebrook, Nichol,
  Bamford, \& Budavari}]{baldry2006}
Baldry, I.~K., Balogh, M.~L., Bower, R.~G., {et~al.} 2006, Monthly Notices of
  the Royal Astronomical Society, 373, 469,
  \dodoi{10.1111/j.1365-2966.2006.11081.x}

\bibitem[{Baronchelli {et~al.}(2020)Baronchelli, Nandra, \&
  Buchner}]{baronchelli2020}
Baronchelli, L., Nandra, K., \& Buchner, J. 2020, Monthly Notices of the Royal
  Astronomical Society, 498, 5284, \dodoi{10.1093/mnras/staa2684}

\bibitem[{Behroozi {et~al.}(2019)Behroozi, Wechsler, Hearin, \&
  Conroy}]{behroozi2019}
Behroozi, P., Wechsler, R.~H., Hearin, A.~P., \& Conroy, C. 2019, Monthly
  Notices of the Royal Astronomical Society, 1134,
  \dodoi{10.1093/mnras/stz1182}

\bibitem[{Benson(2012)}]{benson2012}
Benson, A.~J. 2012, New Astronomy, 17, 175,
  \dodoi{10.1016/j.newast.2011.07.004}

\bibitem[{Bernardi {et~al.}(2017)Bernardi, Meert, Sheth, Fischer,
  {Huertas-Company}, Maraston, Shankar, \& Vikram}]{bernardi2017}
Bernardi, M., Meert, A., Sheth, R.~K., {et~al.} 2017, Monthly Notices of the
  Royal Astronomical Society, 467, 2217, \dodoi{10.1093/mnras/stx176}

\bibitem[{Blanton \& Moustakas(2009)}]{blanton2009}
Blanton, M.~R., \& Moustakas, J. 2009, Annual Review of Astronomy and
  Astrophysics, 47, 159, \dodoi{10.1146/annurev-astro-082708-101734}

\bibitem[{{Blanton} {et~al.}(2003){Blanton}, {Hogg}, {Bahcall}, {Brinkmann},
  {Britton}, {Connolly}, {Csabai}, {Fukugita}, {Loveday}, {Meiksin}, {Munn},
  {Nichol}, {Okamura}, {Quinn}, {Schneider}, {Shimasaku}, {Strauss}, {Tegmark},
  {Vogeley}, \& {Weinberg}}]{blanton2003}
{Blanton}, M.~R., {Hogg}, D.~W., {Bahcall}, N.~A., {et~al.} 2003, \apj, 592,
  819, \dodoi{10.1086/375776}

\bibitem[{{Bolton} \& {Schlegel}(2010)}]{bolton2010}
{Bolton}, A.~S., \& {Schlegel}, D.~J. 2010, \pasp, 122, 248,
  \dodoi{10.1086/651008}

\bibitem[{{Brodzeller et al.}(2023)}]{redrockqso2023}
{Brodzeller et al.} 2023

\bibitem[{Chabrier(2003)}]{chabrier2003}
Chabrier, G. 2003, Publications of the Astronomical Society of the Pacific,
  115, 763, \dodoi{10.1086/376392}

\bibitem[{{Charlot} \& {Fall}(2000)}]{charlot2000}
{Charlot}, S., \& {Fall}, S.~M. 2000, \apj, 539, 718, \dodoi{10.1086/309250}

\bibitem[{{Chaussidon} {et~al.}(2023){Chaussidon}, {Y{\`e}che},
  {Palanque-Delabrouille}, {Alexander}, {Yang}, {Ahlen}, {Bailey}, {Brooks},
  {Cai}, {Chabanier}, {Davis}, {Dawson}, {de laMacorra}, {Dey}, {Dey},
  {Eftekharzadeh}, {Eisenstein}, {Fanning}, {Font-Ribera}, {Gazta{\~n}aga}, {A
  Gontcho}, {Gonzalez-Morales}, {Guy}, {Herrera-Alcantar}, {Honscheid},
  {Ishak}, {Jiang}, {Juneau}, {Kehoe}, {Kisner}, {Kov{\'a}cs}, {Kremin}, {Lan},
  {Landriau}, {Le Guillou}, {Levi}, {Magneville}, {Martini}, {Meisner},
  {Moustakas}, {Mu{\~n}oz-Guti{\'e}rrez}, {Myers}, {Newman}, {Nie}, {Percival},
  {Poppett}, {Prada}, {Raichoor}, {Ravoux}, {Ross}, {Schlafly}, {Schlegel},
  {Tan}, {Tarl{\'e}}, {Zhou}, {Zhou}, \& {Zou}}]{chaussidon2023}
{Chaussidon}, E., {Y{\`e}che}, C., {Palanque-Delabrouille}, N., {et~al.} 2023,
  \apj, 944, 107, \dodoi{10.3847/1538-4357/acb3c2}

\bibitem[{Choi {et~al.}(2016)Choi, Dotter, Conroy, Cantiello, Paxton, \&
  Johnson}]{choi2016}
Choi, J., Dotter, A., Conroy, C., {et~al.} 2016, The Astrophysical Journal,
  823, 102, \dodoi{10.3847/0004-637X/823/2/102}

\bibitem[{Coil {et~al.}(2011)Coil, Blanton, Burles, Cool, Eisenstein,
  Moustakas, Wong, Zhu, Aird, Bernstein, Bolton, \& Hogg}]{coil2011}
Coil, A.~L., Blanton, M.~R., Burles, S.~M., {et~al.} 2011, The Astrophysical
  Journal, 741, 8, \dodoi{10.1088/0004-637X/741/1/8}

\bibitem[{Conroy {et~al.}(2009)Conroy, Gunn, \& White}]{conroy2009}
Conroy, C., Gunn, J.~E., \& White, M. 2009, The Astrophysical Journal, 699,
  486, \dodoi{10.1088/0004-637X/699/1/486}

\bibitem[{Conroy {et~al.}(2010)Conroy, White, \& Gunn}]{conroy2010b}
Conroy, C., White, M., \& Gunn, J.~E. 2010, The Astrophysical Journal, 708, 58,
  \dodoi{10.1088/0004-637X/708/1/58}

\bibitem[{{Cooper} {et~al.}(2022){Cooper}, {Koposov}, {Allende Prieto},
  {Manser}, {Kizhuprakkat}, {Myers}, {Dey}, {Gaensicke}, {Li}, {Rockosi},
  {Valluri}, {Najita}, {Deason}, {Raichoor}, {Wang}, {Ting}, {Kim}, {Carrillo},
  {Wang}, {Beraldo e Silva}, {Han}, {Ding}, {Sanchez-Conde}, {Aguilar},
  {Ahlen}, {Bailey}, {Belokurov}, {Brooks}, {Cunha}, {Dawson}, {de la Macorra},
  {Doel}, {Eisenstein}, {Fagrelius}, {Fanning}, {Font-Ribera}, {Forero-Romero},
  {Gaztanaga}, {Gontcho}, {Guy}, {Honscheid}, {Kehoe}, {Kisner}, {Kremin},
  {Landriau}, {Levi}, {Martini}, {Meisner}, {Miquel}, {Moustakas}, {Nie},
  {Palanque-Delabrouille}, {Percival}, {Poppett}, {Prada}, {Rehemtulla},
  {Schlafly}, {Schlegel}, {Schubnell}, {Sharples}, {Tarle}, {Wechsler},
  {Weinberg}, {Zhou}, \& {Zou}}]{cooper2022}
{Cooper}, A.~P., {Koposov}, S.~E., {Allende Prieto}, C., {et~al.} 2022, arXiv
  e-prints, arXiv:2208.08514.
\newblock \doarXiv{2208.08514}

\bibitem[{Daddi {et~al.}(2007)Daddi, Dickinson, Morrison, Chary, Cimatti,
  Elbaz, Frayer, Renzini, Pope, Alexander, Bauer, Giavalisco, Huynh, Kurk, \&
  Mignoli}]{daddi2007}
Daddi, E., Dickinson, M., Morrison, G., {et~al.} 2007, The Astrophysical
  Journal, 670, 156, \dodoi{10.1086/521818}

\bibitem[{{Darragh-Ford} {et~al.}(2022){Darragh-Ford}, {Wu}, {Mao}, {Wechsler},
  {Geha}, {Forero-Romero}, {Hahn}, {Kallivayalil}, {Moustakas}, {Nadler},
  {Nowotka}, {Peek}, {Tollerud}, {Weiner}, {Aguilar}, {Ahlen}, {Brooks},
  {Cooper}, {de la Macorra}, {Dey}, {Fanning}, {Font-Ribera}, {Gontcho},
  {Honscheid}, {Kisner}, {Kremin}, {Landriau}, {Levi}, {Martini}, {Meisner},
  {Miquel}, {Myers}, {Nie}, {Palanque-Delabrouille}, {Percival}, {Prada},
  {Schlegel}, {Schubnell}, {Tarl{\'e}}, {Vargas-Maga{\~n}a}, {Zhou}, \&
  {Zou}}]{darragh-ford2022}
{Darragh-Ford}, E., {Wu}, J.~F., {Mao}, Y.-Y., {et~al.} 2022, arXiv e-prints,
  arXiv:2212.07433, \dodoi{10.48550/arXiv.2212.07433}

\bibitem[{Dav{\'e} {et~al.}(2017)Dav{\'e}, Rafieferantsoa, \&
  Thompson}]{dave2017a}
Dav{\'e}, R., Rafieferantsoa, M.~H., \& Thompson, R.~J. 2017, arXiv:1704.01135
  [astro-ph].
\newblock \doeprint{1704.01135}

\bibitem[{{DESI Collaboration} {et~al.}(2016{\natexlab{a}}){DESI
  Collaboration}, Aghamousa, Aguilar, Ahlen, Alam, Allen, Prieto, Annis,
  Bailey, Balland, Ballester, Baltay, Beaufore, Bebek, Beers, Bell, Bernal,
  Besuner, Beutler, Blake, Bleuler, Blomqvist, Blum, Bolton, Briceno, Brooks,
  Brownstein, {Buckley-Geer}, Burden, Burtin, Busca, Cahn, Cai, {Cardiel-Sas},
  Carlberg, Carton, Casas, Castander, {Cervantes-Cota}, Claybaugh, Close,
  Coker, Cole, Comparat, Cooper, Cousinou, Crocce, Cuby, Cunningham, Davis,
  Dawson, {de la Macorra}, De~Vicente, Delubac, Derwent, Dey, Dhungana, Ding,
  Doel, Duan, Ealet, Edelstein, Eftekharzadeh, Eisenstein, Elliott, Escoffier,
  Evatt, Fagrelius, Fan, Fanning, Farahi, Farihi, Favole, Feng, Fernandez,
  Findlay, Finkbeiner, Fitzpatrick, Flaugher, Flender, {Font-Ribera},
  {Forero-Romero}, Fosalba, Frenk, Fumagalli, Gaensicke, Gallo,
  {Garcia-Bellido}, Gaztanaga, Fusillo, Gerard, Gershkovich, Giannantonio,
  Gillet, {Gonzalez-de-Rivera}, {Gonzalez-Perez}, Gott, Graur, Gutierrez, Guy,
  Habib, Heetderks, Heetderks, Heitmann, Hellwing, Herrera, Ho, Holland,
  Honscheid, Huff, Hutchinson, Huterer, Hwang, Laguna, Ishikawa, Jacobs,
  Jeffrey, Jelinsky, Jennings, Jiang, Jimenez, Johnson, Joyce, Jullo, Juneau,
  Kama, Karcher, Karkar, Kehoe, Kennamer, Kent, Kilbinger, Kim, Kirkby, Kisner,
  Kitanidis, Kneib, Koposov, Kovacs, Koyama, Kremin, Kron, Kronig,
  {Kueter-Young}, Lacey, Lafever, Lahav, Lambert, Lampton, Landriau, Lang,
  Lauer, Goff, Guillou, Van~Suu, Lee, Lee, Leitner, Lesser, Levi, L'Huillier,
  Li, Liang, Lin, Linder, Loebman, Luki{\'c}, Ma, MacCrann, Magneville,
  Makarem, Manera, Manser, Marshall, Martini, Massey, Matheson, McCauley,
  McDonald, McGreer, Meisner, Metcalfe, Miller, Miquel, Moustakas, Myers, Naik,
  Newman, Nichol, Nicola, {da Costa}, Nie, Niz, Norberg, Nord, Norman, Nugent,
  O'Brien, Oh, Olsen, Padilla, Padmanabhan, Padmanabhan,
  {Palanque-Delabrouille}, Palmese, Pappalardo, P{\^a}ris, Park, Patej,
  Peacock, Peiris, Peng, Percival, Perruchot, Pieri, Pogge, Pollack, Poppett,
  Prada, Prakash, Probst, Rabinowitz, Raichoor, Ree, Refregier, Regal, Reid,
  Reil, Rezaie, Rockosi, Roe, Ronayette, Roodman, Ross, Ross, Rossi, Rozo,
  {Ruhlmann-Kleider}, Rykoff, Sabiu, Samushia, Sanchez, Sanchez, Schlegel,
  Schneider, Schubnell, Secroun, Seljak, Seo, Serrano, Shafieloo, Shan,
  Sharples, Sholl, Shourt, Silber, Silva, Sirk, Slosar, Smith, Smoot, Som,
  Song, Sprayberry, Staten, Stefanik, Tarle, Tie, Tinker, Tojeiro, Valdes,
  Valenzuela, Valluri, {Vargas-Magana}, Verde, Walker, Wang, Wang, Weaver,
  Weaverdyck, Wechsler, Weinberg, White, Yang, Yeche, Zhang, Zhao, Zheng, Zhou,
  Zhou, Zhu, Zou, \& Zu}]{desicollaboration2016}
{DESI Collaboration}, Aghamousa, A., Aguilar, J., {et~al.} 2016{\natexlab{a}},
  arXiv:1611.00036 [astro-ph].
\newblock \doeprint{1611.00036}

\bibitem[{{DESI Collaboration} {et~al.}(2016{\natexlab{b}}){DESI
  Collaboration}, Aghamousa, Aguilar, Ahlen, Alam, Allen, Prieto, Annis,
  Bailey, Balland, Ballester, Baltay, Beaufore, Bebek, Beers, Bell, Bernal,
  Besuner, Beutler, Blake, Bleuler, Blomqvist, Blum, Bolton, Briceno, Brooks,
  Brownstein, {Buckley-Geer}, Burden, Burtin, Busca, Cahn, Cai, {Cardiel-Sas},
  Carlberg, Carton, Casas, Castander, {Cervantes-Cota}, Claybaugh, Close,
  Coker, Cole, Comparat, Cooper, Cousinou, Crocce, Cuby, Cunningham, Davis,
  Dawson, {de la Macorra}, De~Vicente, Delubac, Derwent, Dey, Dhungana, Ding,
  Doel, Duan, Ealet, Edelstein, Eftekharzadeh, Eisenstein, Elliott, Escoffier,
  Evatt, Fagrelius, Fan, Fanning, Farahi, Farihi, Favole, Feng, Fernandez,
  Findlay, Finkbeiner, Fitzpatrick, Flaugher, Flender, {Font-Ribera},
  {Forero-Romero}, Fosalba, Frenk, Fumagalli, Gaensicke, Gallo,
  {Garcia-Bellido}, Gaztanaga, Fusillo, Gerard, Gershkovich, Giannantonio,
  Gillet, {Gonzalez-de-Rivera}, {Gonzalez-Perez}, Gott, Graur, Gutierrez, Guy,
  Habib, Heetderks, Heetderks, Heitmann, Hellwing, Herrera, Ho, Holland,
  Honscheid, Huff, Hutchinson, Huterer, Hwang, Laguna, Ishikawa, Jacobs,
  Jeffrey, Jelinsky, Jennings, Jiang, Jimenez, Johnson, Joyce, Jullo, Juneau,
  Kama, Karcher, Karkar, Kehoe, Kennamer, Kent, Kilbinger, Kim, Kirkby, Kisner,
  Kitanidis, Kneib, Koposov, Kovacs, Koyama, Kremin, Kron, Kronig,
  {Kueter-Young}, Lacey, Lafever, Lahav, Lambert, Lampton, Landriau, Lang,
  Lauer, Goff, Guillou, Van~Suu, Lee, Lee, Leitner, Lesser, Levi, L'Huillier,
  Li, Liang, Lin, Linder, Loebman, Luki{\'c}, Ma, MacCrann, Magneville,
  Makarem, Manera, Manser, Marshall, Martini, Massey, Matheson, McCauley,
  McDonald, McGreer, Meisner, Metcalfe, Miller, Miquel, Moustakas, Myers, Naik,
  Newman, Nichol, Nicola, {da Costa}, Nie, Niz, Norberg, Nord, Norman, Nugent,
  O'Brien, Oh, Olsen, Padilla, Padmanabhan, Padmanabhan,
  {Palanque-Delabrouille}, Palmese, Pappalardo, P{\^a}ris, Park, Patej,
  Peacock, Peiris, Peng, Percival, Perruchot, Pieri, Pogge, Pollack, Poppett,
  Prada, Prakash, Probst, Rabinowitz, Raichoor, Ree, Refregier, Regal, Reid,
  Reil, Rezaie, Rockosi, Roe, Ronayette, Roodman, Ross, Ross, Rossi, Rozo,
  {Ruhlmann-Kleider}, Rykoff, Sabiu, Samushia, Sanchez, Sanchez, Schlegel,
  Schneider, Schubnell, Secroun, Seljak, Seo, Serrano, Shafieloo, Shan,
  Sharples, Sholl, Shourt, Silber, Silva, Sirk, Slosar, Smith, Smoot, Som,
  Song, Sprayberry, Staten, Stefanik, Tarle, Tie, Tinker, Tojeiro, Valdes,
  Valenzuela, Valluri, {Vargas-Magana}, Verde, Walker, Wang, Wang, Weaver,
  Weaverdyck, Wechsler, Weinberg, White, Yang, Yeche, Zhang, Zhao, Zheng, Zhou,
  Zhou, Zhu, Zou, \& Zu}]{desicollaboration2016a}
---. 2016{\natexlab{b}}, arXiv:1611.00037 [astro-ph].
\newblock \doeprint{1611.00037}

\bibitem[{{DESI Collaboration} {et~al.}(2023{\natexlab{a}}){DESI
  Collaboration}, {Adame}, {Aguilar}, {Ahlen}, {Alam}, {Aldering}, {Alexander},
  {Alfarsy}, {Allende Prieto}, {Alvarez}, {Alves}, {Anand}, {Andrade-Oliveira},
  {Armengaud}, {Asorey}, {Avila}, {Aviles}, {Bailey},
  {Balaguera-Antol{\'\i}nez}, {Ballester}, {Baltay}, {Bault}, {Bautista},
  {Behera}, {Beltran}, {BenZvi}, {Beraldo e Silva}, {Bermejo-Climent}, {Berti},
  {Besuner}, {Beutler}, {Bianchi}, {Blake}, {Blum}, {Bolton}, {Brieden},
  {Brodzeller}, {Brooks}, {Brown}, {Buckley-Geer}, {Burtin}, {Cabayol-Garcia},
  {Cai}, {Canning}, {Cardiel-Sas}, {Carnero Rosell}, {Castander},
  {Cervantes-Cota}, {Chabanier}, {Chaussidon}, {Chaves-Montero}, {Chen},
  {Chuang}, {Claybaugh}, {Cole}, {Cooper}, {Cuceu}, {Davis}, {Dawson}, {de
  Belsunce}, {de la Cruz}, {de la Macorra}, {de Mattia}, {Demina},
  {Demirbozan}, {DeRose}, {Dey}, {Dey}, {Dhungana}, {Ding}, {Ding}, {Doel},
  {Doshi}, {Douglass}, {Edge}, {Eftekharzadeh}, {Eisenstein}, {Elliott},
  {Escoffier}, {Fagrelius}, {Fan}, {Fanning}, {Fawcett}, {Ferraro}, {Ereza},
  {Flaugher}, {Font-Ribera}, {Forero-S{\'a}nchez}, {Forero-Romero}, {Frenk},
  {G{\"a}nsicke}, {Garc{\'\i}a}, {Garc{\'\i}a-Bellido}, {Garcia-Quintero},
  {Garrison}, {Gil-Mar{\'\i}n}, {Golden-Marx}, {Gontcho}, {Gonzalez-Morales},
  {Gonzalez-Perez}, {Gordon}, {Graur}, {Green}, {Gruen}, {Guy}, {Hadzhiyska},
  {Hahn}, {Han}, {Hanif}, {Herrera-Alcantar}, {Honscheid}, {Hou}, {Howlett},
  {Huterer}, {Ir{\v{s}}i{\v{c}}}, {Ishak}, {Jana}, {Jiang}, {Jimenez}, {Jing},
  {Joudaki}, {Jullo}, {Juneau}, {Kizhuprakkat}, {Kara{\c{c}}ayl{\i}}, {Karim},
  {Kehoe}, {Kent}, {Khederlarian}, {Kim}, {Kirkby}, {Kisner}, {Kitaura},
  {Kneib}, {Koposov}, {Kov{\'a}cs}, {Kremin}, {Krolewski}, {L'Huillier},
  {Lambert}, {Lamman}, {Lan}, {Landriau}, {Lang}, {Lange}, {Lasker}, {Le
  Guillou}, {Leauthaud}, {Levi}, {Li}, {Linder}, {Lyons}, {Magneville},
  {Manera}, {Manser}, {Margala}, {Martini}, {McDonald}, {Medina},
  {Medina-Varela}, {Meisner}, {Mena-Fern{\'a}ndez}, {Meneses-Rizo}, {Mezcua},
  {Miquel}, {Montero-Camacho}, {Moon}, {Moore}, {Moustakas}, {Mueller},
  {Mundet}, {Mu{\~n}oz-Guti{\'e}rrez}, {Myers}, {Nadathur}, {Napolitano},
  {Neveux}, {Newman}, {Nie}, {Niz}, {Norberg}, {Noriega}, {Paillas},
  {Palanque-Delabrouille}, {Palmese}, {Zhiwei}, {Parkinson}, {Penmetsa},
  {Percival}, {P{\'e}rez-Fern{\'a}ndez}, {P{\'e}rez-R{\`a}fols}, {Pieri},
  {Poppett}, {Porredon}, {Prada}, {Pucha}, {Raichoor},
  {Ram{\'\i}rez-P{\'e}rez}, {Ramirez-Solano}, {Rashkovetskyi}, {Ravoux},
  {Rocher}, {Rockosi}, {Ross}, {Rossi}, {Ruggeri}, {Ruhlmann-Kleider}, {Sabiu},
  {Said}, {Saintonge}, {Samushia}, {Sanchez}, {Saulder}, {Schaan}, {Schlafly},
  {Schlegel}, {Scholte}, {Schubnell}, {Seo}, {Shafieloo}, {Sharples}, {Sheu},
  {Silber}, {Sinigaglia}, {Siudek}, {Slepian}, {Smith}, {Sprayberry},
  {Stephey}, {Su{\'a}rez-P{\'e}rez}, {Sun}, {Tan}, {Tarl{\'e}}, {Tojeiro},
  {Ure{\~n}a-L{\'o}pez}, {Vaisakh}, {Valcin}, {Valdes}, {Valluri},
  {Vargas-Maga{\~n}a}, {Variu}, {Verde}, {Walther}, {Wang}, {Wang}, {Weaver},
  {Weaverdyck}, {Wechsler}, {White}, {Xie}, {Yang}, {Y{\`e}che}, {Yu}, {Yuan},
  {Zhang}, {Zhang}, {Zhao}, {Zheng}, {Zhou}, {Zhou}, {Zou}, {Zou}, \&
  {Zu}}]{sv}
{DESI Collaboration}, {Adame}, A.~G., {Aguilar}, J., {et~al.}
  2023{\natexlab{a}}, arXiv e-prints, arXiv:2306.06307,
  \dodoi{10.48550/arXiv.2306.06307}

\bibitem[{{DESI Collaboration} {et~al.}(2023{\natexlab{b}}){DESI
  Collaboration}, {Adame}, {Aguilar}, {Ahlen}, {Alam}, {Aldering}, {Alexander},
  {Alfarsy}, {Allende Prieto}, {Alvarez}, {Alves}, {Anand}, {Andrade-Oliveira},
  {Armengaud}, {Asorey}, {Avila}, {Aviles}, {Bailey},
  {Balaguera-Antol{\'\i}nez}, {Ballester}, {Baltay}, {Bault}, {Bautista},
  {Behera}, {Beltran}, {BenZvi}, {Beraldo e Silva}, {Bermejo-Climent}, {Berti},
  {Besuner}, {Beutler}, {Bianchi}, {Blake}, {Blum}, {Bolton}, {Brieden},
  {Brodzeller}, {Brooks}, {Brown}, {Buckley-Geer}, {Burtin}, {Cabayol-Garcia},
  {Cai}, {Canning}, {Cardiel-Sas}, {Carnero Rosell}, {Castander},
  {Cervantes-Cota}, {Chabanier}, {Chaussidon}, {Chaves-Montero}, {Chen},
  {Chuang}, {Claybaugh}, {Cole}, {Cooper}, {Cuceu}, {Davis}, {Dawson}, {de
  Belsunce}, {de la Cruz}, {de la Macorra}, {de Mattia}, {Demina},
  {Demirbozan}, {DeRose}, {Dey}, {Dey}, {Dhungana}, {Ding}, {Ding}, {Doel},
  {Doshi}, {Douglass}, {Edge}, {Eftekharzadeh}, {Eisenstein}, {Elliott},
  {Escoffier}, {Fagrelius}, {Fan}, {Fanning}, {Fawcett}, {Ferraro}, {Ereza},
  {Flaugher}, {Font-Ribera}, {Forero-S{\'a}nchez}, {Forero-Romero}, {Frenk},
  {G{\"a}nsicke}, {Garc{\'\i}a}, {Garc{\'\i}a-Bellido}, {Garcia-Quintero},
  {Garrison}, {Gil-Mar{\'\i}n}, {Golden-Marx}, {Gontcho}, {Gonzalez-Morales},
  {Gonzalez-Perez}, {Gordon}, {Graur}, {Green}, {Gruen}, {Guy}, {Hadzhiyska},
  {Hahn}, {Han}, {Hanif}, {Herrera-Alcantar}, {Honscheid}, {Hou}, {Howlett},
  {Huterer}, {Ir{\v{s}}i{\v{c}}}, {Ishak}, {Jacques}, {Jana}, {Jiang},
  {Jimenez}, {Jing}, {Joudaki}, {Jullo}, {Juneau}, {Kizhuprakkat},
  {Kara{\c{c}}ayl{\i}}, {Karim}, {Kehoe}, {Kent}, {Khederlarian}, {Kim},
  {Kirkby}, {Kisner}, {Kitaura}, {Kneib}, {Koposov}, {Kov{\'a}cs}, {Kremin},
  {Krolewski}, {L'Huillier}, {Lambert}, {Lamman}, {Lan}, {Landriau}, {Lang},
  {Lange}, {Lasker}, {Le Guillou}, {Leauthaud}, {Levi}, {Li}, {Linder},
  {Lyons}, {Magneville}, {Manera}, {Manser}, {Margala}, {Martini}, {McDonald},
  {Medina}, {Medina-Varela}, {Meisner}, {Mena-Fern{\'a}ndez}, {Meneses-Rizo},
  {Mezcua}, {Miquel}, {Montero-Camacho}, {Moon}, {Moore}, {Moustakas},
  {Mueller}, {Mundet}, {Mu{\~n}oz-Guti{\'e}rrez}, {Myers}, {Nadathur},
  {Napolitano}, {Neveux}, {Newman}, {Nie}, {Nikutta}, {Niz}, {Norberg},
  {Noriega}, {Paillas}, {Palanque-Delabrouille}, {Palmese}, {Zhiwei},
  {Parkinson}, {Penmetsa}, {Percival}, {P{\'e}rez-Fern{\'a}ndez},
  {P{\'e}rez-R{\`a}fols}, {Pieri}, {Poppett}, {Porredon}, {Pothier}, {Prada},
  {Pucha}, {Raichoor}, {Ram{\'\i}rez-P{\'e}rez}, {Ramirez-Solano},
  {Rashkovetskyi}, {Ravoux}, {Rocher}, {Rockosi}, {Ross}, {Rossi}, {Ruggeri},
  {Ruhlmann-Kleider}, {Sabiu}, {Said}, {Saintonge}, {Samushia}, {Sanchez},
  {Saulder}, {Schaan}, {Schlafly}, {Schlegel}, {Scholte}, {Schubnell}, {Seo},
  {Shafieloo}, {Sharples}, {Sheu}, {Silber}, {Sinigaglia}, {Siudek}, {Slepian},
  {Smith}, {Sprayberry}, {Stephey}, {Su{\'a}rez-P{\'e}rez}, {Sun}, {Tan},
  {Tarl{\'e}}, {Tojeiro}, {Ure{\~n}a-L{\'o}pez}, {Vaisakh}, {Valcin}, {Valdes},
  {Valluri}, {Vargas-Maga{\~n}a}, {Variu}, {Verde}, {Walther}, {Wang}, {Wang},
  {Weaver}, {Weaverdyck}, {Wechsler}, {White}, {Xie}, {Yang}, {Y{\`e}che},
  {Yu}, {Yuan}, {Zhang}, {Zhang}, {Zhao}, {Zheng}, {Zhou}, {Zhou}, {Zou},
  {Zou}, \& {Zu}}]{edr}
---. 2023{\natexlab{b}}, arXiv e-prints, arXiv:2306.06308,
  \dodoi{10.48550/arXiv.2306.06308}

\bibitem[{{Dey} {et~al.}(2019){Dey}, {Schlegel}, {Lang}, {Blum}, {Burleigh},
  {Fan}, {Findlay}, {Finkbeiner}, {Herrera}, {Juneau}, {Landriau}, {Levi},
  {McGreer}, {Meisner}, {Myers}, {Moustakas}, {Nugent}, {Patej}, {Schlafly},
  {Walker}, {Valdes}, {Weaver}, {Y{\`e}che}, {Zou}, {Zhou}, {Abareshi},
  {Abbott}, {Abolfathi}, {Aguilera}, {Alam}, {Allen}, {Alvarez}, {Annis},
  {Ansarinejad}, {Aubert}, {Beechert}, {Bell}, {BenZvi}, {Beutler}, {Bielby},
  {Bolton}, {Brice{\~n}o}, {Buckley-Geer}, {Butler}, {Calamida}, {Carlberg},
  {Carter}, {Casas}, {Castander}, {Choi}, {Comparat}, {Cukanovaite}, {Delubac},
  {DeVries}, {Dey}, {Dhungana}, {Dickinson}, {Ding}, {Donaldson}, {Duan},
  {Duckworth}, {Eftekharzadeh}, {Eisenstein}, {Etourneau}, {Fagrelius},
  {Farihi}, {Fitzpatrick}, {Font-Ribera}, {Fulmer}, {G{\"a}nsicke},
  {Gaztanaga}, {George}, {Gerdes}, {Gontcho}, {Gorgoni}, {Green}, {Guy},
  {Harmer}, {Hernandez}, {Honscheid}, {Huang}, {James}, {Jannuzi}, {Jiang},
  {Joyce}, {Karcher}, {Karkar}, {Kehoe}, {Kneib}, {Kueter-Young}, {Lan},
  {Lauer}, {Le Guillou}, {Le Van Suu}, {Lee}, {Lesser}, {Perreault Levasseur},
  {Li}, {Mann}, {Marshall}, {Mart{\'\i}nez-V{\'a}zquez}, {Martini}, {du Mas des
  Bourboux}, {McManus}, {Meier}, {M{\'e}nard}, {Metcalfe},
  {Mu{\~n}oz-Guti{\'e}rrez}, {Najita}, {Napier}, {Narayan}, {Newman}, {Nie},
  {Nord}, {Norman}, {Olsen}, {Paat}, {Palanque-Delabrouille}, {Peng},
  {Poppett}, {Poremba}, {Prakash}, {Rabinowitz}, {Raichoor}, {Rezaie},
  {Robertson}, {Roe}, {Ross}, {Ross}, {Rudnick}, {Safonova}, {Saha},
  {S{\'a}nchez}, {Savary}, {Schweiker}, {Scott}, {Seo}, {Shan}, {Silva},
  {Slepian}, {Soto}, {Sprayberry}, {Staten}, {Stillman}, {Stupak}, {Summers},
  {Sien Tie}, {Tirado}, {Vargas-Maga{\~n}a}, {Vivas}, {Wechsler}, {Williams},
  {Yang}, {Yang}, {Yapici}, {Zaritsky}, {Zenteno}, {Zhang}, {Zhang}, {Zhou}, \&
  {Zhou}}]{dey2019}
{Dey}, A., {Schlegel}, D.~J., {Lang}, D., {et~al.} 2019, \aj, 157, 168,
  \dodoi{10.3847/1538-3881/ab089d}

\bibitem[{Dickey {et~al.}(2021)Dickey, Starkenburg, Geha, Hahn,
  {Angl{\'e}s-Alc{\'a}zar}, Choi, Dav{\'e}, Genel, Iyer, Maller, Mandelker,
  Somerville, \& Yung}]{dickey2021}
Dickey, C.~M., Starkenburg, T.~K., Geha, M., {et~al.} 2021, The Astrophysical
  Journal, 915, 53, \dodoi{10.3847/1538-4357/abc014}

\bibitem[{Donnari {et~al.}(2021)Donnari, Pillepich, Nelson, Marinacci,
  Vogelsberger, \& Hernquist}]{donnari2021}
Donnari, M., Pillepich, A., Nelson, D., {et~al.} 2021, Monthly Notices of the
  Royal Astronomical Society, 506, 4760, \dodoi{10.1093/mnras/stab1950}

\bibitem[{Donnari {et~al.}(2019)Donnari, Pillepich, Nelson, Vogelsberger,
  Genel, Weinberger, Marinacci, Springel, \& Hernquist}]{donnari2019}
---. 2019, Monthly Notices of the Royal Astronomical Society, 485, 4817,
  \dodoi{10.1093/mnras/stz712}

\bibitem[{Dotter(2016)}]{dotter2016}
Dotter, A. 2016, The Astrophysical Journal Supplement Series, 222, 8,
  \dodoi{10.3847/0067-0049/222/1/8}

\bibitem[{Driver {et~al.}(2011)Driver, Hill, Kelvin, Robotham, Liske, Norberg,
  Baldry, Bamford, Hopkins, Loveday, Peacock, Andrae, {Bland-Hawthorn}, Brough,
  Brown, Cameron, Ching, Colless, Conselice, Croom, Cross, {de Propris}, Dye,
  Drinkwater, Ellis, Graham, Grootes, Gunawardhana, Jones, {van Kampen},
  Maraston, Nichol, Parkinson, Phillipps, Pimbblet, Popescu, Prescott,
  Roseboom, Sadler, Sansom, Sharp, Smith, Taylor, Thomas, Tuffs, Wijesinghe,
  Dunne, Frenk, Jarvis, Madore, Meyer, Seibert, {Staveley-Smith}, Sutherland,
  \& Warren}]{driver2011}
Driver, S.~P., Hill, D.~T., Kelvin, L.~S., {et~al.} 2011, Monthly Notices of
  the Royal Astronomical Society, 413, 971,
  \dodoi{10.1111/j.1365-2966.2010.18188.x}

\bibitem[{Driver {et~al.}(2022)Driver, Bellstedt, Robotham, Baldry, Davies,
  Liske, Obreschkow, Taylor, Wright, Alpaslan, Bamford, Bauer,
  {Bland-Hawthorn}, Bilicki, Bravo, Brough, Casura, Cluver, Colless, Conselice,
  Croom, {de Jong}, D'Eugenio, Propris, Dogruel, Drinkwater, Dvornik, Farrow,
  Frenk, Giblin, Graham, Grootes, Gunawardhana, Hashemizadeh, H{\"a}u{\ss}ler,
  Heymans, Hildebrandt, Holwerda, Hopkins, Jarrett, Jones, Kelvin, Koushan,
  Kuijken, {Lara-L{\'o}pez}, Lange, {L{\'o}pez-S{\'a}nchez}, Loveday, Mahajan,
  Meyer, Moffett, Napolitano, Norberg, Owers, Radovich, Raouf, Peacock,
  Phillipps, Pimbblet, Popescu, Said, Sansom, Seibert, Sutherland, Thorne,
  Tuffs, Turner, van~der Wel, van Kampen, \& Wilkins}]{driver2022}
Driver, S.~P., Bellstedt, S., Robotham, A. S.~G., {et~al.} 2022, Monthly
  Notices of the Royal Astronomical Society, stac472,
  \dodoi{10.1093/mnras/stac472}

\bibitem[{{Foreman-Mackey} {et~al.}(2014){Foreman-Mackey}, Hogg, \&
  Morton}]{foreman-mackey2014}
{Foreman-Mackey}, D., Hogg, D.~W., \& Morton, T.~D. 2014, The Astrophysical
  Journal, 795, 64, \dodoi{10.1088/0004-637X/795/1/64}

\bibitem[{{Franx} {et~al.}(2008){Franx}, {van Dokkum}, {F{\"o}rster Schreiber},
  {Wuyts}, {Labb{\'e}}, \& {Toft}}]{franx2008}
{Franx}, M., {van Dokkum}, P.~G., {F{\"o}rster Schreiber}, N.~M., {et~al.}
  2008, \apj, 688, 770, \dodoi{10.1086/592431}

\bibitem[{Genel {et~al.}(2014)Genel, Vogelsberger, Springel, Sijacki, Nelson,
  Snyder, {Rodriguez-Gomez}, Torrey, \& Hernquist}]{genel2014}
Genel, S., Vogelsberger, M., Springel, V., {et~al.} 2014, Monthly Notices of
  the Royal Astronomical Society, 445, 175, \dodoi{10.1093/mnras/stu1654}

\bibitem[{{Guy} {et~al.}(2023){Guy}, {Bailey}, {Kremin}, {Alam}, {Alexander},
  {Allende Prieto}, {BenZvi}, {Bolton}, {Brooks}, {Chaussidon}, {Cooper},
  {Dawson}, {de la Macorra}, {Dey}, {Dey}, {Dhungana}, {Eisenstein},
  {Font-Ribera}, {Forero-Romero}, {Gazta{\~n}aga}, {Gontcho A Gontcho},
  {Green}, {Honscheid}, {Ishak}, {Kehoe}, {Kirkby}, {Kisner}, {Koposov}, {Lan},
  {Landriau}, {Le Guillou}, {Levi}, {Magneville}, {Manser}, {Martini},
  {Meisner}, {Miquel}, {Moustakas}, {Myers}, {Newman}, {Nie},
  {Palanque-Delabrouille}, {Percival}, {Poppett}, {Prada}, {Raichoor},
  {Ravoux}, {Ross}, {Schlafly}, {Schlegel}, {Schubnell}, {Sharples},
  {Tarl{\'e}}, {Weaver}, {Y{\'e}che}, {Zhou}, {Zhou}, \& {Zou}}]{guy2023}
{Guy}, J., {Bailey}, S., {Kremin}, A., {et~al.} 2023, \aj, 165, 144,
  \dodoi{10.3847/1538-3881/acb212}

\bibitem[{Hahn {et~al.}(2017)Hahn, Tinker, \& Wetzel}]{hahn2017}
Hahn, C., Tinker, J.~L., \& Wetzel, A.~R. 2017, The Astrophysical Journal, 841,
  6, \dodoi{10.3847/1538-4357/aa6d6b}

\bibitem[{Hahn {et~al.}(2015)Hahn, Blanton, Moustakas, Coil, Cool, Eisenstein,
  Skibba, Wong, \& Zhu}]{hahn2015}
Hahn, C., Blanton, M.~R., Moustakas, J., {et~al.} 2015, The Astrophysical
  Journal, 806, 162, \dodoi{10.1088/0004-637X/806/2/162}

\bibitem[{Hahn {et~al.}(2019)Hahn, Starkenburg, Choi, Dav{\'e}, Dickey, Geha,
  Genel, Hayward, Maller, Mandyam, Pandya, Popping, Rafieferantsoa, Somerville,
  \& Tinker}]{hahn2019}
Hahn, C., Starkenburg, T.~K., Choi, E., {et~al.} 2019, The Astrophysical
  Journal, 872, 160, \dodoi{10.3847/1538-4357/aafedd}

\bibitem[{Hahn {et~al.}(2021)Hahn, Starkenburg, {Angl{\'e}s-Alc{\'a}zar}, Choi,
  Dav{\'e}, Dickey, Iyer, Maller, Somerville, Tinker, \& Yung}]{hahn2021}
Hahn, C., Starkenburg, T.~K., {Angl{\'e}s-Alc{\'a}zar}, D., {et~al.} 2021, {{IQ
  Collaboratory III}}: {{The Empirical Dust Attenuation Framework}} -- {{Taking
  Hydrodynamical Simulations}} with a {{Grain}} of {{Dust}}

\bibitem[{Hahn {et~al.}(2022{\natexlab{a}})Hahn, Kwon, Tojeiro, Siudek,
  Canning, Mezcua, Tinker, Brooks, Doel, Fanning, Gazta{\~n}aga, Kehoe,
  Landriau, Meisner, Moustakas, Poppett, Tarle, Weiner, \& Zou}]{hahn2022}
Hahn, C., Kwon, K.~J., Tojeiro, R., {et~al.} 2022{\natexlab{a}}, The {{DESI
  PRObabilistic Value-Added Bright Galaxy Survey}} ({{PROVABGS}}) {{Mock
  Challenge}}

\bibitem[{Hahn {et~al.}(2022{\natexlab{b}})Hahn, Wilson, {Ruiz-Macias}, Cole,
  Weinberg, Moustakas, Kremin, Tinker, Smith, Wechsler, Ahlen, Alam, Bailey,
  Brooks, Cooper, Davis, Dawson, Dey, Dey, Eftekharzadeh, Eisenstein, Fanning,
  {Forero-Romero}, Frenk, Gazta{\~n}aga, Gontcho, Guy, Honscheid, Ishak,
  Juneau, Kehoe, Kisner, Lan, Landriau, Le~Guillou, Levi, Magneville, Martini,
  Meisner, Myers, Nie, Norberg, {Palanque-Delabrouille}, Percival, Poppett,
  Prada, Raichoor, Ross, Safonova, Saulder, Schlafly, Schlegel, {Sierra-Porta},
  Tarle, Weaver, Y{\`e}che, Zarrouk, Zhou, Zhou, \& Zou}]{hahn2022c}
Hahn, C., Wilson, M.~J., {Ruiz-Macias}, O., {et~al.} 2022{\natexlab{b}}, {{DESI
  Bright Galaxy Survey}}: {{Final Target Selection}}, {{Design}}, and
  {{Validation}}

\bibitem[{Henriques {et~al.}(2009)Henriques, Thomas, Oliver, \&
  Roseboom}]{henriques2009}
Henriques, B. M.~B., Thomas, P.~A., Oliver, S., \& Roseboom, I. 2009, Monthly
  Notices of the Royal Astronomical Society, 396, 535,
  \dodoi{10.1111/j.1365-2966.2009.14730.x}

\bibitem[{Henriques {et~al.}(2015)Henriques, White, Thomas, Angulo, Guo,
  Lemson, Springel, \& Overzier}]{henriques2015}
Henriques, B. M.~B., White, S. D.~M., Thomas, P.~A., {et~al.} 2015, Monthly
  Notices of the Royal Astronomical Society, 451, 2663,
  \dodoi{10.1093/mnras/stv705}

\bibitem[{Hogg {et~al.}(2010)Hogg, Myers, \& Bovy}]{hogg2010}
Hogg, D.~W., Myers, A.~D., \& Bovy, J. 2010, The Astrophysical Journal, 725,
  2166, \dodoi{10.1088/0004-637X/725/2/2166}

\bibitem[{{Ilbert} {et~al.}(2013){Ilbert}, {McCracken}, {Le F{\`e}vre},
  {Capak}, {Dunlop}, {Karim}, {Renzini}, {Caputi}, {Boissier}, {Arnouts},
  {Aussel}, {Comparat}, {Guo}, {Hudelot}, {Kartaltepe}, {Kneib}, {Krogager},
  {Le Floc'h}, {Lilly}, {Mellier}, {Milvang-Jensen}, {Moutard}, {Onodera},
  {Richard}, {Salvato}, {Sanders}, {Scoville}, {Silverman}, {Taniguchi},
  {Tasca}, {Thomas}, {Toft}, {Tresse}, {Vergani}, {Wolk}, \&
  {Zirm}}]{ilbert2013}
{Ilbert}, O., {McCracken}, H.~J., {Le F{\`e}vre}, O., {et~al.} 2013, \aap, 556,
  A55, \dodoi{10.1051/0004-6361/201321100}

\bibitem[{Iovino {et~al.}(2010)Iovino, Cucciati, Scodeggio, Knobel, Kova{\v c},
  Lilly, Bolzonella, Tasca, Zamorani, Zucca, Caputi, Pozzetti, Oesch,
  Lamareille, Halliday, Bardelli, Finoguenov, Guzzo, Kampczyk, Maier, Tanaka,
  Vergani, Carollo, Contini, Kneib, Le~F{\`e}vre, Mainieri, Renzini, Bongiorno,
  Coppa, {de la Torre}, {de Ravel}, Franzetti, Garilli, Le~Borgne, Le~Brun,
  Mignoli, Pell{\`o}, Peng, {Perez-Montero}, Ricciardelli, Silverman, Tresse,
  Abbas, Bottini, Cappi, Cassata, Cimatti, Koekemoer, Leauthaud, Maccagni,
  Marinoni, McCracken, Memeo, Meneux, Porciani, Scaramella, Schiminovich, \&
  Scoville}]{iovino2010}
Iovino, A., Cucciati, O., Scodeggio, M., {et~al.} 2010, Astronomy and
  Astrophysics, 509, A40, \dodoi{10.1051/0004-6361/200912558}

\bibitem[{Jamieson {et~al.}(2022)Jamieson, Li, {Alves de Oliveira},
  {Villaescusa-Navarro}, Ho, \& Spergel}]{jamieson2022}
Jamieson, D., Li, Y., {Alves de Oliveira}, R., {et~al.} 2022, Field {{Level
  Neural Network Emulator}} for {{Cosmological N-body Simulations}}

\bibitem[{Karamanis \& Beutler(2020)}]{karamanis2020}
Karamanis, M., \& Beutler, F. 2020, arXiv e-prints, arXiv:2002.06212

\bibitem[{Kingma \& Ba(2017)}]{kingma2017}
Kingma, D.~P., \& Ba, J. 2017, arXiv:1412.6980 [cs].
\newblock \doeprint{1412.6980}

\bibitem[{{Kirkby et al.}(2023)}]{kirbky2023}
{Kirkby et al.} 2023

\bibitem[{Kwon {et~al.}(2022)Kwon, Hahn, \& Alsing}]{kwon2022}
Kwon, K.~J., Hahn, C., \& Alsing, J. 2022, Neural {{Stellar Population
  Synthesis Emulator}} for the {{DESI PROVABGS}}

\bibitem[{{Lan} {et~al.}(2023){Lan}, {Tojeiro}, {Armengaud}, {Prochaska},
  {Davis}, {Alexander}, {Raichoor}, {Zhou}, {Y{\`e}che}, {Balland}, {BenZvi},
  {Berti}, {Canning}, {Carr}, {Chittenden}, {Cole}, {Cousinou}, {Dawson},
  {Dey}, {Douglass}, {Edge}, {Escoffier}, {Glanville}, {A Gontcho}, {Guy},
  {Hahn}, {Howlett}, {Hwang}, {Jiang}, {Kov{\'a}cs}, {Mezcua}, {Moore},
  {Nadathur}, {Oh}, {Parkinson}, {Rocher}, {Ross}, {Ruhlmann-Kleider}, {Sabiu},
  {Said}, {Saulder}, {Sierra-Porta}, {Weiner}, {Yu}, {Zarrouk}, {Zhang}, {Zou},
  {Ahlen}, {Bailey}, {Brooks}, {Cooper}, {de la Macorra}, {Dey}, {Dhungana},
  {Doel}, {Eftekharzadeh}, {Fanning}, {Font-Ribera}, {Garrison},
  {Gazta{\~n}aga}, {Kehoe}, {Kisner}, {Kremin}, {Landriau}, {Le Guillou},
  {Levi}, {Magneville}, {Meisner}, {Miquel}, {Moustakas}, {Myers}, {Newman},
  {Nie}, {Palanque-Delabrouille}, {Percival}, {Poppett}, {Prada}, {Schubnell},
  {Tarl{\'e}}, {Weaver}, {Zhang}, \& {Zhou}}]{vi}
{Lan}, T.-W., {Tojeiro}, R., {Armengaud}, E., {et~al.} 2023, \apj, 943, 68,
  \dodoi{10.3847/1538-4357/aca5fa}

\bibitem[{Leja {et~al.}(2019)Leja, Speagle, Johnson, Conroy, {van Dokkum}, \&
  Franx}]{leja2019a}
Leja, J., Speagle, J.~S., Johnson, B.~D., {et~al.} 2019, arXiv,
  arXiv:1910.04168

\bibitem[{Lejeune {et~al.}(1997)Lejeune, Cuisinier, \& Buser}]{lejeune1997}
Lejeune, T., Cuisinier, F., \& Buser, R. 1997, A \&amp; A Supplement series,
  Vol. 125, October II 1997, p.229-246., 125, 229, \dodoi{10.1051/aas:1997373}

\bibitem[{Lejeune {et~al.}(1998)Lejeune, Cuisinier, \& Buser}]{lejeune1998}
---. 1998, Astronomy and Astrophysics Supplement, v.130, p.65-75, 130, 65,
  \dodoi{10.1051/aas:1998405}

\bibitem[{{Levi} {et~al.}(2013){Levi}, {Bebek}, {Beers}, {Blum}, {Cahn},
  {Eisenstein}, {Flaugher}, {Honscheid}, {Kron}, {Lahav}, {McDonald}, {Roe},
  {Schlegel}, \& {representing the DESI collaboration}}]{levi2013}
{Levi}, M., {Bebek}, C., {Beers}, T., {et~al.} 2013, arXiv e-prints,
  arXiv:1308.0847.
\newblock \doarXiv{1308.0847}

\bibitem[{Li \& White(2009)}]{li2009}
Li, C., \& White, S. D.~M. 2009, Monthly Notices of the Royal Astronomical
  Society, 398, 2177, \dodoi{10.1111/j.1365-2966.2009.15268.x}

\bibitem[{Lu {et~al.}(2014)Lu, Wechsler, Somerville, Croton, Porter, Primack,
  Behroozi, Ferguson, Koo, Guo, Safarzadeh, Finlator, Castellano, White,
  Sommariva, \& Moody}]{lu2014}
Lu, Y., Wechsler, R.~H., Somerville, R.~S., {et~al.} 2014, The Astrophysical
  Journal, 795, 123, \dodoi{10.1088/0004-637X/795/2/123}

\bibitem[{Malz \& Hogg(2020)}]{malz2020}
Malz, A.~I., \& Hogg, D.~W. 2020, How to Obtain the Redshift Distribution from
  Probabilistic Redshift Estimates

\bibitem[{{Mannucci} {et~al.}(2010){Mannucci}, {Cresci}, {Maiolino}, {Marconi},
  \& {Gnerucci}}]{mannucci2010}
{Mannucci}, F., {Cresci}, G., {Maiolino}, R., {Marconi}, A., \& {Gnerucci}, A.
  2010, \mnras, 408, 2115, \dodoi{10.1111/j.1365-2966.2010.17291.x}

\bibitem[{Marchesini {et~al.}(2009)Marchesini, {van Dokkum},
  F{\"o}rster~Schreiber, Franx, Labb{\'e}, \& Wuyts}]{marchesini2009}
Marchesini, D., {van Dokkum}, P.~G., F{\"o}rster~Schreiber, N.~M., {et~al.}
  2009, The Astrophysical Journal, 701, 1765,
  \dodoi{10.1088/0004-637X/701/2/1765}

\bibitem[{McLachlan \& Peel(2000)}]{mclachlan2000}
McLachlan, G., \& Peel, D. 2000, Finite {{Mixture Models}}
  ({Wiley-Interscience})

\bibitem[{{Miller} {et~al.}(2023){Miller}, {Doel}, {Gutierrez}, {Besuner},
  {Brooks}, {Gallo}, {Heetderks}, {Jelinsky}, {Kent}, {Lampton}, {Levi},
  {Liang}, {Meisner}, {Sholl}, {Silber}, {Sprayberry}, {Aguilar}, {de la
  Macorra}, {Eisenstein}, {Fanning}, {Font-Ribera}, {Gaztanaga}, {Gontcho},
  {Honscheid}, {Jimenez}, {Joyce}, {Kehoe}, {Kisner}, {Kremin}, {Landriau}, {Le
  Guillou}, {Magneville}, {Martini}, {Miquel}, {Moustakas}, {Nie}, {Percival},
  {Poppett}, {Prada}, {Rossi}, {Schlegel}, {Schubnell}, {Seo}, {Sharples},
  {Tarle}, {Vargas-Magana}, \& {Zhou}}]{miller2023}
{Miller}, T.~N., {Doel}, P., {Gutierrez}, G., {et~al.} 2023, arXiv e-prints,
  arXiv:2306.06310, \dodoi{10.48550/arXiv.2306.06310}

\bibitem[{Moustakas {et~al.}(2013)Moustakas, Coil, Aird, Blanton, Cool,
  Eisenstein, Mendez, Wong, Zhu, \& Arnouts}]{moustakas2013}
Moustakas, J., Coil, A.~L., Aird, J., {et~al.} 2013, The Astrophysical Journal,
  767, 50, \dodoi{10.1088/0004-637X/767/1/50}

\bibitem[{{Muzzin} {et~al.}(2013){Muzzin}, {Marchesini}, {Stefanon}, {Franx},
  {McCracken}, {Milvang-Jensen}, {Dunlop}, {Fynbo}, {Brammer}, {Labb{\'e}}, \&
  {van Dokkum}}]{muzzin2013}
{Muzzin}, A., {Marchesini}, D., {Stefanon}, M., {et~al.} 2013, \apj, 777, 18,
  \dodoi{10.1088/0004-637X/777/1/18}

\bibitem[{{Myers} {et~al.}(2023){Myers}, {Moustakas}, {Bailey}, {Weaver},
  {Cooper}, {Forero-Romero}, {Abolfathi}, {Alexander}, {Brooks}, {Chaussidon},
  {Chuang}, {Dawson}, {Dey}, {Dey}, {Dhungana}, {Doel}, {Fanning},
  {Gazta{\~n}aga}, {A Gontcho}, {Gonzalez-Morales}, {Hahn}, {Herrera-Alcantar},
  {Honscheid}, {Ishak}, {Karim}, {Kirkby}, {Kisner}, {Koposov}, {Kremin},
  {Lan}, {Landriau}, {Lang}, {Levi}, {Magneville}, {Napolitano}, {Martini},
  {Meisner}, {Newman}, {Palanque-Delabrouille}, {Percival}, {Poppett}, {Prada},
  {Raichoor}, {Ross}, {Schlafly}, {Schlegel}, {Schubnell}, {Tan}, {Tarle},
  {Wilson}, {Y{\`e}che}, {Zhou}, {Zhou}, \& {Zou}}]{myers2023}
{Myers}, A.~D., {Moustakas}, J., {Bailey}, S., {et~al.} 2023, \aj, 165, 50,
  \dodoi{10.3847/1538-3881/aca5f9}

\bibitem[{Nelson {et~al.}(2015)Nelson, Pillepich, Genel, Vogelsberger,
  Springel, Torrey, {Rodriguez-Gomez}, Sijacki, Snyder, Griffen, Marinacci,
  Blecha, Sales, Xu, \& Hernquist}]{nelson2015}
Nelson, D., Pillepich, A., Genel, S., {et~al.} 2015, Astronomy and Computing,
  13, 12, \dodoi{10.1016/j.ascom.2015.09.003}

\bibitem[{Noeske {et~al.}(2007)Noeske, Weiner, Faber, Papovich, Koo,
  Somerville, Bundy, Conselice, Newman, Schiminovich, Le~Floc'h, Coil, Rieke,
  Lotz, Primack, Barmby, Cooper, Davis, Ellis, Fazio, Guhathakurta, Huang,
  Kassin, Martin, Phillips, Rich, Small, Willmer, \& Wilson}]{noeske2007}
Noeske, K.~G., Weiner, B.~J., Faber, S.~M., {et~al.} 2007, The Astrophysical
  Journal Letters, 660, L43, \dodoi{10.1086/517926}

\bibitem[{Norberg {et~al.}(2009)Norberg, Baugh, Gazta{\~n}aga, \&
  Croton}]{norberg2009}
Norberg, P., Baugh, C.~M., Gazta{\~n}aga, E., \& Croton, D.~J. 2009, Monthly
  Notices of the Royal Astronomical Society, 396, 19,
  \dodoi{10.1111/j.1365-2966.2009.14389.x}

\bibitem[{Pacifici {et~al.}(2023)Pacifici, Iyer, Mobasher, {da Cunha},
  Acquaviva, Burgarella, Calistro~Rivera, Carnall, Chang, Chartab, Cooke,
  Fairhurst, Kartaltepe, Leja, Ma{\l}ek, Salmon, Torelli, {Vidal-Garc{\'i}a},
  Boquien, Brammer, Brown, Capak, Chevallard, Circosta, Croton, Davidzon,
  Dickinson, Duncan, Faber, Ferguson, Fontana, Guo, Haeussler, Hemmati,
  Jafariyazani, Kassin, Larson, Lee, Mantha, Marchi, Nayyeri, Newman, Pandya,
  Pforr, Reddy, Sanders, Shah, Shahidi, Stevans, Triani, Tyler, Vanderhoof, {de
  la Vega}, Wang, \& Weston}]{pacifici2023}
Pacifici, C., Iyer, K.~G., Mobasher, B., {et~al.} 2023, The Astrophysical
  Journal, 944, 141, \dodoi{10.3847/1538-4357/acacff}

\bibitem[{Paxton {et~al.}(2011)Paxton, Bildsten, Dotter, Herwig, Lesaffre, \&
  Timmes}]{paxton2011}
Paxton, B., Bildsten, L., Dotter, A., {et~al.} 2011, The Astrophysical Journal
  Supplement Series, 192, 3, \dodoi{10.1088/0067-0049/192/1/3}

\bibitem[{Paxton {et~al.}(2013)Paxton, Cantiello, Arras, Bildsten, Brown,
  Dotter, Mankovich, Montgomery, Stello, Timmes, \& Townsend}]{paxton2013}
Paxton, B., Cantiello, M., Arras, P., {et~al.} 2013, The Astrophysical Journal
  Supplement Series, 208, 4, \dodoi{10.1088/0067-0049/208/1/4}

\bibitem[{Paxton {et~al.}(2015)Paxton, Marchant, Schwab, Bauer, Bildsten,
  Cantiello, Dessart, Farmer, Hu, Langer, Townsend, Townsley, \&
  Timmes}]{paxton2015}
Paxton, B., Marchant, P., Schwab, J., {et~al.} 2015, The Astrophysical Journal
  Supplement Series, 220, 15, \dodoi{10.1088/0067-0049/220/1/15}

\bibitem[{Peng {et~al.}(2010)Peng, Lilly, Kova{\v c}, Bolzonella, Pozzetti,
  Renzini, Zamorani, Ilbert, Knobel, Iovino, Maier, Cucciati, Tasca, Carollo,
  Silverman, Kampczyk, {de Ravel}, Sanders, Scoville, Contini, Mainieri,
  Scodeggio, Kneib, Le~F{\`e}vre, Bardelli, Bongiorno, Caputi, Coppa, {de la
  Torre}, Franzetti, Garilli, Lamareille, Le~Borgne, Le~Brun, Mignoli,
  Perez~Montero, Pello, Ricciardelli, Tanaka, Tresse, Vergani, Welikala, Zucca,
  Oesch, Abbas, Barnes, Bordoloi, Bottini, Cappi, Cassata, Cimatti, Fumana,
  Hasinger, Koekemoer, Leauthaud, Maccagni, Marinoni, McCracken, Memeo, Meneux,
  Nair, Porciani, Presotto, \& Scaramella}]{peng2010}
Peng, Y.-j., Lilly, S.~J., Kova{\v c}, K., {et~al.} 2010, The Astrophysical
  Journal, 721, 193, \dodoi{10.1088/0004-637X/721/1/193}

\bibitem[{{Planck Collaboration} {et~al.}(2014){Planck Collaboration}, Ade,
  Aghanim, {Armitage-Caplan}, Arnaud, Ashdown, {Atrio-Barandela}, Aumont,
  Baccigalupi, Banday, Barreiro, Bartlett, Battaner, Benabed, Beno{\^i}t,
  {Benoit-L{\'e}vy}, Bernard, Bersanelli, Bielewicz, Bobin, Bock, Bonaldi,
  Bond, Borrill, Bouchet, Bridges, Bucher, Burigana, Butler, Calabrese,
  Cappellini, Cardoso, Catalano, Challinor, Chamballu, Chary, Chen, Chiang,
  Chiang, Christensen, Church, Clements, Colombi, Colombo, Couchot, Coulais,
  Crill, Curto, Cuttaia, Danese, Davies, Davis, {de Bernardis}, {de Rosa}, {de
  Zotti}, Delabrouille, Delouis, D{\'e}sert, Dickinson, Diego, Dolag, Dole,
  Donzelli, Dor{\'e}, Douspis, Dunkley, Dupac, Efstathiou, Elsner, En{\ss}lin,
  Eriksen, Finelli, Forni, Frailis, Fraisse, Franceschi, Gaier, Galeotta,
  Galli, Ganga, Giard, Giardino, {Giraud-H{\'e}raud}, Gjerl{\o}w,
  {Gonz{\'a}lez-Nuevo}, G{\'o}rski, Gratton, Gregorio, Gruppuso, Gudmundsson,
  Haissinski, Hamann, Hansen, Hanson, Harrison, {Henrot-Versill{\'e}},
  {Hern{\'a}ndez-Monteagudo}, Herranz, Hildebrandt, Hivon, Hobson, Holmes,
  Hornstrup, Hou, Hovest, Huffenberger, Jaffe, Jaffe, Jewell, Jones, Juvela,
  Keih{\"a}nen, Keskitalo, Kisner, Kneissl, Knoche, Knox, Kunz, {Kurki-Suonio},
  Lagache, L{\"a}hteenm{\"a}ki, Lamarre, Lasenby, Lattanzi, Laureijs, Lawrence,
  Leach, Leahy, Leonardi, {Le{\'o}n-Tavares}, Lesgourgues, Lewis, Liguori,
  Lilje, {Linden-V{\o}rnle}, {L{\'o}pez-Caniego}, Lubin,
  {Mac{\'i}as-P{\'e}rez}, Maffei, Maino, Mandolesi, Maris, Marshall, Martin,
  {Mart{\'i}nez-Gonz{\'a}lez}, Masi, Massardi, Matarrese, Matthai, Mazzotta,
  Meinhold, Melchiorri, Melin, Mendes, Menegoni, Mennella, Migliaccio, Millea,
  Mitra, {Miville-Desch{\^e}nes}, Moneti, Montier, Morgante, Mortlock, Moss,
  Munshi, Murphy, Naselsky, Nati, Natoli, Netterfield, {N{\o}rgaard-Nielsen},
  Noviello, Novikov, Novikov, O'Dwyer, Osborne, Oxborrow, Paci, Pagano, Pajot,
  Paladini, Paoletti, Partridge, Pasian, Patanchon, Pearson, Pearson, Peiris,
  Perdereau, Perotto, Perrotta, Pettorino, Piacentini, Piat, Pierpaoli,
  Pietrobon, Plaszczynski, Platania, Pointecouteau, Polenta, Ponthieu, Popa,
  Poutanen, Pratt, Pr{\'e}zeau, Prunet, Puget, Rachen, Reach, Rebolo, Reinecke,
  Remazeilles, Renault, Ricciardi, Riller, Ristorcelli, Rocha, Rosset, Roudier,
  {Rowan-Robinson}, {Rubi{\~n}o-Mart{\'i}n}, Rusholme, Sandri, Santos,
  Savelainen, Savini, Scott, Seiffert, Shellard, Spencer, Starck, Stolyarov,
  Stompor, Sudiwala, Sunyaev, Sureau, Sutton, {Suur-Uski}, Sygnet, Tauber,
  Tavagnacco, Terenzi, Toffolatti, Tomasi, Tristram, Tucci, Tuovinen,
  T{\"u}rler, Umana, Valenziano, Valiviita, Van~Tent, Vielva, Villa, Vittorio,
  Wade, Wandelt, Wehus, White, White, Wilkinson, Yvon, Zacchei, \&
  Zonca}]{planckcollaboration2014a}
{Planck Collaboration}, Ade, P. a.~R., Aghanim, N., {et~al.} 2014, Astronomy
  \&amp; Astrophysics, Volume 571, id.A16,
  {$<$}NUMPAGES{$>$}66{$<$}/NUMPAGES{$>$} pp., 571, A16,
  \dodoi{10.1051/0004-6361/201321591}

\bibitem[{{Pozzetti} {et~al.}(2010){Pozzetti}, {Bolzonella}, {Zucca},
  {Zamorani}, {Lilly}, {Renzini}, {Moresco}, {Mignoli}, {Cassata}, {Tasca},
  {Lamareille}, {Maier}, {Meneux}, {Halliday}, {Oesch}, {Vergani}, {Caputi},
  {Kova{\v{c}}}, {Cimatti}, {Cucciati}, {Iovino}, {Peng}, {Carollo}, {Contini},
  {Kneib}, {Le F{\'e}vre}, {Mainieri}, {Scodeggio}, {Bardelli}, {Bongiorno},
  {Coppa}, {de la Torre}, {de Ravel}, {Franzetti}, {Garilli}, {Kampczyk},
  {Knobel}, {Le Borgne}, {Le Brun}, {Pell{\`o}}, {Perez Montero},
  {Ricciardelli}, {Silverman}, {Tanaka}, {Tresse}, {Abbas}, {Bottini}, {Cappi},
  {Guzzo}, {Koekemoer}, {Leauthaud}, {Maccagni}, {Marinoni}, {McCracken},
  {Memeo}, {Porciani}, {Scaramella}, {Scarlata}, \& {Scoville}}]{pozzetii2010}
{Pozzetti}, L., {Bolzonella}, M., {Zucca}, E., {et~al.} 2010, \aap, 523, A13,
  \dodoi{10.1051/0004-6361/200913020}

\bibitem[{Press {et~al.}(1992)Press, Teukolsky, Vetterling, \&
  Flannery}]{press1992}
Press, W.~H., Teukolsky, S.~A., Vetterling, W.~T., \& Flannery, B.~P. 1992,
  Numerical {{Recipes}} in {{C}} ({{2Nd Ed}}.): {{The Art}} of {{Scientific
  Computing}} ({New York, NY, USA}: {Cambridge University Press})

\bibitem[{{Raichoor} {et~al.}(2020){Raichoor}, {Eisenstein}, {Karim}, {Newman},
  {Moustakas}, {Brooks}, {Dawson}, {Dey}, {Duan}, {Eftekharzadeh},
  {Gazta{\~n}aga}, {Kehoe}, {Landriau}, {Lang}, {Lee}, {Levi}, {Meisner},
  {Myers}, {Palanque-Delabrouille}, {Poppett}, {Prada}, {Ross}, {Schlegel},
  {Schubnell}, {Staten}, {Tarl{\'e}}, {Tojeiro}, {Y{\`e}che}, \&
  {Zhou}}]{raichoor2020}
{Raichoor}, A., {Eisenstein}, D.~J., {Karim}, T., {et~al.} 2020, Research Notes
  of the American Astronomical Society, 4, 180,
  \dodoi{10.3847/2515-5172/abc078}

\bibitem[{{Raichoor} {et~al.}(2023){Raichoor}, {Moustakas}, {Newman}, {Karim},
  {Ahlen}, {Alam}, {Bailey}, {Brooks}, {Dawson}, {de la Macorra}, {de Mattia},
  {Dey}, {Dey}, {Dhungana}, {Eftekharzadeh}, {Eisenstein}, {Fanning},
  {Font-Ribera}, {Garc{\'\i}a-Bellido}, {Gazta{\~n}aga}, {A Gontcho}, {Guy},
  {Honscheid}, {Ishak}, {Kehoe}, {Kisner}, {Kremin}, {Lan}, {Landriau}, {Le
  Guillou}, {Levi}, {Magneville}, {Manera}, {Martini}, {Meisner}, {Myers},
  {Nie}, {Palanque-Delabrouille}, {Percival}, {Poppett}, {Prada}, {Ross},
  {Ruhlmann-Kleider}, {Sabiu}, {Schlafly}, {Schlegel}, {Tarl{\'e}}, {Weaver},
  {Y{\`e}che}, {Zhou}, {Zhou}, \& {Zou}}]{raichoor2023}
{Raichoor}, A., {Moustakas}, J., {Newman}, J.~A., {et~al.} 2023, \aj, 165, 126,
  \dodoi{10.3847/1538-3881/acb213}

\bibitem[{{Raichoor et al.}(2023)}]{fba}
{Raichoor et al.} 2023

\bibitem[{{Ruiz-Macias} {et~al.}(2020){Ruiz-Macias}, {Zarrouk}, {Cole},
  {Norberg}, {Baugh}, {Brooks}, {Dey}, {Duan}, {Eftekharzadeh}, {Eisenstein},
  {Forero-Romero}, {Gazta{\~n}aga}, {Hahn}, {Kehoe}, {Landriau}, {Lang},
  {Levi}, {Lucey}, {Meisner}, {Moustakas}, {Myers}, {Palanque-Delabrouille},
  {Poppett}, {Prada}, {Raichoor}, {Schlegel}, {Schubnell}, {Tarl{\'e}},
  {Weinberg}, {Wilson}, \& {Y{\`e}che}}]{ruiz2020}
{Ruiz-Macias}, O., {Zarrouk}, P., {Cole}, S., {et~al.} 2020, Research Notes of
  the American Astronomical Society, 4, 187, \dodoi{10.3847/2515-5172/abc25a}

\bibitem[{{Saintonge} \& {Catinella}(2022)}]{saintonge2022}
{Saintonge}, A., \& {Catinella}, B. 2022, \araa, 60, 319,
  \dodoi{10.1146/annurev-astro-021022-043545}

\bibitem[{Salim {et~al.}(2007)Salim, Rich, Charlot, Brinchmann, Johnson,
  Schiminovich, Seibert, Mallery, Heckman, Forster, Friedman, Martin,
  Morrissey, Neff, Small, Wyder, Bianchi, Donas, Lee, Madore, Milliard, Szalay,
  Welsh, \& Yi}]{salim2007}
Salim, S., Rich, R.~M., Charlot, S., {et~al.} 2007, The Astrophysical Journal
  Supplement Series, 173, 267, \dodoi{10.1086/519218}

\bibitem[{{S{\'a}nchez-Bl{\'a}zquez} {et~al.}(2006){S{\'a}nchez-Bl{\'a}zquez},
  Peletier, {Jim{\'e}nez-Vicente}, Cardiel, Cenarro, {Falc{\'o}n-Barroso},
  Gorgas, Selam, \& Vazdekis}]{sanchez-blazquez2006}
{S{\'a}nchez-Bl{\'a}zquez}, P., Peletier, R.~F., {Jim{\'e}nez-Vicente}, J.,
  {et~al.} 2006, Monthly Notices of the Royal Astronomical Society, 371, 703,
  \dodoi{10.1111/j.1365-2966.2006.10699.x}

\bibitem[{{Santini} {et~al.}(2022){Santini}, {Castellano}, {Fontana},
  {Fortuni}, {Menci}, {Merlin}, {Pagul}, {Testa}, {Calabr{\`o}}, {Paris}, \&
  {Pentericci}}]{santini2022}
{Santini}, P., {Castellano}, M., {Fontana}, A., {et~al.} 2022, \apj, 940, 135,
  \dodoi{10.3847/1538-4357/ac9a48}

\bibitem[{{Schlafly} {et~al.}(2023){Schlafly}, {Kirkby}, {Schlegel}, {Myers},
  {Raichoor}, {Dawson}, {Aguilar}, {Allende Prieto}, {Bailey}, {BenZvi},
  {Bermejo-Climent}, {Brooks}, {de la Macorra}, {Dey}, {Doel}, {Fanning},
  {Font-Ribera}, {Forero-Romero}, {Garc{\'\i}a-Bellido}, {Gontcho}, {Guy},
  {Hahn}, {Honscheid}, {Ishak}, {Juneau}, {Kehoe}, {Kisner}, {Kremin},
  {Landriau}, {Lang}, {Lasker}, {Levi}, {Magneville}, {Manser}, {Martini},
  {Meisner}, {Miquel}, {Moustakas}, {Newman}, {Nie}, {Palanque-Delabrouille},
  {Percival}, {Poppett}, {Rockosi}, {Ross}, {Rossi}, {Tarl{\'e}}, {Weaver},
  {Y{\`e}che}, \& {Zhou}}]{ops}
{Schlafly}, E.~F., {Kirkby}, D., {Schlegel}, D.~J., {et~al.} 2023, arXiv
  e-prints, arXiv:2306.06309, \dodoi{10.48550/arXiv.2306.06309}

\bibitem[{{Schlegel et al.}(2023)}]{schlegel2023}
{Schlegel et al.} 2023

\bibitem[{{Schubnell} {et~al.}(2016){Schubnell}, {Ameel}, {Besuner},
  {Gershkovich}, {Heetderks}, {Hoerler}, {Kneib}, {Heetderks}, {Silber},
  {Tarl{\'e}}, \& {Weaverdyck}}]{schubnell2016}
{Schubnell}, M., {Ameel}, J., {Besuner}, R.~W., {et~al.} 2016, in Society of
  Photo-Optical Instrumentation Engineers (SPIE) Conference Series, Vol. 9908,
  Ground-based and Airborne Instrumentation for Astronomy VI, ed. C.~J.
  {Evans}, L.~{Simard}, \& H.~{Takami}, 990892, \dodoi{10.1117/12.2233370}

\bibitem[{{Silber} {et~al.}(2023){Silber}, {Fagrelius}, {Fanning}, {Schubnell},
  {Aguilar}, {Ahlen}, {Ameel}, {Ballester}, {Baltay}, {Bebek}, {Benton Beard},
  {Besuner}, {Cardiel-Sas}, {Casas}, {Castander}, {Claybaugh}, {Dobson},
  {Duan}, {Dunlop}, {Edelstein}, {Emmet}, {Elliott}, {Evatt}, {Gershkovich},
  {Guy}, {Harris}, {Heetderks}, {Heetderks}, {Honscheid}, {Illa}, {Jelinsky},
  {Jelinsky}, {Jimenez}, {Karcher}, {Kent}, {Kirkby}, {Kneib}, {Lambert},
  {Lampton}, {Leitner}, {Levi}, {McCauley}, {Meisner}, {Miller}, {Miquel},
  {Mundet}, {Poppett}, {Rabinowitz}, {Reil}, {Roman}, {Schlegel}, {Serrano},
  {Van Shourt}, {Sprayberry}, {Tarl{\'e}}, {Tie}, {Weaverdyck}, {Zhang},
  {Azzaro}, {Bailey}, {Becerril}, {Blackwell}, {Bouri}, {Brooks},
  {Buckley-Geer}, {Castro}, {Derwent}, {Dey}, {Dhungana}, {Doel}, {Eisenstein},
  {Fahim}, {Garcia-Bellido}, {Gazta{\~n}aga}, {A Gontcho}, {Gutierrez},
  {H{\"o}rler}, {Kehoe}, {Kisner}, {Kremin}, {Kronig}, {Landriau}, {Le
  Guillou}, {Martini}, {Moustakas}, {Palanque-Delabrouille}, {Peng},
  {Percival}, {Prada}, {Allende Prieto}, {de Rivera}, {Sanchez}, {Sanchez},
  {Sharples}, {Soares-Santos}, {Schlafly}, {Weaver}, {Zhou}, {Zhu}, {Zou}, \&
  {DESI Collaboration}}]{silber2023}
{Silber}, J.~H., {Fagrelius}, P., {Fanning}, K., {et~al.} 2023, \aj, 165, 9,
  \dodoi{10.3847/1538-3881/ac9ab1}

\bibitem[{{Smith} {et~al.}(2019){Smith}, {He}, {Cole}, {Stothert}, {Norberg},
  {Baugh}, {Bianchi}, {Wilson}, {Brooks}, {Forero-Romero}, {Moustakas},
  {Percival}, {Tarle}, \& {Wechsler}}]{smith2019}
{Smith}, A., {He}, J.-h., {Cole}, S., {et~al.} 2019, \mnras, 484, 1285,
  \dodoi{10.1093/mnras/stz059}

\bibitem[{Somerville \& Dav{\'e}(2015)}]{somerville2015a}
Somerville, R.~S., \& Dav{\'e}, R. 2015, Annual Review of Astronomy and
  Astrophysics, 53, 51, \dodoi{10.1146/annurev-astro-082812-140951}

\bibitem[{Speagle {et~al.}(2014)Speagle, Steinhardt, Capak, \&
  Silverman}]{speagle2014}
Speagle, J.~S., Steinhardt, C.~L., Capak, P.~L., \& Silverman, J.~D. 2014, The
  Astrophysical Journal Supplement Series, 214, 15,
  \dodoi{10.1088/0067-0049/214/2/15}

\bibitem[{{Tacconi} {et~al.}(2020){Tacconi}, {Genzel}, \&
  {Sternberg}}]{tacconi2020}
{Tacconi}, L.~J., {Genzel}, R., \& {Sternberg}, A. 2020, \araa, 58, 157,
  \dodoi{10.1146/annurev-astro-082812-141034}

\bibitem[{Tinker {et~al.}(2017)Tinker, Wetzel, Conroy, \& Mao}]{tinker2017}
Tinker, J.~L., Wetzel, A.~R., Conroy, C., \& Mao, Y.-Y. 2017, Monthly Notices
  of the Royal Astronomical Society, 472, 2504, \dodoi{10.1093/mnras/stx2066}

\bibitem[{Trayford {et~al.}(2017)Trayford, Camps, Theuns, Baes, Bower, Crain,
  Gunawardhana, Schaller, Schaye, \& Frenk}]{trayford2017}
Trayford, J.~W., Camps, P., Theuns, T., {et~al.} 2017, Monthly Notices of the
  Royal Astronomical Society, 470, 771, \dodoi{10.1093/mnras/stx1051}

\bibitem[{{Tremonti} {et~al.}(2004){Tremonti}, {Heckman}, {Kauffmann},
  {Brinchmann}, {Charlot}, {White}, {Seibert}, {Peng}, {Schlegel}, {Uomoto},
  {Fukugita}, \& {Brinkmann}}]{tremonti2004}
{Tremonti}, C.~A., {Heckman}, T.~M., {Kauffmann}, G., {et~al.} 2004, \apj, 613,
  898, \dodoi{10.1086/423264}

\bibitem[{{van der Wel} {et~al.}(2014){van der Wel}, {Franx}, {van Dokkum},
  {Skelton}, {Momcheva}, {Whitaker}, {Brammer}, {Bell}, {Rix}, {Wuyts},
  {Ferguson}, {Holden}, {Barro}, {Koekemoer}, {Chang}, {McGrath},
  {H{\"a}ussler}, {Dekel}, {Behroozi}, {Fumagalli}, {Leja}, {Lundgren},
  {Maseda}, {Nelson}, {Wake}, {Patel}, {Labb{\'e}}, {Faber}, {Grogin}, \&
  {Kocevski}}]{vanderwel2014}
{van der Wel}, A., {Franx}, M., {van Dokkum}, P.~G., {et~al.} 2014, \apj, 788,
  28, \dodoi{10.1088/0004-637X/788/1/28}

\bibitem[{{Villaescusa-Navarro} {et~al.}(2022){Villaescusa-Navarro}, Genel,
  {Angl{\'e}s-Alc{\'a}zar}, Perez, {Villanueva-Domingo}, Wadekar, Shao,
  Mohammad, Hassan, Moser, Lau, Valle, Nicola, Thiele, Jo, Philcox,
  Oppenheimer, Tillman, Hahn, Kaushal, Pisani, Gebhardt, Delgado, Caliendo,
  Kreisch, Wong, Coulton, Eickenberg, Parimbelli, Ni, Steinwandel, La~Torre,
  Dave, Battaglia, Nagai, Spergel, Hernquist, Burkhart, Narayanan, Wandelt,
  Somerville, Bryan, Viel, Li, Irsic, Kraljic, \&
  Vogelsberger}]{villaescusa-navarro2022a}
{Villaescusa-Navarro}, F., Genel, S., {Angl{\'e}s-Alc{\'a}zar}, D., {et~al.}
  2022, arXiv:2201.01300 [astro-ph].
\newblock \doeprint{2201.01300}

\bibitem[{Vogelsberger {et~al.}(2014)Vogelsberger, Genel, Springel, Torrey,
  Sijacki, Xu, Snyder, Nelson, \& Hernquist}]{vogelsberger2014}
Vogelsberger, M., Genel, S., Springel, V., {et~al.} 2014, Monthly Notices of
  the Royal Astronomical Society, 444, 1518, \dodoi{10.1093/mnras/stu1536}

\bibitem[{Westera {et~al.}(2002)Westera, Lejeune, Buser, Cuisinier, \&
  Bruzual}]{westera2002}
Westera, P., Lejeune, T., Buser, R., Cuisinier, F., \& Bruzual, G. 2002,
  Astronomy and Astrophysics, 381, 524, \dodoi{10.1051/0004-6361:20011493}

\bibitem[{Wetzel {et~al.}(2013)Wetzel, Tinker, Conroy, \& {van den
  Bosch}}]{wetzel2013}
Wetzel, A.~R., Tinker, J.~L., Conroy, C., \& {van den Bosch}, F.~C. 2013,
  Monthly Notices of the Royal Astronomical Society, 432, 336,
  \dodoi{10.1093/mnras/stt469}

\bibitem[{{Y{\`e}che} {et~al.}(2020){Y{\`e}che}, {Palanque-Delabrouille},
  {Claveau}, {Brooks}, {Chaussidon}, {Davis}, {Dawson}, {Dey}, {Duan},
  {Eftekharzadeh}, {Eisenstein}, {Gazta{\~n}aga}, {Kehoe}, {Landriau}, {Lang},
  {Levi}, {Meisner}, {Myers}, {Newman}, {Poppett}, {Prada}, {Raichoor},
  {Schlegel}, {Schubnell}, {Staten}, {Tarl{\'e}}, \& {Zhou}}]{yeche2020}
{Y{\`e}che}, C., {Palanque-Delabrouille}, N., {Claveau}, C.-A., {et~al.} 2020,
  Research Notes of the American Astronomical Society, 4, 179,
  \dodoi{10.3847/2515-5172/abc01a}

\bibitem[{York {et~al.}(2000)York, Adelman, Anderson, Anderson, Annis, Bahcall,
  Bakken, Barkhouser, Bastian, Berman, Boroski, Bracker, Briegel, Briggs,
  Brinkmann, Brunner, Burles, Carey, Carr, Castander, Chen, Colestock,
  Connolly, Crocker, Csabai, Czarapata, Davis, Doi, Dombeck, Eisenstein,
  Ellman, Elms, Evans, Fan, Federwitz, Fiscelli, Friedman, Frieman, Fukugita,
  Gillespie, Gunn, Gurbani, {de Haas}, Haldeman, Harris, Hayes, Heckman,
  Hennessy, Hindsley, Holm, Holmgren, Huang, Hull, Husby, Ichikawa, Ichikawa,
  Ivezi{\'c}, Kent, Kim, Kinney, Klaene, Kleinman, Kleinman, Knapp, Korienek,
  Kron, Kunszt, Lamb, Lee, Leger, Limmongkol, Lindenmeyer, Long, Loomis,
  Loveday, Lucinio, Lupton, MacKinnon, Mannery, Mantsch, Margon, McGehee,
  McKay, Meiksin, Merelli, Monet, Munn, Narayanan, Nash, Neilsen, Neswold,
  Newberg, Nichol, Nicinski, Nonino, Okada, Okamura, Ostriker, Owen, Pauls,
  Peoples, Peterson, Petravick, Pier, Pope, Pordes, Prosapio, Rechenmacher,
  Quinn, Richards, Richmond, Rivetta, Rockosi, Ruthmansdorfer, Sandford,
  Schlegel, Schneider, Sekiguchi, Sergey, Shimasaku, Siegmund, Smee, Smith,
  Snedden, Stone, Stoughton, Strauss, Stubbs, SubbaRao, Szalay, Szapudi,
  Szokoly, Thakar, Tremonti, Tucker, Uomoto, Vanden~Berk, Vogeley, Waddell,
  Wang, Watanabe, Weinberg, Yanny, Yasuda, \& {SDSS Collaboration}}]{york2000}
York, D.~G., Adelman, J., Anderson, Jr., J.~E., {et~al.} 2000, The Astronomical
  Journal, 120, 1579, \dodoi{10.1086/301513}

\bibitem[{{Zhou} {et~al.}(2020){Zhou}, {Newman}, {Dawson}, {Eisenstein},
  {Brooks}, {Dey}, {Dey}, {Duan}, {Eftekharzadeh}, {Gazta{\~n}aga}, {Kehoe},
  {Landriau}, {Levi}, {Licquia}, {Meisner}, {Moustakas}, {Myers},
  {Palanque-Delabrouille}, {Poppett}, {Prada}, {Raichoor}, {Schlegel},
  {Schubnell}, {Staten}, {Tarl{\'e}}, \& {Y{\`e}che}}]{zhou2020}
{Zhou}, R., {Newman}, J.~A., {Dawson}, K.~S., {et~al.} 2020, Research Notes of
  the American Astronomical Society, 4, 181, \dodoi{10.3847/2515-5172/abc0f4}

\bibitem[{{Zhou} {et~al.}(2023){Zhou}, {Dey}, {Newman}, {Eisenstein}, {Dawson},
  {Bailey}, {Berti}, {Guy}, {Lan}, {Zou}, {Aguilar}, {Ahlen}, {Alam}, {Brooks},
  {de la Macorra}, {Dey}, {Dhungana}, {Fanning}, {Font-Ribera}, {Gontcho},
  {Honscheid}, {Ishak}, {Kisner}, {Kov{\'a}cs}, {Kremin}, {Landriau}, {Levi},
  {Magneville}, {Manera}, {Martini}, {Meisner}, {Miquel}, {Moustakas}, {Myers},
  {Nie}, {Palanque-Delabrouille}, {Percival}, {Poppett}, {Prada}, {Raichoor},
  {Ross}, {Schlafly}, {Schlegel}, {Schubnell}, {Tarl{\'e}}, {Weaver},
  {Wechsler}, {Y{\'e}che}, \& {Zhou}}]{zhou2023}
{Zhou}, R., {Dey}, B., {Newman}, J.~A., {et~al.} 2023, \aj, 165, 58,
  \dodoi{10.3847/1538-3881/aca5fb}

\bibitem[{{Zou} {et~al.}(2017){Zou}, {Zhou}, {Fan}, {Zhang}, {Zhou}, {Nie},
  {Peng}, {McGreer}, {Jiang}, {Dey}, {Fan}, {He}, {Jiang}, {Lang}, {Lesser},
  {Ma}, {Mao}, {Schlegel}, \& {Wang}}]{zou2017}
{Zou}, H., {Zhou}, X., {Fan}, X., {et~al.} 2017, \pasp, 129, 064101,
  \dodoi{10.1088/1538-3873/aa65ba}

\end{thebibliography}
\end{document}